\documentclass{article}

\usepackage[preprint]{neurips_2023}

\newtheorem{definition}{Definition}

\usepackage[utf8]{inputenc} 
\usepackage[T1]{fontenc}    
\usepackage{hyperref}       
\usepackage{url}            
\usepackage{booktabs}       
\usepackage{amsfonts}       
\usepackage{nicefrac}       
\usepackage{microtype}      
\usepackage{xcolor}         
\usepackage{amsmath}
\usepackage{stmaryrd}
\usepackage{amssymb}
\usepackage{graphicx}
\usepackage{bbm}
\usepackage{ragged2e}

\def\argmax{\mathop{\textstyle{\textrm{argmax}}}}

\title{Multi-agent Reinforcement Learning: A Comprehensive Survey}

\usepackage{authblk}

\author[1]{Dom Huh}
\author[1,3]{Prasant Mohapatra}
\affil[1]{University of California, Davis \authorcr
  \tt \{dhuh, pmohapatra\}@ucdavis.edu}
\affil[3]{University of South Florida}

\begin{document}

\maketitle

\begin{abstract}
Multi-agent systems (MAS) are widely prevalent and crucially important in numerous real-world applications, where multiple agents must make decisions to achieve their objectives in a shared environment. Despite their ubiquity, the development of intelligent decision-making agents in MAS poses several open challenges to their effective implementation. This survey examines these challenges, placing an emphasis on studying seminal concepts from game theory (GT) and machine learning (ML) and connecting them to recent advancements in multi-agent reinforcement learning (MARL), i.e. the research of data-driven decision-making within MAS. Therefore, the objective of this survey is to provide a comprehensive perspective along the various dimensions of MARL, shedding light on the unique opportunities that are presented in MARL applications while highlighting the inherent challenges that accompany this potential. Therefore, we hope that our work will not only contribute to the field by analyzing the current landscape of MARL but also motivate future directions with insights for deeper integration of concepts from related domains of GT and ML. With this in mind, this work delves into a detailed exploration of recent and past efforts of MARL and its related fields and describes prior solutions that were proposed and their limitations, as well as their applications.
\end{abstract}

\section{Introduction}
\label{sec:introduction}
Multi-agent reinforcement learning (MARL) has long been recognized as a pivotal domain in artificial intelligence (AI), promising dynamic solutions for complex tasks within multi-agent systems (MAS) that involve multiple goal-oriented decision-making, i.e. control, agents. The importance of devising such solutions is evident, as it enables the realization of a wide array of real-world applications, where the consideration of the existence of other intelligent agents is required. The process of developing these agents is largely centered around facilitating the emergence of not only decision-making abilities but also an adept understanding of social dynamics in a data-driven manner. Hence, with proper modeling and learning methods, these agents strive to leverage the multi-agent nature of their shared environment to achieve their individual and collective goals.

Alongside this focus on social behaviors and their connection to an agent's decision-making capabilities, the motivation to integrate concepts from related domains such as game theory (GT) and machine learning (ML) becomes critical, as GT and ML provide a rich background and broader perspective to the problem posed by MARL. However, it remains equally important to concurrently study the distinctive challenges that arise under the MARL paradigm  \cite{Stone2000MAS, Bernstein2013Complexity} to fully understand the field's unique intricacies and promote potential breakthroughs.

In the past decade, there has been a significant interest in MARL efforts, poised to endow the desirable behavioral qualities that characterize intelligent social agents through data-driven learning processes \cite{Gronauer2022survey}. These learning processes have ranged from assuming classical perspectives of full rationality to embodying models of bounded rationality \cite{Shoham2007MultiagentLearning}, wherein agents progressively refine their behaviors in a myopic and iterative manner over time and experience. A common theme of this survey is emphasizing the increasing incorporation of the realism and its complexities that pervades real-world MAS applications into our MARL solutions. And despite the many successes in MARL in these prior efforts, many open challenges persist, paving the way for future research endeavors. The central purpose of this survey is to provide a comprehensive view of these efforts, concretely staging the current state of MARL research from a holistic perspective.

The structure of this survey is as follows. In Section \ref{section:background}, we define the problem statement of learning optimal control within MAS, and discuss seminal concepts from foundational fields of MARL such as GT and ML -- all of which provide the foundation to the recent ideas studied in MARL. In Section \ref{section:learning in mas}, we discuss the unique benefits and challenges of learning in a MAS and explore the learning pathologies that plague the MARL paradigms. Lastly, in Section \ref{section: prospects}, the prospects of MARL are studied, such as MARL-specific simulation, training paradigms, communication methods, the challenges of multi-agent credit assignment and ad-hoc team-play, social learning, and agent modeling, and a detailed discussion regarding the recent efforts associated to these prospects.

\section{Related Surveys}
There exists a rich literature surveying various research topics in MARL, from general overviews \cite{Stone2000MAS, Shoham2003MultiAgentRL, Busoniu2008ComprehensiveSurvey, Bloembergen2015, Gronauer2022survey} to specific topics such as dealing with non-stationarity \cite{Hernandez2017Survey}, independent learners \cite{matignon2012independent}, multi-agent learning \cite{Panait2005CooperativeML, Tuyls2012MAL}, communication \cite{Luong2019Networking, zhu2024survey}, ad-hoc team-play \cite{mirsky2022survey}, agent modeling \cite{Baarslag2016, Albrecht_2018}, safety \cite{Lasota2017ASO} and knowledge transfer \cite{silva2018transfer}. However, many of these perspectives are constrained, excluding important aspects that come only from GT and ML viewpoints, thereby lacking the context needed to fully realize the current developments and limitations of MARL.

In contrast, our article presents a more modern and comprehensive survey that aims to provide a holistic view of the challenges inherent and unique to learning control within multi-agent environments. Additionally, we provide context to these challenges by weaving together the perspectives from GT and ML into a new unified view to present a novel understanding of the distinctive nature of the MARL problem.

\section{Background} \label{section:background}
In this section, we formalize the basis of the MARL problem statement and define its learning goals and solution concepts. We additionally explore related fields, i.e. GT and ML, that contribute to a deeper understanding of the study of MARL.

\subsection{Multi-agent Environment} \label{subsection: MAS}
\begin{figure}
    \centering
    \includegraphics[width=0.4\textwidth]{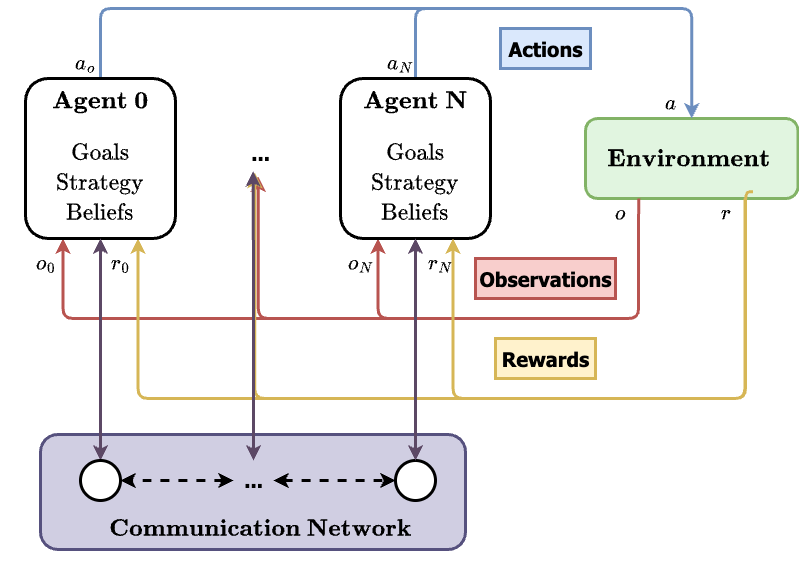}
    \caption{A visualization of a generalized multi-agent system following the iterative control process.}
    \label{fig:mas}
\end{figure}

A MAS consists of a population of decision-making agents that exist within a shared environment, as illustrated in Figure \ref{fig:mas}. These agents observe their environment and communicate with one another to perform actions that align with their objectives. This observation and communication amongst agents is constrained by decentralization.

\begin{definition}[Decentralization]
In a decentralized setting, each agent is capable of perceiving its environment, communicating with other agents, and taking action autonomously.
\end{definition}

There exist two forms of decentralization. Natural decentralization is the limitations imposed by physical realities, like communication range, while artificial decentralization involves specific requirements to improve tractability, such as communication bandwidth \cite{whiteson2020talk}. More generally, each agent is defined and inhibited by a set constraint.

The agents can devise strategies to make decisions to reach their defined goals. This causal association between stimuli, action, and objective is what we refer to as the agent's behavior \cite{Matari1994LearningTB}. The environment then responds to the agent's actions, transitioning the current setting to the next state and providing the agents with feedback signals. This response from the environment is referred to as the model. This process between the agents and their environment forms a closed-looped interaction that iterates until a terminal condition is met.

The MAS setting described here is widely prevalent in real-world applications, encompassing autonomous vehicles \cite{shalevshwartz2016safe, zhou2020smarts}, internet marketing \cite{Jin2018bidding}, multi-robot control \cite{Matari1997ReinforcementLI}, networking applications such as optimizing communication networks \cite{Luong2019Networking} and traffic control \cite{Calvo2018HeterogeneousMD, chu2019multiagent}, and multiplayer game playing \cite{samvelyan19smac}.

\subsection{Stochastic Game}\label{subsection:stochastic game}
\begin{figure}
    \centering
    \includegraphics[width=0.4\textwidth]{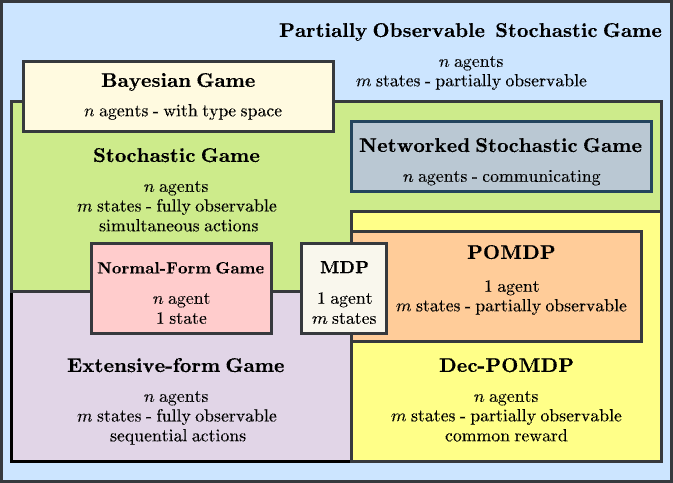}
    \caption{Models of Games: The overview of different models of multi-agent interactions is illustrated, from normal-form games to variations of stochastic games. We note that Markov Decision Processes (MDP) and POMDP are not models of a game but are included for a complete illustration. The following figure was adapted from \cite{Albrecht2024Book}.}
    \label{fig:gamemodels}
\end{figure}

We now introduce a formal representation of the MAS setting described in Section \ref{subsection: MAS}, called the stochastic game \cite{Shapley1953StochasticGame, Bowling2001Analysis}, where the term ``game" refers to the interactions between strategic agents. This framework serves as the basis for a wide range of multi-agent applications  \cite{Buoniu2010MultiagentRL}, and is related to other models of games, as seen in Figure \ref{fig:gamemodels}.

\begin{definition}[Stochastic Game]
A stochastic game is a 5-tuple $(N, S, A, r, \mathcal{T})$ where these elements are defined as:
\begin{itemize}
    \item $N$ is the set of $n$ agents.
    \item $S$ is the set of (global) states.
    \item $A = A_0 \times A_1 \times \dots \times  A_n$ is the joint action space, where $A_i$ is the action space of agent $i$.
    \item $r = r_0 \times r_1 \times \dots \times r_n$ is the joint reward function.
    \item $\mathcal{T}: S \times A \times S \mapsto P(S)$ is the state transition operator which maps a state-action pair to the probability of next states.
\end{itemize}
\end{definition}

The state defines the global setting and configuration of the environment.  To transition from one state to another, each agent $i$ uses their policy $\pi_i: S \mapsto A_i$ --- a functional representation of a composite of the agent's behaviors that is expressed as a mapping from its perceived state to action --- to make decisions, otherwise known as their strategy. Generally, a policy returns a probability distribution over the action space conditioned on the state. $$\mathop{\sum}_{a\in A} \pi(a|s)= 1$$
The joint policy $\pi: S \mapsto A$ is defined as a mapping to the joint action space, commonly achieved by concatenating the local actions computed by each agent's policies. We introduce the concept of an information set as an aggregate state, where it encapsulates all information available to an agent during its decision-making process.

The state transition function $\mathcal{T}$ returns the probability of reaching certain states from a given state-action pair. As each state holds the Markov property, i.e. every state is sufficient to infer the future, the state transition function is sufficiently conditioned on the current state and action.
\begin{equation}
    \mathcal{T} (s,a,s') = P(s_{t+1} = s'| s_t = s, a_t = a)
\end{equation}
Using the stochastic game framework as a basis, we can build and introduce additional concepts to align with various real-world scenarios.

\paragraph{Sequential and Macro-Actions} 
In many multi-agent applications, agents may perform actions that occur asynchronously, meaning either agents take turns performing their actions, or their actions take different duration of time. In cases where agents take turns, we distinguish this form of decision-making as sequential and is represented as an extensive-form game, otherwise, we refer to settings where agents' actions are done at the same time as simultaneous and use the stochastic game framework. In the latter case where actions have varying duration, we turn to an alternate framework called the macro-action stochastic game \cite{amato2019modeling, xiao2022asynchronous}. This framework removes the assumption of synchronized primitive-action execution, which assumes that all actions take the same period of time. Hence, the notion of macro-actions is introduced, where macro-actions are temporally-extended actions that can vary in duration. Consequently, this also introduces a notion of macro-observation.

\paragraph{Imperfect Information} It may be intractable for agents to directly perceive the (global) state of the environment. Instead, agents may only have access to observations. Thereby, we introduce the joint observation space $O = O_0 \times \dots \times O_n$ that accounts for the notion of imperfect information \cite{Shoham2008MAS}. Observations do not necessarily satisfy the Markovian property and may contain limited insights regarding the true state of the environment. In fact, even joint observational histories $h= \{o_0,o_1,\dots o_t\}$ can be insufficient to infer the state $s_t$ of the environment. However, the contrapositive is true, where the emission transition function $\mathcal{E}(o,s)$ defines the mapping from states to corresponding joint observations that can be potentially induced.
\begin{equation}
    \mathcal{E} (o,s) = P(o|s)
\end{equation}

The concept of imperfect information encapsulates the limits of information agents are restricted to, and help enforce the constraints of decentralization. Imperfect information is strongly related to two other important ideas of information limitations: partial observability and incomplete information. Partial observability constrains agents' access to a subset of the state that can be obscured by noise. In settings with incomplete information, agents do not have common knowledge about the game that is being played, which leads to uncertainties regarding various aspects of the state. A framework that takes these concepts of information limitation into account is called partially-observable stochastic games (POSG), and if agents hold a belief, or type, over these uncertainties, this is called a Bayesian game \cite{harsanyi1967games}.

\paragraph{Reward Function} Within all models of decision-making, the reward function plays an central role in guiding the behaviors of agents. The reward function provides immediate feedback to agents, evaluating the state and/or action of the agents with respects to their objective in the form of a scalar value. The common goal in optimizing control in MAS is to maximize this reward signal, thereby instilling a degree of rationality.

\begin{definition} (Rationality)
    Rational agents take actions that will maximize the expected cumulative reward, known as the expected return, they receive over time.
\end{definition}

In certain multi-agent applications, it can be difficult to define individualized reward functions for each agent, but it may be trivial to define a single reward function for all agents. This difficulty relates to the nature of the task, often derived from an inability to define proper multi-agent credit assignment. In such applications, we use the decentralized MDP (Dec-MDP), where a centralized reward function returns a collective score shared by all agents.

\paragraph{Nature of Interaction} An important factor to keep in mind is the relationships between the agents' actions and decisions and how they interact with one another, which we refer to as the nature of interactions. In each setting, the interplay between each agent's behaviors can lead to different dynamics and outcomes and may require varying considerations, notably to the definition of the reward function.

\begin{itemize}
    \item Cooperative setting - All agents share the aligning goal.
    \item Adversarial setting - Agents have dichotomous goals.
    \item Mixed setting - Agents have varying goals.
\end{itemize}

In a cooperative setting, all agents share aligning goals, where the reward function is typically designed to promote collaboration and joint success. In an adversarial setting, groups of agents may have dichotomous goals, meaning agents that aim to maximize their own reward can impede the achievement of other agents' goals. In a mixed setting, agents have varying goals, and this introduces additional complexities as agents may have aligning, conflicting, or overlapping interests at various points in time. These interactions can be realized with inherent and artificial structures placed within the task definition.


There exists a deep interconnection between the reward function and the nature of interactions, as the reward signals directly guide the emergent behaviors and outcomes within a MAS. For instances, the rewards of each agent may be computed by a function of not only their own policies but also the policies of others. This concept of multi-agent reward dependency is formalized with aggregative games.

\paragraph{Social Context}
The purpose of social context is to ensure mutually consistency between agents' actions within an shared environment, and is defined by social convention and role assignment \cite{Busoniu2008ComprehensiveSurvey}. Social conventions are established through social constructs to often define preferences for joint action profiles and is used to resolve conflicting action selections. Role assignments restrict and define the actions available for each agent as well as influence their rationality and objectives. The properties and conditions of these elements are largely considered task-specific, however, all MARL solutions must take into account these unique characteristics and demands of each situation.

\paragraph{Networked Games} To tractably perform certain multi-agent tasks, agents must communicate with one another. This is often achieved through defined communication channels and form a communication network $\mathcal{G}(s)$ that is associated to each state $s$. This element $\mathcal{G}(s) = (V,E)$, where vertices $V$ correspond to the agents and an edge $(i, j) \in E$ exists if agents $i$ and $j$ can communicate, is appended to the stochastic game framework, forming the networked stochastic game.

\paragraph{Coordination} 
The outcome of an agent's actions can often depend on the actions of other agents. This dependency of behaviors and the corresponding consequences of their actions between agents is encapsulated by the concept of coordination. There are two common approaches of embedding the idea of coordination into a game paradigm, through a coordination graph or interactions.
\begin{itemize}
    \item Coordination graphs specify the coordination dependencies between agents' actions in form of a graph, where the nodes represent agents and the edges represent these dependencies \cite{nair2005, guestrin2001, kok2005utile}. This concept is expressed with action-graph games \cite{jiang2011action}. With coordination graphs, we can factorize the global utility into a set of local payoff and utility functions. Each payoff function is associated with a subset of agents and can be interpreted as (hyper)-edges in a graph where the nodes are agents \cite{amato2014scalable}, whereas the utility functions reflect individual agent's utility. With this factorization, maximum-a-posteriori estimation techniques, such as variable elimination \cite{guestrin2001}, max-plus \cite{Vlassis2004Maxplus}, or Q-learning \cite{kok2005utile}, can be used to compute the optimal joint action. More recent extensions leverage deep neural networks to model different components of the factorized value function \cite{böhmer2020deep} or the coordination graph itself \cite{li2021deep}.
    \item Similar to coordination graphs, interactions broadly define specific conditions in which coordination should occur \cite{koller2003multi}. However, unlike coordination graphs, we can specify restriction of these interactions for only defined for certain states or state-actions \cite{melo2009learning, melo2011decentralized, de2010learning}. Additionally, these interactions are classified as strategic compliments or substitute \cite{monaco2016games}, where interactions can produce mutual reinforcements or discouragements.
\end{itemize} 
Thereby, the concept of coordination is illustrated naturally through ``social" networks, thereby often studied within network games as defined in classic GT. We note that network games are not networked stochastic game, where the former specifies coordination dependencies and the latter defines communication channels.

\paragraph{Return} A state-action trajectory $\tau = \{s_0,a_0,s_1,a_1, \dots, s_T\}$ is sampled from the dynamics model distribution $p_{\pi}(\tau)$ following a joint policy $\pi$, where:
\begin{equation}
    p_{\pi}(\tau) = P(s_0) \prod_{t=0}^T \pi(a_t|s_t) \mathcal{T}(s_t,a_t,s_{t+1})
\end{equation}
With this trajectory, we can compute an important quantity called the return, which calculates the future utility of a given state. Formally, the return $G_{(i,t)}$ at each time step for an agent $i$, sometimes called the agent's gain, is the cumulative future discounted reward.
\begin{equation}
    G_{(i,t)}(\tau) = \mathop{\sum}_{t'=t}^{T} \gamma^{t'} r_i(s_{t'}, a_{i,t'})
\end{equation}
where $\gamma \in [0,1]$ is the discount factor that enforces a diminishing value for more distant rewards.

\paragraph{Value Function} The value function $V_{\pi_i}(s_t|\pi)$ and the Q-value function $Q_{\pi_i}(s_t,a_{i,t}|\pi)$ map the state $s_t$ and state-action pair $(s_t,a_{i,t})$ to the expected return for an agent $i$ given some joint policy $\pi$, and are commonly used to develop solutions for optimizing controls in MAS.
\begin{equation}
    V_{\pi_i}(s_t|\pi) = \mathbb{E}_{\tau \sim p_{\pi}(\tau|s_t)}[G_{(i,t)}(\tau)]
\end{equation}
\begin{equation}
    Q_{\pi_i}(s_t,a_{i,t}|\pi) = \mathbb{E}_{\tau \sim p_{\pi}(\tau|s_t,a_{i,t})}[G_{(i,t)}(\tau)]
\end{equation}
\begin{equation}
    V_{\pi_i}(s_t|\pi) = \mathbb{E}_{a \sim \pi(a|s_t)}[Q_{\pi_i}(s_t,a_{i,t}|\pi)]
\end{equation}
The advantage function $A_{\pi_i}(s_t,a_{i,t}|\pi)$ measures the benefit of taking action $a_{i,t}$ in state $s_t$ for agent $i$ under the joint policy $\pi$. It quantifies how much better or worse it is to choose action $a_{i,t}$ compared to the average expected return of all actions in that state, similar to the concept of regret \cite{jin2018regret}. Hence, the advantage function is defined as the difference between the Q-value function $Q_{\pi_i}(s_t, a_{i,t}|\pi)$ and the state-value function $V_{\pi_i}(s_t|\pi)$.
\begin{equation}
    A_{\pi_i}(s_t,a_{i,t}|\pi) = Q_{\pi_i}(s_t,a_{i,t}|\pi) - V_{\pi_i}(s_t|\pi)
\end{equation}
\paragraph{MAS Objective} The objective $J_i(\cdot)$ of agent $i$ is expressed as maximizing the expected return, where trajectories $\tau$ are sampled from the dynamics model distribution $p_{\pi}(\tau)$ following a joint policy $\pi$.
\begin{equation}\label{eq:agent_objective}
    J_i(\pi) = \mathbb{E}_{\tau \sim p_{\pi}(\tau)}[G_{(i,0)}(\tau)]
\end{equation}
In practice, the expectation of the return can be approximated using the Monte Carlo sampling or through temporal difference estimation. Such empirical estimates can result in return estimates with high variance, resulting in greater learning complexity, a central and persistent issue in MARL solutions.

We state the general form of the MAS optimization objective, which we describe as maximizing the expected return for all agents. 
\begin{equation}\label{equation: objective}
    \textrm{maximize } J_i(\pi), \forall i \in N
\end{equation}

\subsection{Game Theory}\label{subsection:gt}
The field of GT, often studied within the domains of economics, provides a formal context that analyzes and conceptualizes strategic interactions among multiple agents within a market \cite{Osborne2003AnIT}. The term ``market" will be interchangeably used with the idea of a shared environment. We examine the MAS optimization objective from a game theoretic perspective to understand solution concepts relating to learning goals and social principles used to identify the joint behaviors that are considered desirable or interesting \cite{Myerson1985Intro}. For further study into GT, we refer readers to the following resource \cite{Osborne2003AnIT}.

\paragraph{Utility and Prospect Theory} Under the lens of utility theory \cite{fishburn1979utility, Shoham2008MAS}, we study how individuals make choices by quantifying their preferences, defining an preference relation. We define the utility function $u(\cdot)$ to reflect the individual's subjective evaluation of their overall satisfaction, assigning a numerical value to each outcome and obeys the axioms of preference. Utility theory affirms the sufficiency of scalar value representation of the agent's preferences \cite{Shoham2008MAS}.

In many real-world applications, agents face uncertainty. The causes of this uncertainty stem not only from the environment dynamics but also from the behaviors of other agents. Under such uncertain settings, the agents' decision-making processes should account for these unpredictabilities. We introduce this notion as risk and the domain of prospect theory, a study of decision-making under risk \cite{kahneman2013prospect, tversky1992advances, prashanth2022risk}. By taking risk into account, agents are capable of weighing their intrinsic preferences and objectives against their unknowns, defining the agents' aversion to losses, their reference points, and diminishing sensitivity \cite{kahneman2013prospect}. Hence, through prospect theory, we gain insights into behaviors driven by risk, where ``losses loom larger than gains" thereby illustrating the behaviors that come with loss-averse and gain-seeking decisions, and remains a large focus in behavioral GT \cite{camerer1997progress}. There exists a diminishing sensitivity to these losses and gains, meaning that the impact of a change diminishes with the distance from the reference point. Reference points are a concept introduced by prospect theory that represents the status quo and argues that the value of an agent is defined by the final utility positions, as stated in traditional utility theory, may not paint the whole picture, and instead, it is important to view their value in terms of gains and losses, i.e. the agent's value on a relative scale. However, the behaviors driven by risk defined by prospect theory become less relevant under repeated market settings, as experience within a market reduces uncertainty \cite{loomes2003anomalies}. We note that (expected) utility theory does account for uncertainty in its own manner, where the solutions derived from utility theory are considered risk-neutral but this faces violations in many risk-sensitive games \cite{kahneman2013prospect}.

Generally, risk-sensitive decision-making can be modeled with chance constraints, which are statistical constraints to averse based on a probability of a high utility loss at specified risk-tolerance levels, or value-at-risk (VaR) and expected shortfall (i.e. conditional VaR (CVaR)), which place a similar constraint which is now specified by quantiles of the utility distribution, i.e. the tail risk.

\paragraph{Incentives and Mechanisms} Incentives provide context to the unique behaviors that come along with asymmetries in information and actions between agents in market settings, which are commonly studied under contract theory and auctions \cite{salanie2005economics}. For instance, agents may be unequally informed about the various parts of the market, which can be reflective of the agent's roles or the social context, and this can lead to adverse selections, moral hazards, and nonverifiability \cite{laffont2009theory}. Adverse selection refers to situations where the agent with less information may make decisions that are unfavorable or risky, whereas moral hazard refers to the tendency of agents to take greater risks when they are insured or protected because they know they are shielded from some of the consequences. Lastly, nonverifiability refers to situations where the quality or performance of goods or services exchanged cannot be easily verified by the parties involved. Together, these concepts underscore the critical role of incentives in aligning behaviors and outcomes in markets characterized by information asymmetries. Incentives influence how agents gather and disclose information, manage risks, and fulfill contractual obligations, thereby shaping the efficiency and effectiveness of market interactions.

We now consider how these games and their rules are constructed, i.e. game form, to induce certain desired outcomes and perhaps more importantly, avoid undesirable outcomes. This effort is studied under the domain of mechanism design \cite{hurwicz2006designing} and accounts for both incentive and feasibility constraints, as well as the distribution of agent preferences, particularly distinguishing between whether they are public or private. Mechanism design remains a key focus in GT research and plays a crucial role in both theory and practice, offering frameworks to improve market functionality, regulatory policy, and organizational design in contexts of dynamic markets.


\paragraph{Solution Concepts and Equilibrium} Solution concepts refer to states of equilibria and are described by meaningful or interesting properties used to evaluate joint policies, otherwise known as strategy profiles. Pareto efficiency is a general criterion commonly used to evaluate strategy profiles. A strategy profile $\pi^{'}$ Pareto dominates another strategy profile $\pi$ if:
\begin{align}
    \forall i\in N: u_i(\pi^{'}) \geq u_i(\pi) \nonumber \\
    \exists i\in N: u_i(\pi^{'}) > u_i(\pi)
\end{align}
In other words, no agent using strategy profile $\pi^{'}$ can be better off without making another agent worse off by using $\pi$. We define Pareto improvement as any adjustments to a strategy profile that makes the resulting strategy profile more Pareto efficient, i.e. $\pi\rightarrow\pi^{'}$.
\begin{definition} [Pareto Efficiency]
A strategy profile $\pi^*$ is a Pareto efficient solution if it is not Pareto dominated by any other strategy profiles.
\end{definition}
Pareto efficiency focuses on maximizing overall welfare, which is the sum of the all agent's utilities. However, Pareto efficiency does not emphasize individual rationality or collective stability. Therefore, Pareto efficient solution may not necessarily be a good measure if the truly optimal solution requires agents to act against their self-interest or deviate from some locally desirable strategies. Importantly, Pareto optimality does not address the issue of fairness or equality, as some individuals may benefit more than others in the pursuit of maximizing overall welfare \cite{Shoham2008MAS}.

The concept of best response provides an alternative perspective in analyzing strategic interactions in multi-agent systems that aligns more with concepts of individual rationality and collective stability. Given a strategy profile $\pi= \{\pi_0,\pi_1,\dots, \pi_i, \dots, \pi_N\}$, a best response strategy $\pi'_i$ for agent $i$ is defined by:
\begin{equation}
    \forall \pi_i: u_i(\llbracket \pi_{-i},\pi'_{i}\rrbracket) \geq u_i(\llbracket \pi_{-i},\pi_{i}\rrbracket)
\end{equation}
where $\pi_{-i} = \pi \setminus \pi_i$ denotes the strategy profile without agent $i$ and $\llbracket\pi_1,\pi_2\rrbracket$ is the strategy profile of $\pi_1,\pi_2$. In other words, the best response strategy of an agent is one that maximizes its utility given the strategies chosen by other players.

The solution concept of a Nash equilibrium (NE) $\pi^*$ applies this notion of best response to the collective, where for all agents $i$, $\pi^*_i$ is the best response to $\pi^*_{-i}$.
\begin{equation}
    \forall \pi_i: u(\llbracket\pi_{-i}^*,\pi^*_{i}\rrbracket) \geq u(\llbracket\pi_{-i}^*,\pi_{i}\rrbracket)
\end{equation}
\begin{definition} [Nash Equilibrium]
A Nash equilibrium (NE) defines a state where no individual agent can increase its expected return by unilaterally deviating from their policy \cite{Nash1951NONCOOPERATIVEG}.
\end{definition}
Hence, within a Nash equilibrium, all of the agents' strategy is the best response to the other agents' strategy.

Unfortunately, similar to Pareto efficient solutions, NE is not unique, and determining the differences between sample equilibria in terms of their social behaviors is unclear. In fact, the behaviors exhibited by sample equilibria can vary starkly, where their comparison can be computed using other concepts of efficiency such as coordination ratio \cite{koutsoupias1999worst} or Price of Stability \cite{anshelevich2008price}. While NE remains a popular criterion for multi-agent decision-making under uncertainty, computing this equilibrium may be computationally intractable in complex games. Moreover, achieving desired joint behaviors is not guaranteed through this approach \cite{Shoham2003MultiAgentRL, matignon2012independent}. Nonetheless, NE does represent stable points akin to saddle points within the optimization landscape.

To address the intractability of computing a strict NE, $\epsilon$-Nash equilibrium relaxes the requirements by allowing the agent to deviate if it improves its expected returns by more than some value $\epsilon$.
\begin{equation}
    u(\llbracket\pi_{-i}^*,\pi^*_{i}\rrbracket) \geq u(\llbracket\pi_{-i}^*,\pi_{i}\rrbracket) - \epsilon,\forall \pi_i
\end{equation}

Correlated equilibrium (CE) is an important generalization of NE that adds the notion of correlating strategies among agents. This considers the existence of signals that coordinate agents' actions \cite{Aumann1974CE}, as well as the introduction of a correlating distribution over the strategies of all agents. In cases where such correlating signals do not necessarily affect the joint strategies and cause no agents to deviate regardless of the information provided by the signals, we call the following optimization state a coarse correlated equilibrium (CCE). As mentioned, NE is a special case of CE, where the correlating distribution of the agents' strategies is a product of independent distributions.

The concept of perfect equilibrium \cite{Selten1988PE} refines the idea of NE in a different manner --- by imposing additional requirements of consistency and mutual optimality. This solution concept describes an optimization state where agents' strategies are mutually consistent and take into account the possibility of off-equilibrium actions, requiring agents to choose strategies that are robust against such deviations. In fact, numerous additional refinements and solution concepts exist that take into account asymmetric roles and information \cite{bacsar1998dynamic}, situations involving imperfect information \cite{Shoham2008MAS}, or even games with sequential dynamics \cite{kreps1982sequential} and non-stationary considerations \cite{daskalakis2022complexity, kim2022influencing}.

\paragraph{Equilibrium Analysis and Computation} The process of computing the optimal decisions for multiple agents is referred to as equilibrium computation \cite{Daskalakis2022EC} and has been studied under the framework of optimization and variational inequalities \cite{harker1990finite, kovalev2023optimalalgorithmsdecentralizedstochastic}. Equilibrium computation aims to find specific points of interest, known as equilibria, within an optimization landscape that spans the strategy space of multiple agents. Formally, equilibria represent stable or ``optimal" solutions where the dynamical system reaches a balanced or steady state. Equilibria can be described as being local or global, indicating whether the state is a locally optimal solution or the best solution across the entire landscape. The two important properties we must consider for equilibrium computation are their existence and tractability. The existence of an equilibrium largely depends on the utility function of each agent, and whether it is concave with respects to their actions, where in nonconcave games, the existence of equilibrium is at risk. The tractability of equilibrium refers to the complexity of solving for an equilibrium and this remains a significant challenge to be compute efficient even in nonconcave games.

\paragraph{Equilibrium Complexity} \label{par: complexity} Even with the existences of an equilibrium solution \cite{Nash1951NONCOOPERATIVEG}, the complexity of computing this equilibrium must be discussed. This discussion is most aptly had under the pretense of a total search problem (i.e. PPAD) rather than as a decision problem (i.e. P/NP). Although, computing an equilibrium can be classified an NP-hard decision problem \cite{Gilboa1989Complexity, Conitzer2008Complexity, Nisan2007AlgGT, Albrecht2024Book}.

\begin{definition}(PPAD)
    Polynomial parity argument for directed graphs (PPAD) consists of problems that can be reduced to the following (End-Of-The-Line) problem:
    
    \textit{Given a directed graph with vertices have at most one predecessor and/or one successor, and a source vertex $s$, where $s$ has no predecessor, find a vertex $t$ with no predecessor or no successor, such that $s\neq t$. Let there be a polynomial-time function that returns the predecessor and successor of all vertices in this graph.}

    where PPAD-completeness is shown by reducing the End-Of-The-Line problem. To support the hardness of PPAD problems, \cite{Daskalakis2009Complexity} proposes the following question: How can one hope to devise an efficient algorithm that telescopes exponentially long paths in every implicitly given graph?
\end{definition}

Computing the Nash equilibrium is proven to be a PPAD-complete \cite{Daskalakis2009Complexity, chen2007settling}, further affirming its difficulty and potential intractability.

\paragraph{Learning Dynamics} While these solution concepts of equilibrium give context for stability and optimality, they do not provide insights into the process of reaching such states. The concept of learning dynamics bridges this disconnect by detailing the procedure for reaching equilibrium. Additionally, learning dynamics also helps analyze the transition into equilibrium itself, which can be equally or more important to gain insights into task-specific understandings, and the processes that come after achieving equilibrium, where within that steady state, whether continual lifelong learning and further adaptations may come into the equation. In general, there exist two fundamental learning dynamics within the process of equilibrium computation: best-response dynamics and no-regret dynamics.

Best-response dynamics directly optimize to converge to an NE or one of its refinements. A seminal algorithm of best-response learning dynamics is fictitious play (FP), where all agents iteratively compute the best response to other agents' strategies, or most specifically, the uniform distribution over the past strategies of the other agents \cite{Robinson1951FicitiousPlay}.

We summarize the seminal realizations and recent advancements in best-response dynamics learning:
\begin{itemize}
    \item There are improvements to the idea of FP that include improved robustness to perturbations in the form of stochastic FP \cite{fudenberg1993learning} as well as uncertainty within these models of other agents and update these beliefs through Bayesian updating, known as rational and Bayesian learning \cite{Albrecht2024Book, Jordan1991, kalai1993rational}. Specifically, Bayesian learning utilizes the value of information \cite{chalkiadakis2003coordination}, which considers how actions will influence future behaviors and devise more accurate beliefs. 
    \item FP algorithms are typically used in normal-form games but can be realized in extension-form games using behavioral strategies, known as extensive-form fictitious play (XFP), or through approximate best response and sample-based learning, i.e. ML approach to XFP, known as Fictitious Self-Play \cite{heinrich2015fictitious}.
    \item Double oracle approach adopts the same iterative procedure as fictitious play, but instead, agents now compute a meta-NE to a restricted game that is maintained and expanded by the past best-response strategies \cite{mcmahan2003planning, adam2021double}.
    \item Policy Space Response Oracles, Deep Cognitive Hierarchy  \cite{lanctot2017unified}, and Extensive-Form Double Oracle  \cite{mcaleer2021xdo} incorporate and build upon the double oracle algorithm with the use of ML techniques, and address computational issues that exist when extending double oracle to extensive form games, promote generalization and prevent overfitting to specific equilibrium \cite{bighashdel2024policy, lanctot2017unified}.
    \item Value iteration \cite{Shapley1953StochasticGame} can be applied over the joint-action space, resulting in solutions such as Minimax Q-learning \cite{Littman1994Minimaxq}, Nash Q-Learning \cite{Hu2003NashQ}, Correlated Q-learning \cite{Greenwald2003CEQ}, and Friend-or-Foe Q-learning \cite{littman2001ffq}.
    \item Replicator dynamics \cite{maynard1976evolution} adapts the concept of evolutionary dynamics to achieve best-response policies \cite{tuyls2006evolutionary}.
    \item Infinitesimal gradient ascent (IGA) \cite{Singh2000iga} utilizes gradient learning to optimize agent's policy with respect to their utility. Similar methods and their extensions to improve convergence and other theoretical properties include using variable step sizes \cite{Bowling2002Wolf, Bowling2004GigaWolf}, proximal point optimization \cite{mokhtari2019unified}, momentum \cite{gidel2019negative}, extra-gradient\cite{Korpelevich1976TheEM}, and optimistic gradients \cite{rakhlin2013optimization, daskalakis2018limit, wei2021linear}.
\end{itemize}

In no-regret learning dynamics, the aim is instead to minimize regret, a measure of how much an agent would have gained in utility if they had chosen a different strategy often in retrospect \cite{hart2000simple}. Intuitively, regret can be quantified by the average cost between the utility of the best possible strategy profile and the actual utility of the chosen strategy profile. This notion of regret is called external regret, where comparisons of decisions are performed using an expert offline strategy. On the other hand, internal, or swap, regret takes a more online approach, compared to a modified strategy that swaps out certain actions with others from the original strategy. There exists a well-known connection between no-regret and Nash equilibrium \cite{Zinkevich2007regret}, where an $\epsilon$-Nash equilibrium is a profile that achieves an upper bound regret of $\leq \epsilon$. Usually, the tractability of computing and verifying best-response can be compromised with games with high complexity, and therefore no-regret dynamics approaches are an attractive alternative as they scale very well when using domain-specific abstractions, such as in Poker AI applications \cite{Zinkevich2007regret, tammelin2014solving}.
We summarize the seminal realizations and recent advancements in no-regret dynamics learning:
\begin{itemize}
    \item Regret matching serves as the foundation for regret minimization algorithms, and is achieved through repeated self-play that computes the strategy iterates based on a distribution of normalized positive regret \cite{blackwell1956analog, hart2000simple}. Convergence to a stable solution, i.e. sample equilibrium, is achieved by taking the average overall strategy iterates.
    \item Counterfactual regret minimization (CFR) extends regret matching to extensive form games with counterfactual regret, which accounts for the sequential nature of actions in extensive form games \cite{Zinkevich2007regret}. To ensure sufficient coverage over the game tree for CFR updates, a sampling approach using external or chance sampling \cite{lanctot2013monte} or a more extensive search, like in CFR+ \cite{tammelin2014solving, bowling2015heads}, must be considered. Recent advancements, such as CFR+ and advantage-based regret minimization \cite{jin2018regret}, have demonstrated performance improvements with resetting and positive clipping negative cumulative regret to zero, which can be a form of ``optimism under uncertainty".
    \item Follow-the-Leader (FTL) is another popular regret-minimizing technique studied more in online learning, and FTL constructs online strategies to follow the actions with minimal loss over past rounds, i.e. the best ``expert" \cite{shalev2012online}. However, a naive implementation of FTL is unstable, where a natural solution to this instability is to append a time-varying regularization term, known as Follow-the-Regularized-Leader (FoReL). Importantly, the choice of regularization term leads to different regret bounds. Another method to address this instability and avoid overfitting from FoReL is to use the subgradient method with the proximal term as the Bregman divergence, known as mirror descent \cite{orabona2023modern}. For both FoReL and mirror descent methods, continued efforts to seek improvements are made to the regret bounds, with techniques as simple as gradient clipping \cite{pmlr-v99-cutkosky19a}.
    \item Multiplicative weights update/hedge algorithm generalizes FTL, as it now maintains weights that are updated with a non-infinite learning rate rather than selecting the single ``expert" \cite{freund1999adaptive, Arora2012}. The learning rate can be adaptive to ensure consistent low regret on easy and hard instances with the doubling trick or using both FTL and hedge periodically \cite{de2014follow}.
    \item IGA and its extensions achieve both best-response and no-regret under certain conditions \cite{Zinkevich2003giga}, and with the advent of deep learning in MAS applications, gradient methods have become quite popular.
\end{itemize}

\paragraph{Connection to MARL} The literature on GT is extensive, with ongoing efforts investigating diverse challenges in strategic interactions among multiple agents. However, the significance of MARL lies in the integration of data-driven considerations into these strategic decision-making processes. So, unlike traditional GT, MARL leverages insights from data and statistics to manage complex markets where agents' behaviors can be shaped by data-driven models. This approach enables MARL to effectively address dynamic, uncertain, and large-scale scenarios that traditional GT often finds challenging to handle.

\subsection{Machine Learning}\label{subsection:deep learning}
The field of ML represents a crucial domain within AI, with the objective of constructing data-driven solutions that excel in pattern recognition using statistical models \cite{james2013introduction}. Central to ML are two primary tasks: data collection and data analysis. Data collection defines the process of gathering and managing data which constructs the target data distribution, whereas the task of data analysis encompasses the methods of statistical inference using this data distribution. Hence, the data plays a pivotal role in the ML process, defining the underlying distribution on which ML models rely their statistics upon.

This significance has become glaringly apparent in real-world ML applications, where the integrity of data is often tested by strategic agents aiming to manipulate statistics or exploit the predictive models themselves \cite{zrnic2022leads}. Unfortunately, this multi-agent dynamic is canonically neglected in ML literature but has more recently gained traction in the emerging fields of strategic classification and adversarial ML, which develop solutions regarding the cycle between the development and deployment of ML models and the post-hoc response of strategic agents influencing the model, which can potentially adversarially ``attack" the model and ``pollute" the data. Another popular stream of research that blends GT into ML is in generative modeling, notably with the training of generative-adversarial networks (GAN). GANs employ a min-max optimization scheme to train an generative model with an discriminator model that classifies between true and generated data. However, an naive implementation without proper GT considerations leads to chaotic and oscillatory learning \cite{daskalakis2018traininggansoptimism}. In essence, the efficacy of real-world ML hinges on robust data practices and the ability to navigate the complex landscape of strategic interactions in diverse scenarios.

\subsubsection{Deep Learning}
Deep learning is a popular and general approach in contemporary AI research, with its foundations relying on the use of artificial neural networks (ANNs). ANNs have showcased remarkable proficiency in a wide variety of general applications \cite{Goodfellow2016DL}.

Concretely, ANNs define a set of parameters $\theta$ that act as a function that processes and transforms data using layers of linear and non-linear operations. We optimize these parameters with respect to a defined cost function $J(\theta)$ using methods of statistical learning and numerical optimization, such as stochastic gradient descent (SGD) \cite{james2013introduction}.
\begin{equation}
    \theta = \theta - \nabla_\theta J(\theta)
\end{equation}
A key attribute of a deep learning approach is its ability to develop a scalable end-to-end solution that can capture complex distributions underlying many real-world applications. Additionally, we avoid the need to handcraft features and instead allow the optimization to learn curated latent representation tailored toward the task at hand in a data-driven manner. Hence, the emphasis is on extracting patterns, relationships, and insights directly from data, rather than relying heavily on predefined rules or models, although such efforts are not necessarily orthogonal.

In the context of MARL, integrating these deep learning methods into MARL solutions has shown promise for developing expressive and adaptive multi-agent systems for complex tasks, where agent's behaviors are defined using a composite of ANN representations and optimized in an end-to-end manner. For example, we can represent the elements of a stochastic game, such as the policy, value function, or model, using ANNs to optimize the joint behavior. The central constraints in such methods reside with the need for large-scale data collection and the high computational cost required for training.

\subsubsection{Reinforcement Learning} \label{subsection:reinforcement learning}
The domain of reinforcement learning (RL) focuses on learning how to make decisions guided by reward signals \cite{Sutton2018RL} and serves as the general foundation of learning controls. A caveat of traditional RL research is that it is normally studied under a single-agent setting, i.e. using the MDP framework.

\paragraph{Predictions and Control} We define the process of sequential decision-making as control. Control contrasts the concept of predictions, as predictions are static inferences that have no causal relationship to future inputs and predictions. Formally, predictions are independent and identically distributed (IID), whereas no such assumptions are made for control. Within RL frameworks, control is learned within an environment whose dynamics the agent has limited knowledge of \cite{recht2019tour}, however predictions still play a major role in this capability, whether it is incorporated implicitly or explicitly, to ensure that the agent has an understanding of the world it resides.

\begin{figure}
    \centering
    \includegraphics[width=0.35\textwidth]{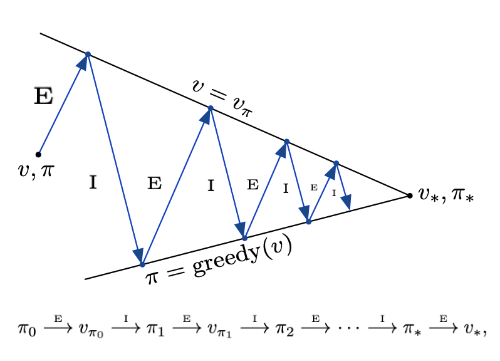}
    \caption{Policy Iteration: The process of policy iteration consists of an iterative cycle of policy evaluation (shown as $\xrightarrow{\text{E}}$) and policy improvement (shown as $\xrightarrow{\text{I}}$). Policy evaluation computes the value function for current policy whereas policy improvement updates current policy with respect to evaluated value function. The following figure was taken and modified from \cite{Sutton2018RL}.}
    \label{fig:policyiteration}
\end{figure}

\paragraph{Value Function Approximation} A value-based approach for RL estimates the value function to approximate the optimal policy. Hence, value-based solutions often do not define an explicit policy and instead, use an implicit policy computed using the value function, $\pi(a|s) \propto Q(s, a)$. These Q-learning methods \cite{Watkins1989Qlearning, watkins1992q} utilize approximate dynamic programming by applying the concept of policy iterations, as shown in Figure \ref{fig:policyiteration}, with the Bellman (expectation) equations \cite{bellman1957dynamic}, a set of recursive definitions of the value functions.
\begin{align} \label{equation: bellman equation}
Q_{\pi}(s_t,a_t) =& \mathbb{E}_{\tau \sim p_{\pi}(\tau|s_t,a_t)}[G_{t}(\tau)]  \nonumber\\
=& r(s_t,a_t) + \nonumber \\ & \mathbb{E}_{s_{t+1}\sim \mathcal{T}(s_t,a_t,s_{t+1}), a_{t+1}\sim \pi(a|s_{t+1})}[Q_{\pi}(s_{t+1},a_{t+1})]
\end{align}
We define the Bellman backup operator $\mathcal{B}_\pi$ on our value function estimate $Q$ such that $Q = \mathcal{B}_\pi Q$ is equivalent to Equation \ref{equation: bellman equation}. By iteratively applying the Bellman backup operator $\mathcal{B}_\pi$, $\mathcal{B}_\pi$ is shown to be a contraction mapping, thereby according to the Banach fixed-point theorem, $Q$ is guaranteed to converge to a unique fixed point that corresponds to the true Q-function $Q_\pi$ for the policy $\pi$ \cite{Sutton2018RL}.

By employing an optimal (implicit) policy denoted as $\pi^{*}(a|s) = \delta(\argmax_{a\in A} Q(s, a))$, the need for policy evaluation, a process that scales poorly with the state-action space, can be circumvented. This approach, known as value iteration, leverages the Bellman optimality equation in the form of the Bellman backup $\mathcal{B}_{\pi}$ to minimize the Bellman residual error.
\begin{equation}
    Q_{\pi^*}(s_t,u_t) = r(s_t,u_t) + \mathbb{E}_{s_{t+1}\sim \mathcal{T}(s_t,u_t,s_{t+1})}[\max_{a\in A}Q_{\pi^*}(s_{t+1}, a)]
\end{equation}

Value function approximation methods have come with several improvements over the past decade, including:
\begin{itemize}
    \item Extension to deep neural networks representation for continuous state space and usage of experience replay to stabilize off-policy learning \cite{mnih2013playing}.
    \item Distributional value function representation \cite{bellemare17c51, dabney2018distributional, dabney2018implicit, yang2019fully} for more expressive value approximations.
    \item Mitigate overoptimism/maximization bias with double learning \cite{van2016deep}, clipped double Q-learning, and target policy smoothing \cite{fujimoto2018td3}.
    \item Utilization of a dueling architecture \cite{wang2016dueling} to help figure out which states are valuable without having to learn the effect of each action for each state.
    \item Improvements to experience replay to prioritize important transitions \cite{schaul2015prioritized}, to perform better with sparse rewards \cite{andrychowicz2017hindsight}, and to work with recurrent networks \cite{hausknecht2015deep, kapturowski2018recurrent}.
    \item Improved exploration through randomized modeling \cite{fortunato2017noisy}, intrinsic rewards \cite{badia2020up, badia2020agent57}, maximum entropy framework \cite{haarnoja2017reinforcement}.
    \item Handle continuous action spaces by assuming a deterministic policy \cite{lillicrap2015ddpg}, parameterizing the Q-function as a well-defined convex function \cite{gu2016continuous}, or utilizing sampling methods \cite{kalashnikov2018qt} in order to compute the Bellman update. 
    \item Learn on datasets of offline experiences while managing distributional mismatch and continuing improvements beyond behaviors seen in dataset \cite{Levine2020OfflineRL} by appending a behavior regularization term to the standard RL training \cite{fujimoto2021minimalist}, regularizing overestimation with conservatism \cite{kumar2020conservative}, utilizing probabilistic regression \cite{kostrikov2021offline, Ma2021CODRL}, and a model estimate \cite{yu2022combo}. This application is known as batch or offline RL.
    \item Integration of the listed improvements through the years \cite{hessel2017rainbow, kapturowski2022humanlevel}.
\end{itemize}

Value-based approaches are commonly used in MARL algorithms, as they offer a powerful and expressive solution to learning and producing a control policy in an off-policy approach, thereby being sample-efficient.

\paragraph{Policy Gradient} A policy-based approach of RL directly optimizes the objective stated in Equation \ref{equation: objective}. A common approach is to perform gradient ascent along the objective using the policy gradient, pioneered by the REINFORCE (REward Increment $=$ Nonnegative Factor $\times$ Offset Reinforcement $\times$ Characteristic Eligibility) algorithm \cite{Williams1988Reinforce, Williams1992PG}.
\begin{align}\label{equation: policy gradient}
    \nabla_{\theta_\pi} J(\pi) &= \nabla_{\theta_\pi}\mathbb{E}_{\tau \sim p_{\pi}(\tau)}[G_{0}(\tau) - b(\tau)] \nonumber \\
    &= \mathbb{E}_{\tau \sim p_{\pi}(\tau)}[\sum_{t=0}^T (G_{t}(\tau) - b(\tau)) \nabla_{\theta^\pi} \log(\pi(a_t|s_t))]
\end{align}
where $\theta_\pi$ represents the parameters of the policy $\pi$ and $b(\cdot)$ is the reinforcement baseline function used to stabilize the approximated reinforcement $G_t(\tau)$\footnote{We highlight that the return $G_t(\tau)$ here removed the notation of which agent it is referring to, as we are only considering one agent in this scenario.}. As the expectation is computed through Monte Carlo sampling, it can often lead to high variance despite providing an unbiased estimate of the expected return. Hence, as noted by \cite{Williams1992PG, schulman2018highdimensional}, the choice of $b(\cdot)$ can improve convergence and performance by mitigating this variance.

Notable improvements to this vanilla policy gradient consist of:
\begin{itemize}
    \item Learn on data from various policies, i.e. an off-policy approach, with policy gradient methods by using importance sampling corrections \cite{Jie2010offpolicypg, degris2012off} that enables training on a more diverse dataset in a sample-efficient manner.
    \item Mitigate dominating gradients on the parameter level by reparameterizing loss under a probabilistic manifold using Fisher information matrix \cite{kakade2001natural} with natural policy gradient, or Kronecker-factored approximated curvature \cite{wu2017scalable}.
    \item Utilize surrogate objective that guarantees monotonic improvements, conditioned on a trust-region optimization \cite{Kakade2002ApproximatelyOA, schulman2017trust, schulman2017proximal}.
\end{itemize}

Unlike value-based approaches, policy-based approaches are not commonly practiced in MARL, largely due to the sample inefficiencies from its on-policy requirements and high variance. However, in cases where samples can be generated cheaply, policy-based approaches can perform very well.

\paragraph{Actor-Critic Methods} Actor-critic algorithms provide an integration of policy-based and value-based methods \cite{Sutton1999AC}. Instead of estimating the value function of an implicit policy, we model the value function for an explicitly defined policy thereby minimizing Bellman residual using Equation \ref{equation: bellman equation} and learning from a policy gradient. With access to a value estimate, temporal difference (TD) learning or $n$-step learning can be utilized instead of Monte Carlo sampling to approximate the offset reinforcement. While these bootstrap methods do introduce bias, the variance is considerately lowered as updates are no longer dependent on the entire trajectories but only on a subset of the trajectory \cite{Sutton2018RL}. Additionally, the value function can also improve the quality of the policy gradient by stabilizing the return estimate with a new baseline \cite{schulman2018highdimensional}.

Much of the improvements to the actor-critic methods follow the advancements from value-based and policy-based approaches, and also consist of:
\begin{itemize}
    \item The use of generalized baseline functions \cite{schulman2018highdimensional} and action-dependent control variate \cite{gu2017qprop, liu2018actiondepedent, tucker2018mirage} to stabilize the variance of policy gradient.
    \item Distributed training paradigms for actor-critic methods through asynchronous learning between multiple parallel actors \cite{mnih2016asynchronous, espeholt2018impala}.
    \item Stabilized learning of solution with shared parameters between actor and critic by using phasic learning and representation regularization \cite{cobbe2020phasic, Huh2021MixAM}.
    \item Extending efforts from policy-based and value-based methods to an actor-critic framework, such as experience replay \cite{wang2016sample} and soft Q-learning \cite{haarnoja2018soft, haarnoja2019soft}.
\end{itemize}

\paragraph{Model-based Approaches} With a model-based approach, an explicit estimate of the transition dynamics $\mathcal{T}$ and/or reward probabilities is maintained. These estimates can be used for planning without an explicit policy or for a more sample-efficient policy improvement \cite{sutton1990dyna}. Methods that do not maintain such estimates are referred to as a model-free approach. While model-based approaches have shown success in highly complex tasks, these methods can be more challenging as accurately and comprehensively representing the transition dynamics and reward function, especially over long time horizons, is non-trivial and difficult.

There have been notable improvements in model-based approaches, which include:
\begin{itemize}
    \item Integrate model learning and policy optimization with deep learning methods \cite{ha2018worldmodels, weber2018imaginationaugmented, hafner2020dream, hafner2022mastering, hafner2023mastering}.
    \item Mitigate distributional mismatch using online data collection \cite{ross2011reduction}, uncertainty estimation \cite{deisenroth2011pilco, feinberg2018modelbased} via Bayesian neural network \cite{blundell2015weight}, bootstrap ensembling \cite{chua2018deep, kurutach2018modelensemble, clavera2018modelbased, luo2021algorithmic, Buckman2018smve}, or dropout \cite{gal2017concrete}.
    \item Model-based learning on imperfect information \cite{Watter2015Embed, zhang2019solar}.
\end{itemize}

\paragraph{Successor Representation} Successor representation (SR)
is an alternative approach to the model-based and model-free approach, where it disentangles the state transitions from reward estimation by maintaining a state occupancy function $M(\cdot)$ for a given policy $\pi$ \cite{Dayan1993SR}.
\begin{equation}
    M_\pi(s,s') = \mathbb{E}_{\tau \sim p_\pi(\tau|s_0 = s)}[\sum_{t=0}^T \gamma^t \mathbbm{1}(s_t = s') ]
\end{equation}
where $\mathbbm{1}$ is the characteristic function. The state occupancy function captures the notion of environmental affordance, caching statistics relating to which future states are possible from a given state. Similar to the Bellman equations, we can derive a recursive definition for the state occupancy function.
$$M_\pi(s,s') = \mathbbm{1}(s_t = s') + \gamma \mathbb{E}_{s_{t+1} \sim p_\pi(\tau|s_t = s)}[M(s_{t+1},s')]$$
This factorization provides a compact and structured representation of the model for efficient and adaptive learning in complex environments \cite{kulkarni2016deep}. Generally, SR methods alleviate the cost of learning a complex environment model while still being adaptive to distal reward changes, given the disentanglement of the reward function from the state transition.

Recently, SR has reemerged with a new utility in RL that can help capture different aspects of the RL problem, despite the initial idea proposed several decades ago. Notably, a grounded formulation to extend SR with deep learning methods \cite{blier2021learning, touati2021learning} and the utilization of passive data to learn latent and useful features with SR \cite{ghosh2023reinforcement} are some recent applications that have captivated some new interest in this technique.

\begin{table*}
  \centering
  \begin{tabular}{l|c}
    \textbf{Challenges} & \textbf{Pathologies} \\ \hline
    Computational complexity & Stochasticity, Deception  \\
    Non-stationarity & Moving-target problem  \\
    Coordination & Miscoordination, Relative Overgeneralization, Alter-Exploration Problem  \\
    Performance Evaluation &  n/a \\
  \end{tabular}
  \caption{A table of MARL challenges and their learning pathologies.}
  \label{tab:1}
\end{table*}

\paragraph{Foundation Model for Control} Foundation models are ML models that are pre-trained on large-scale data to be adapted for diverse downstream tasks and have largely been successful and practiced in recent ML research \cite{bommasani2022opportunities}. This sentiment for foundation models is echoed in RL applications, where leveraging pre-trained models can also offer advantages, especially considering RL's training inefficiencies. However, questions regarding its effective implementation remain at large.

The recent efforts of developing foundational models for RL has been in general navigation \cite{shah2022gnm, shah2023vint, baker2022video}, robotic manipulation \cite{walke2023bridgedata} and other broad robotic applications \cite{embodimentcollaboration2023open}. These works have largely been motivated by practices in large-language model training methodologies and offline RL \cite{chebotar2023qtransformer} and creating multi-modal prompting to integrate successes from the research of computer vision and natural language. The key insight with foundation models is recognizing that across tasks, there are shared traits and skills that are required in many of these tasks that do not necessarily require re-learning for each task. Instead, learning these aspects in a unified manner rather than individually and myopically within a single task can lead to not only an amortized cost of learning but also more robust and superior behaviors.

\section{Learning in a Multi-agent Environment} \label{section:learning in mas}
The transition from single-agent RL to learning within a multi-agent stochastic game setting promises numerous opportunities but consequently is tied to challenging difficulties that require paradigm-altering considerations \cite{Resnick2005Thesis}. In contrast to single-agent control systems, where one agent interacts with its environment, a multi-agent setting involves managing the decision-making and learning processes of multiple entities that can interact with each other and their shared environment. Facilitating stable, adaptive, and social behaviors within a multi-agent control learning process is non-trivial, and comes with several intricacies that converge to necessitate desired solutions. In this section, we cover the positive and negative aspects that are unique to MARL, largely highlighting its potential and consequences in the hopes of providing meaningful context and a deeper look into these concepts.

\subsection{Benefits of MARL}\label{subsection:benefit of marl}
MARL directly addresses the optimization problem of developing multiple decision-making agents within a shared environment. As a result, these algorithms are specifically designed to consider the complexities inherent to multi-agent dynamics during training. Therefore, the optimization can produce valuable behaviors that span from adaptive social learning \cite{ndousse2021emergent} to the emergence of coalitions that enable intelligent interactions and improve robustness against dynamic changes within the population \cite{Busoniu2008ComprehensiveSurvey}.

This perspective of approaching control optimization that MARL studies also enables the realization of applications that could not be accurately modeled in single-agent RL alone \cite{Matari1994LearningTB}, particularly in scenarios that involve multiple adaptive agents that can influence the environment or other agents \cite{lanctot2017unified}. The context of training under MAS features additionally introduces the chance to capitalize on the unique prospects and structures of the multi-agent nature, which can enhance the efficiency and depth of training through various means discussed in the later parts of this survey.

\subsection{Challenges of MARL}\label{subsection:challenges of marl}
The task of MARL is not only faced with the already challenging objective of optimizing control but is further riddled with the inclusion of multiple interacting agents. Throughout this paper, we investigate four central challenges --- computational complexity, non-stationarity, coordination, and performance evaluation --- that are interdependent and coupled to various learning pathologies.

\begin{definition} [Learning Pathology]
    Learning pathology refers to undesirable or sub-optimal behaviors that can emerge within MARL learning dynamics. As listed in \cite{palmer2020thesis}, these include stochasticity, deception, the moving-target problem, miscoordination, relative overgeneralization, and the alter-exploration problem.
\end{definition}

Much of the problematic intricacies of MARL stem from a mixture of these challenges, as they tend to be deeply interconnected.

\subsubsection{Computational Complexity} As mentioned in Section \ref{par: complexity}, optimizing for an equilibrium solution is theoretically demonstrated as challenging, where the cost of modeling multiple agents can easily scale intractably. This is seen as the joint state-action space scales exponentially with respect to the number of agents \cite{qu2021scalable}, the complexity of the optimization problem and the computational cost required to render the optimization impractical to learn any useful joint behaviors. Other key factors, such as the pathologies of stochasticity and deception, play a major role in contributing to this complexity.

\paragraph{Stochasticity} In many real-world MAS applications, stochastic rewards or transitions pervade, arising from factors such as noise or unobservable elements in the state space. When a game is inherently stochastic, the challenge of identifying the sources responsible for the stochasticity can lead to learning instabilities. This, in turn, requires agents to accumulate more experience to discern and adapt to these sources of uncertainty \cite{palmer2020thesis}.

\paragraph{Deception} The pathology of deception involves a game that contains certain states that have a high local reward but low return. In other words, there exist settings where a trade-off should be made from immediate reward for long-term success. For many MARL algorithms, especially those derived from maximum-based learners \cite{Kapetanakis2003coordinate, matignon2012independent}, over-estimation is a key challenge that leads to early convergence to sub-optimal equilibrium. 

In general, to address the challenge of computational complexity in MARL applications, it becomes even more crucial to design our optimization model carefully such that the modeled components are not susceptible to fail to converge or to find good approximate equilibrium states during learning. With these considerations, we need to balance the trade-offs between solution complexity, computational efficiency, and learning performance to effectively devise a tractable MARL solution.

\subsubsection{Non-stationarity} Informally, a stationary process can be classified by its underlying distribution's invariance under time shifts. In a multi-agent setting, the decisions of each agent impact the transition dynamics of the environment. Consequently, from the perspective of each agent, the other agents' behaviors are inherent components of the dynamics. Therefore, whenever these external agents adapt and alter their behaviors, the underlying model distribution in the perspective of the agent also changes, rendering it non-stationary \cite{Hernandez2017Survey}. Importantly, this non-stationarity arises not from a stochastic process that can be easily approximated, such as white-noise Gaussian, but rather from the structured learning process of the external agents \cite{Daskalakis2022EC}.

This non-stationarity is a critical deviation from the fundamental assumption made by conventional single-agent RL algorithms \cite{Choi1999Nonstationarity}. The absence of the stationary property destabilizes the optimization process and contributes to the pathology of the moving-target problem. This problem arises from the fact that what an agent has learned and needs to learn is dependent on other agents' evolving behaviors \cite{Tuyls2012MAL}. Hence, the learning landscape the agents are optimizing over is in flux at each update step. This setting can make learning infeasible for agents to properly converge to a stable behavior \cite{papoudakis2019dealing}. Therefore, it is essential to consider extensions of methods that can effectively account for this non-stationarity to develop stable algorithms for MARL.

\subsubsection{Coordination} One of the unique aspects of developing a multi-agent solution is the ability of the agents to work together to achieve their goals. As each agent makes decisions based on its local observations in a shared environment, they can heavily benefit from coordinating their actions to achieve a joint strategy that maximizes the collective return and avoids unintended interference to mitigate diminishing returns \cite{Cai2007collision}.

However, achieving successful coordination is difficult, notably when agents have limited information about the environment and the behaviors of other agents. Addressing this challenge of coordination has been a significant research question in MARL that studies how agents can engage with one another in various manners depending on their settings while also effectively succeeding in their local tasks. Specifically, these efforts towards successful coordination are directly associated with the learning pathologies of miscoordination and relative overgeneralization.

\begin{table}[]
    \centering
    \begin{tabular}{c|ccc}
        & $a$ & $b$ & $c$\\ \hline
        $a$ & $1$ & $-2$ & $0$ \\ 
        $b$ & $-2$& $1$  & $0$ \\
        $c$ & $0$ & $0$  & $0.5$
    \end{tabular}
    \caption{Payoff Matrix of a Climbing Game: Motivated by the climbing game proposed in \cite{Kapetanakis2003coordinate}, this game consists of two agents, each with three possible actions $(a,b,c)$.}
    \label{tab:bimatrix}
\end{table}

\paragraph{Miscoordination} The pathology of miscoordination is also known as the Pareto-selection problem and can be observed when two or more incompatible Pareto-optimal equilibria are present. As a consequence, the agents can potentially choose an action from different equilibria due to improper coordination and therefore harming their performance. For example, given a bi-matrix game defined in Table \ref{tab:bimatrix}, there exists two equilibria, $(a, a), (b,b)$, where either both agents choose action $a$ or $b$. However, these equilibria are incompatible, since the strategy profiles $(a,b)$ and $(b, a)$ come from either equilibria, however, if they are intermixed, neither resulting strategies are optimal.

\begin{definition} [Incompatiable Equilibria]
    Two equilibria $\pi$ and $\pi^{'}$ are incompatible if and only if,
    $$ \exists_{i} \pi_{i} \neq \pi^{'}_{i}, J_i([\pi_i,\pi^{'}_{-i}]) < J_i(\pi^{'})$$
    where $[\pi_i,\pi^{'}_{-i}]$ signifies a strategy profile using agent $i$ action from $\pi$ and the other agent's action from $\pi^{'}$.
\end{definition}

\paragraph{Relative Overgeneralization} The pathology of relative overgeneralization occurs in games where, as a result of a shadowed equilibrium, the agents converge upon a sub-optimal Nash Equilibrium that is Pareto-dominated by at least one other Nash Equilibrium.
\begin{definition} [Shadowed Equilibrium]
An equilibrium $\pi$ is shadowed by another one $\pi^{'}$ if there exists an agent that receives a low return by unilaterally deviating from this equilibrium and if this return is lower than the minimal return when deviating from the other equilibrium \cite{palmer2020thesis}. 
$$\exists_{i} \exists_{\bar{\pi}} J([\bar{\pi}_{i},\pi_{-i}]) < \min_j J([\bar{\pi}_{j},\pi^{'}_{-j}])$$
\end{definition}

\begin{figure}[b]
    \centering
    \includegraphics[width=0.75\linewidth]{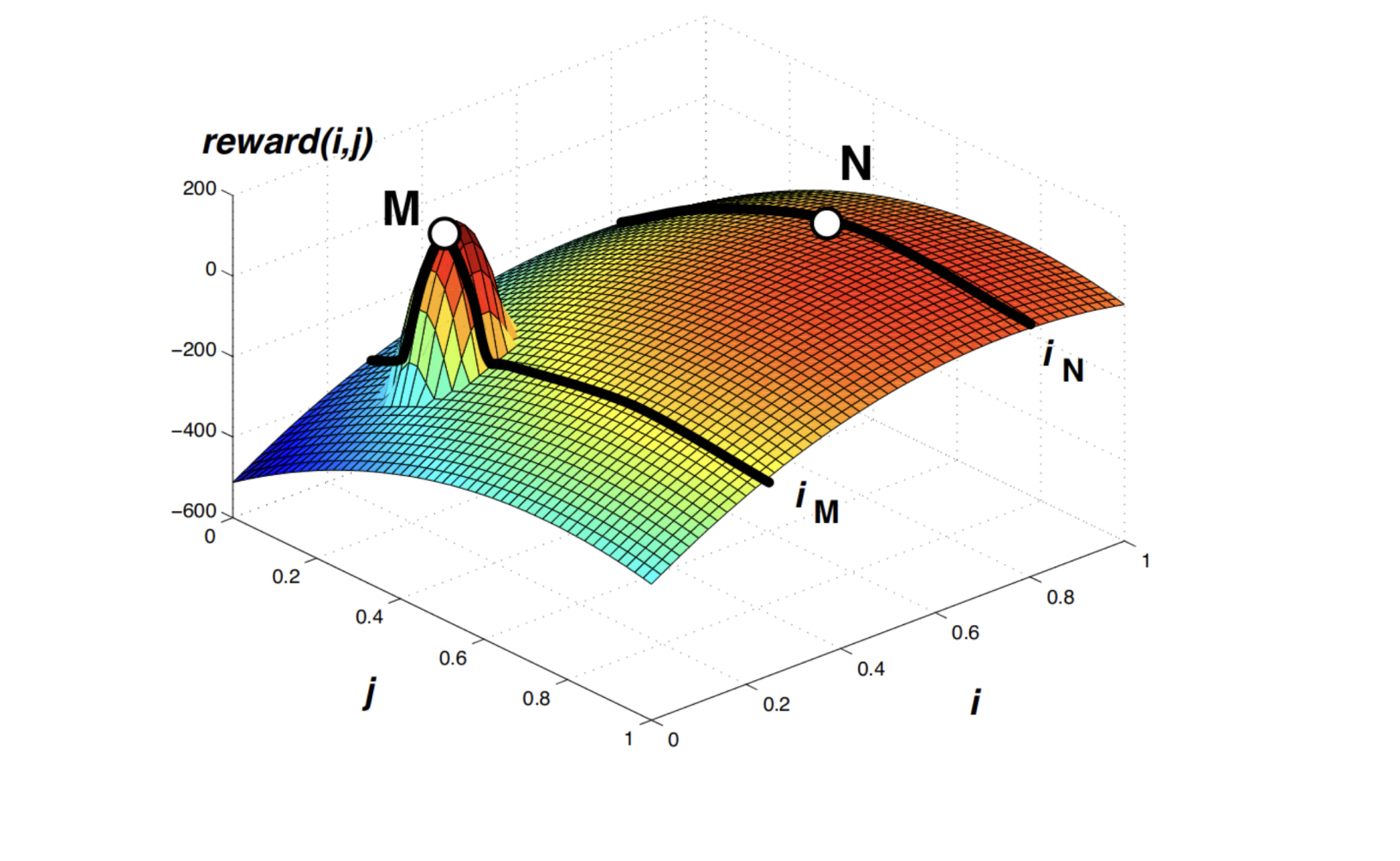}
    \caption{Visualization of relative overgeneralization in a two-player game from \cite{Wei2016lenient}.}
    \label{fig:relativeovergeneralization}
\end{figure}

For instance, in the bi-matrix game shown in Table \ref{tab:bimatrix}, while $(a, a),(b,b)$ are both Pareto-optimal equilibria, a miscoordination penalty of $-2$ is associated with both of them. No such penalty exists with action $c$. Hence, both equilibria are shadowed by $(c,c)$, as the expected gain if one agent deviates unilaterally from either equilibrium is inferior to the lowest expected gain if one agent deviates unilaterally from $(c,c)$. Hence, agents can be drawn to sub-optimal but wide peaks in the return space due to a greater likelihood of achieving beneficial collaboration. This derives as a form of action shadowing.

\begin{definition} [Action Shadowing]
Action shadowing is a phenomenon wherein one individual action seems more favorable than another, despite the potential greater return of the second action \cite{fulda2007shadowing}. Concretely, a sub-optimal policy may result in a higher average payoff when paired with arbitrary actions chosen by other agents, leading to utility values of optimal actions being underestimated.
\end{definition}

The pathology of relative overgeneralization is visualized in Figure \ref{fig:relativeovergeneralization} over a continuous action space. The $x$ and $y$ axes represent the actions of agents $i$ and $j$ respectively. The $z$ axis represents the reward for each joint action. The reward space is structured where action $i_M$ can lead to the optimal reward, however, due to miscoordination being less severely punished for actions approaching $i_N$, the agents are drawn towards the
sub-optimal Nash equilibrium.

\paragraph{Exploration} 
The challenge of exploration in MARL describes the issue of how to effectively explore unknown environments to collect valuable experiences that benefit the agents' learning the most \cite{Sutton2018RL}. To address this challenge, a balance between exploration and exploitation must be struck, where agents must decide whether it is more valuable to take the actions that they know would lead to good returns or take the actions they have not tried yet that may lead to even greater returns or at the very least reduce the uncertainty regarding those actions. An improper balance on either side can result in an incomplete coverage over the state-action space, leading to sub-optimal convergence. This challenge is especially exacerbated in intricate environments with sparse and delayed reward signals, noisy transitions, long horizons, and non-stationary dynamics \cite{Hao_2023}. In addition, in light of the inherent delicacy involved in optimizing and coordinating multi-agent systems under the influence of shadowed equilibria and miscoordination penalties, such exploration can increase the likelihood of deceptive transitions and introduce instabilities within the learning dynamics \cite{matignon2012independent}. For instance, other agents may adapt to these exploratory actions too abruptly even though they may lead to a shadowing equilibrium. We call this pathology the alter-exploration problem. Another key issue noted by prior efforts is the lazy agent problem \cite{liu2023lazy} when certain agents learn a good policy but some agents have less incentive to continue to explore and learn themselves, as their actions may negatively affect the already high-performing agents. For example, as discussed in \cite{sunehag2017valuedecomposition}, consider the scenario of training a soccer team with the number of goals as the team's reward signal. If certain players are more proficient scorers than others, it becomes evident that when the less skilled player takes a shot, the outcome is less favorable on average. Consequently, the weaker player learns to avoid taking shots \cite{Hausknecht2016thesis}.

Traditionally, simple exploration methods, such as $\epsilon$-greedy \cite{Sutton2018RL} or noise perturbation \cite{fujimoto2018addressing}, can be employed for random action selection, however, such naive methods can lead to unintentional and indiscriminate exploration which can be inefficient in complex learning tasks with exploration challenges. While exploration remains an open challenge with much room for improvement, there exists more studied and developed exploration methods, as follows:

\begin{itemize}
    \item \textit{Uncertainty-oriented Exploration: } With a lack of knowledge regarding certain actions, agents can incorporate this uncertainty into the decision-making process when tackling this balance between exploration and exploitation. A common heuristic to employ is the principle of ''Optimism in the Face of Uncertainty", where agents are incentivized to explore state-action pairs with high epistemic uncertainty. Epistemic uncertainty represents the errors that arise from insufficient and inaccurate knowledge about the environment whereas aleatoric uncertainty represents the inherent randomness of the environment. Typically, this approach requires some modeling of these uncertainties. A common approach is to parameterize the solution as a distribution \cite{Zhu_2020, sun2021dfac, zhao2022mcmarl} to be able to properly express the stochasticity of the environment and leveraging a classic exploration technique utilizing these estimates.
    \item \textit{Intrinsic Motivation-oriented Exploration: } An alternative approach is to incorporate a meta-task of exploration by introducing and designing intrinsic rewards for agents. These rewards can be to minimize prediction errors regards the environment \cite{zheng2021episodic}, motivated by novelty of states \cite{mahajan2020maven, iqbal2021coordinated, liu2021cooperative} or driven by information gain \cite{houthooft2017vime}. For instance, in \cite{du2019learning}, the individual intrinsic reward is learned and used to update an agent’s policy to maximize the team reward.
    \item \textit{Multi-agent Exploration: } In a multi-agent setting, we face the challenge of not only complexity but also miscoordination (i.e. alter-exploration problem). This issue requires some level of coordinated exploration \cite{Hao_2023}, as exploratory actions of one agent can affect the learning of others \cite{palmer2018lenient}, and in certain multi-agent tasks, efficient exploration requires a degree of global planning as opposed to pure local exploration \cite{Brafman_2003}.
\end{itemize}

\subsubsection{Performance Evaluation}\label{performance_eval} Assessing the performance of a MARL algorithm is a multifaceted challenge. Firstly, we highlight that the success of one agent's policy is intricately linked to the policies of other agents, which renders individual assessments unclear to properly interpret. This is compounded by the problem of establishing unbiased and useful metrics that quantify inherently qualitative social behaviors, such as measuring the quality of communication and coordination \cite{havrylov2017emergence, jaques2019social, bogin2019emergence, lowe2019pitfalls} and determining the appropriate evaluations dependent on the roles assigned to all agents (i.e. oligopoly with leader-follower structure).

Furthermore, we can focus on more egalitarian criteria, ie. concentrating on the agents that are most struggling \cite{Zhang2014} as opposed to an more utilitarian approach such as social welfare. Hence, it also becomes important to understand multi-agent credit assignment and discern the individual impact of each agent in terms of the coalition's utility, especially where there exists only a global reward structure, e.g. within a Dec-MDP. A classic example is the Shapley value, which quantifies and captures the notion of marginal contribution by averaging all possible combinations of the marginalized population's achieved utility \cite{Shapley1952}. More recently, a popular approach in deep MARL is to utilize an advantage function, which compares the current Q-values to a counterfactual \cite{foerster2017counterfactual, li2022dae}, or to utilize value function decomposition \cite{sunehag2017valuedecomposition} that marginalizes the contribution of each agent.

From a theoretical perspective, determining which solution concepts would lead to optimal behaviors and strategies, as studied in game theory literature, remains unclear and often task-specific \cite{bergerson2021multiagent}. Even so, as many studied solution concepts prove prohibitively costly to explicitly measure and optimize with complex and dynamic tasks, the research for definable metrics that capture the nuances of diverse MAS settings persists as an ongoing and open challenge.

\section{Prospects of MARL}\label{section: prospects}
In this section, we cover the unique properties of learning controls in a multi-agent environment that can help promote the benefits or address the underlying issues of the MARL approach \ref{section:learning in mas}.

\subsection{Simulating MARL Tasks} \label{subsection: simulation}
A lingering concern for MARL algorithms is its learning complexity \cite{daskalakis2022complexity}. Often, in order to acquire valuable behaviors, it requires significant computational resources. This constraint arises from various factors, namely sample inefficiency and a brittle optimization landscape \cite{gu2017qprop, vanhasselt2018deep, tucker2018mirage, henderson2019deep}.


Sample efficiency is broken down into two factors: the number of environment interactions required by each agent to learn and generalize and the cost associated with each interaction \cite{Huh2023isaacteams}. Hence, it is important to emphasize the pivotal role that simulators play in this equation, as they can enhance the learning process by yielding higher-quality samples through greater accessibility, stability, accuracy, and precision of the retrieved data.

\begin{figure}
    \centering
    \includegraphics[width=0.75\linewidth]{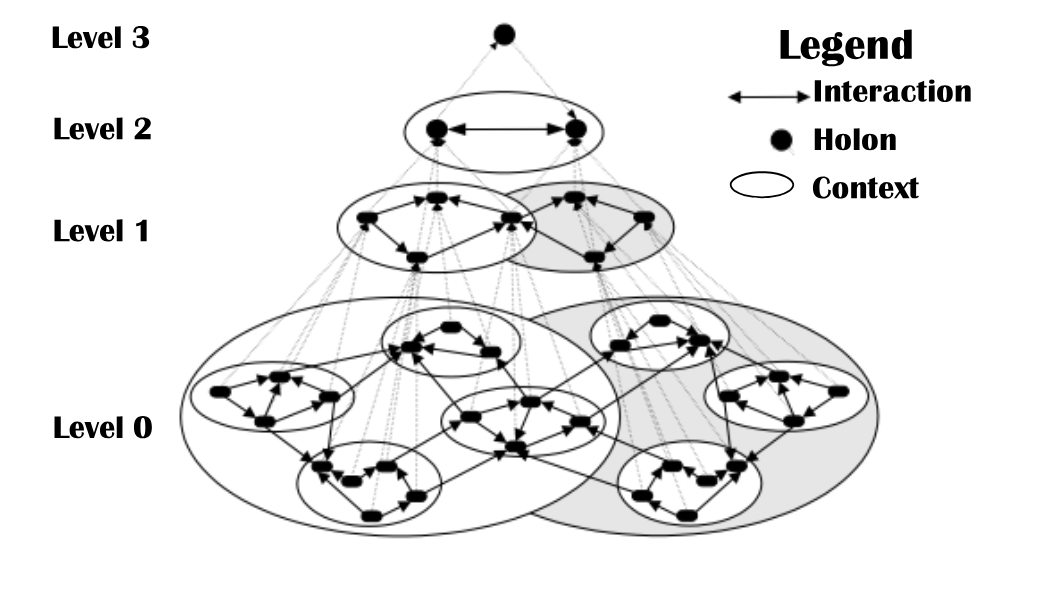}
    \caption{A general diagram of a holonic/multilevel simulation taken from \cite{tchappi2018brief}.}
    \label{fig:holonic}
\end{figure}

\paragraph{Parallelized and Vectorized Processing} Over the past several decades, the processing capabilities of massively parallelizable processors, such as graphic processing units (GPUs) and tensor processing units (TPUs), have advanced significantly. This progress has opened up exciting possibilities for leveraging this technology to simulate highly complex tasks effectively, minimizing the cost of simulating a large number of trajectories. These processors are specifically designed to handle parallel computations efficiently by executing tasks simultaneously on a large scale, enabling simulations to run orders of magnitude faster compared to conventional CPU-based implementations. There have been notable efforts to develop task-specific \cite{Dalton2019atari} and general physics-based simulators, such as IsaacSims \cite{makoviychuk2021isaac}, Brax \cite{freeman2021brax}, and MuJoCo XLA that leverage GPUs. Many of these simulators primarily concentrate on addressing single-agent problems, with some addressing their extension to MAS \cite{chen2022towards, gu2023safe, Huh2023isaacteams, flair2023jaxmarl}. In turn, there exists promising potential to further optimize and extend the usage of vectorized processing to MAS scenarios \cite{lan2021warpdrive, lechner2023gigastep}.

\paragraph{Multilevel Simulation}
Many complex multi-agent tasks can be factorized into a hierarchical structure using a multilevel simulation paradigm, as an effort to manage its complexity. Multilevel simulation introduces several organizing levels that encapsulate various individual components into monolithic abstractions \cite{ghosh1986concept}. These levels are defined in distinctive manners, integrating microscopic to macroscopic attributes \cite{haman2017towards}. Such modeling is defined in a holonic paradigm \cite{tchappi2018brief} (see Figure \ref{fig:holonic}), where holons are used as these abstractions and holons are defined as stable self-similar structures that behave as both an entity and an organization. Holons satisfy three important conditions: holons are stable, autonomous, and cooperative with one another. However, a holonic organization of a task is often difficult to properly define.

\paragraph{Open Source Environments} There exists a wide range of libraries that are used as common benchmarks for MARL research. Table \ref{tab:env} provides descriptions of some of the more widely recognized environments.

\begin{table}[htbp]
    \centering
    \begin{tabular}{p{2in}|p{3in}}
        \hline
        \hline
        \Centering{Name} & \Centering{Description}\\
        \hline

        \Centering{Multi-agent Particle Environment (MPE) \cite{lowe2020multiagent}} &
        \RaggedRight{Various social tasks focused on communication within a particle world setting.}\\
 \hline
        \Centering{StarCraft Multi-Agent Challenge (SMAC)\cite{samvelyan19smac, ellis2022smacv2, phan2023attention}} &
        \RaggedRight{Cooperative StarCraft decentralized micromanagement scenarios.}\\
 \hline
        \Centering{PettingZoo \cite{terry2021pettingzoo}} &
        \RaggedRight{A collection of different MARL tasks and libraries.}\\
 \hline
        \Centering{MA-Gym \cite{magym}} &
        \RaggedRight{Multi-agent tasks in a grid-world setting.}\\
 \hline
        \Centering{MAgent \cite{zheng2017magent}} &
        \RaggedRight{A collection of many-agent tasks.}\\
 \hline
        \Centering{Level-based Foraging (LBF) and Robot Warehouse (RWARE) \cite{christianos2020shared}} &
        \RaggedRight{Customizable grid-world foraging task and simulation warehouse with robots moving and delivering product in gridworld}\\
 \hline
        \Centering{Google Research Football \cite{kurach2020google}} &
        \RaggedRight{Simulated soccer game using physics-based 3D simulator.}\\
\hline
        \Centering{Overcooked \cite{carroll2020utility}} &
        \RaggedRight{Human-AI coordination on multiplayer video-game task.}\\
\hline
        \Centering{Vectorized Multi-Agent Simulator (VMAS) \cite{bettini2022vmas}} &
        \RaggedRight{Various MARL tasks using a vectorized Pytorch-based 2D physics engine}\\
\hline
        \Centering{IsaacTeams \cite{Huh2023isaacteams}} &
        \RaggedRight{Various physics-based MARL tasks using GPU-accelerated IsaacSim platform.}\\
\hline
        \Centering{JaxMARL \cite{flair2023jaxmarl}} &
        \RaggedRight{Various physics-based MAS tasks using GPU-accelerated Brax platform.}\\
        \hline
        \hline
    \end{tabular}
    \caption{List of open-source environments}
    \label{tab:env}
\end{table}

\subsection{MARL Training Schemes} \label{subsection: marl training}
In this part, we explore various training paradigms used in and unique to MARL applications. Specifically, we look into the centralized training paradigms and the use of off-policy learning.

\subsubsection{Centralized Training}
There exists a spectrum of agent representations in training and execution to combat the scalability and complexity issues of learning joint strategies. This spectrum includes three main categories: centralized training and centralized execution (CTCE), decentralized training and decentralized execution (DTDE), and centralized training and decentralized execution (CTDE) \cite{lowe2020multiagent}. The idea of centralization couples components of agents' behaviors to provide a more complete state of information to work with or to decrease the complexity of the task by distributing the workload over multiple agents. Importantly, leveraging some level of centralization poses a good solution for handling non-stationarity, as each agent now will have access to global information to account for the changes in other agents' behaviors.

\paragraph{CTCE} The CTCE paradigm entails a fully centralized approach that involves mapping a collection of local observations from each agent to distributions over individual action spaces. In this case, we essentially reduced the MAS control problem into a single-agent RL optimization over the concatenated observations and combinatorial joint action space. While CTCE provides expressive and complete representations of a MAS \cite{Gupta2017CooperativeMC} and performs well against non-CTCE methods \cite{yu2022surprising}, the assumptions of decentralization are largely compromised. This is because, during execution, decentralized agents must only make decisions based on their local observations and do not have access to the global information it was trained on. Therefore, nontrivial and unnatural adjustments must be made to convert the joint policy from a centralized executor to a decentralized executors, such as masking the other agent's information to prevent information leakage. CTCE approaches also fail to address the curse of dimensionality problem \cite{Gronauer2022survey}, otherwise expressed as the exponential scaling caused by the joint state-action space of MAS.

\paragraph{DTDE} On the other side of the spectrum, DTDE proposes a fully decentralized approach that adheres to all decentralization constraints in all aspects of training and execution. We note that this does not necessarily mean the agents cannot perceive nor is aware of other agent's existence, where this is known as independent learners (IL) vs. joint action learner (JAL) as defined in \cite{Claus1998dynamics}, although IL can be considered an extreme form of DTDE. Prior efforts \cite{Claus1998dynamics, Tan1997independent, Lauer2000, tampuu2015multiagent, Jaderberg_2019} have demonstrated that IL with standard RL algorithms does demonstrate the ability to converge to an equilibrium in particular and fine-tuned settings. A key challenge in the DTDE training scheme, as notably emphasized, is non-stationarity. This challenge is exacerbated by the absence of centralization, leading to a potential loss of mutual information among agents' behaviors, which, in turn, can give rise to various learning pathologies \cite{palmer2020thesis}. To mitigate these pathologies, a widely adopted strategy involves inducing optimism \cite{Matignon2007, palmer2018lenient, lyu2020likelihood}, through hysteric learning or leniency. This approach restricts the reduction of value estimations, thereby alleviating the impact of other agents' exploration strategies and promoting exploration beyond equilibria that can easily trap agents without the added optimism.
\begin{itemize}
    \item Hysteric learning: While DTDE has demonstrated success in deterministic settings, independent learning has struggled to replicate such achievements in stochastic settings. A significant stumbling block has been the tendency to overestimate the value function, a consequence of the inherent stochasticity, resulting in sub-optimal solutions, as stated in \cite{Matignon2007}. To address this issue, hysteric Q-learning \cite{Matignon2007} was introduced to provide an optimistic update function that assigns greater weight to positive experiences, particularly beneficial in cooperative multi-agent scenarios. This is achieved through the use of two learning rates, denoted as $\alpha$ and $\beta$. The larger learning rate, $\alpha$, is applied when updating Q-values following positive value updates, while $\beta$ is utilized otherwise.
    \item Leniency: Alternatively, another method to adjust the degree of optimism during the learning process is leniency \cite{palmer2018lenient}. Leniency effectively allows for the forgiveness or disregard of suboptimal actions taken by teammates that result in low rewards during initial exploration, taking in the form of lenient Q-value updates and lenient-based exploration. Over time, this optimism exhibited by lenient agents is gradually reduced as they encounter and revisit state-action pairs. Consequently, agents become less lenient in situations frequently encountered, while retaining their optimistic outlook in unexplored territories. This shift towards average-based reward learning from maximum-based, helps lenient agents steer clear of suboptimal joint policies, especially in environments where rewards are subject to stochastic fluctuations. Empirically, leniency shows higher learning stability compared to hysteretic learning, primarily due to temperature-enabled leniency at different stages of estimation maturity. The leniency decay allows for a more faithful representation of domain dynamics during later stages of training, where it is probable that teammate policies become stable and near-optimal, assuming the rate of decay is appropriate and value maturity is synchronized across all states.
\end{itemize}
Like CTCE, DTDE also demonstrates a significant issue of scalability, as a distributed solution requires each agent to not only be represented individually but also require their own set of samples for learning. As agents are not granted any access to the global state, which can be pivotal not only for sample efficiency but also for good performance 
\cite{Gupta2017CooperativeMC}.

\paragraph{CTDE} CTDE provides a middle-ground by centralizing certain variables during training that still enable decentralized execution of agents \cite{Kramer2016ctde}. This approach strikes a balance between the advantages of centralization while maintaining the constraints set by natural and artificial decentralization during execution. CTDE is commonly practiced by using a centralized value function.

Centralization of the value function allows all agents access to comprehensive state information during training, without violating decentralization constraints during execution since the value function is not required for decision-making. This facilitates enhanced learning, efficient updates, and coordination between actor and critic, promoting improved policy convergence \cite{foerster2017counterfactual, lowe2020multiagent}. Additionally, single-agent RL algorithms can naturally be extended, similar to DTDE, however, using a shared critic model (i.e. MAPG \cite{samvelyan19smac}, MADDPG \cite{lowe2020multiagent} and MAPPO \cite{yu2022surprising}).

Like CTCE, a centralized value function remains prone to scalability issues. As the number of agents grows, its representation needs to handle a larger or more intricate state space. This can lead to significant computational costs and difficulties in defining the concise state space. Another critical challenge is the centralized-decentralized mismatch. Since the value function is shared among agents, sub-optimal policies from one agent can have a detrimental impact on the policy learning of other agents, causing catastrophic miscoordination \cite{wang2020offpolicy}. Largely, this increased variance in learning a shared critic remains a long-standing challenge.

\paragraph{Parameter Sharing} A commonly used approach to implement centralization is through parameter sharing, where different agents share representation modes \cite{gupta2017learning}. This allows each agent to update the same parameters, potentially leading to a more efficient and richer learning process. In a sense, all agents can aggregate their experience and learn much faster in a more memory-efficient manner \cite{fu2022revisiting}. However, it is important to note that this can more likely lead to homogeneity in behaviors and introduce instability, especially when dealing with highly diverse agents, as it transforms the problem into a difficult multi-task optimization problem. Parameter sharing among such agents, especially if they are heterogeneous, becomes a nontrivial task.

\paragraph{Subtask Sharing} Another approach to centralization involves the use of global subtasks. In many MAS, tasks can be decomposed into subtasks universally defined amongst all agents. As a result, each agent's task can be decomposed, where these global subtasks can be assigned accordingly to each agent. To achieve this, the process of task decomposition needs to consider how the subtasks can be defined properly. While subtasks can be defined using domain knowledge \cite{spanoudakis2010using}, more generalizable decomposition methods, such as RODE \cite{wang2020rode} and LDSA \cite{yang2022ldsa}, have been introduced. The core idea behind both approaches is to learn embeddings over the actions or trajectories and perform clustering to define subtasks. A significant challenge in these approaches is ensuring the subtasks' definitions are distinct \cite{yang2022ldsa}.

\subsubsection{Off-policy Learning}
To improve the sample efficiency of MARL training, agents can learn from experiences that come from different policies in some manner \cite{Sutton2018RL, silver2014dpg}, and this is known as off-policy learning. Typically, off-policy approaches rely on storing samples in an experience replay buffer. However, proper implementation of experience replay in a multi-agent setting is non-trivial due to the non-stationary dynamics of the environment that can render past experiences obsolete, as other agents' behaviors change, learning with their prior behaviors may be out of distribution. Previous efforts \cite{foerster17er} account for these discrepancies through two methods. The first approach utilizes importance sampling to correct the policy updates, however, there remain questions about the tractability of computing the importance weightings and its large and unbounded variance. Extensions to alleviate these issues, including truncation, do reduce variance, however, introduce additional bias. Fingerprinting, on the other hand, appends contextual information regarding the current stage of learning of the agents in the environment to the samples that are stored in the experience replay buffer, to disambiguate the age of the sample. Both approaches prove to stabilize the experience replay sufficiently. 

Aside from the depreciation of samples, other issues have been addressed, such as ensuring concurrency of experience sampling \cite{omidshafiei2017deep} and detecting to manage miscoordination and relative over-generalization with off-policy learning by using variable learning rate to accommodate for exploratory actions \cite{palmer2019negative, lyu2020likelihood}. Traditional experience replay mechanisms, such as priority replay \cite{schaul2016prioritized}, have been experimented with MARL. However, a naive application may deteriorate convergence and performance due to the noisy reward and the continuous behavior changes of coexisting agents, causing a priority bias. Hence, a lenient reward function is modeled \cite{zheng2018weighted} to correct the priority bias.

\subsubsection{Offline Learning} The practice of offline learning with MARL  adopts much of the same ideas from single-agent offline RL, existing in the form of behavior regularization and conservatism \cite{pan2022plan}, but with further considerations required to mitigate underlying issues that come along with the mixture of offline RL and MAS, such as the propagation of extrapolation error \cite{yang2021believe} and agent-wise imbalances within the offline data \cite{tian2023learning}. 

\subsection{Agent Awareness}
Agents can exhibit different degrees of awareness of other agents, which can be classified into three categories: independent, tracking, and agent-aware \cite{Busoniu2008ComprehensiveSurvey}. The selection of an awareness level involves unique considerations, advantages, and challenges, contingent upon the specific nature of the interaction and the task. At each level, a trade-off between stability and adaptability is made.

\begin{definition} [Stability and Adaptability]
Stability focuses on achieving convergence to a stationary policy $\pi$, while adaptability aims to maintain or improve performance in the face of changes in other agents' behaviors. 
\end{definition}

\begin{definition}[Stationary Policy]
A policy $\pi^i$ is stationary, for any state $s^i$ and time-steps $t$ and $t'$, where $t \neq t'$,
$$|\pi^i_t(a|s^i_t) - \pi^i_{t'}(a|s^i_{t'})| \leq \epsilon, \forall a\in A$$
\end{definition}

While stability and adaptability are not necessarily dichotomous objectives, the balance between the two helps illustrate the extent to which coordination is emphasized. Effectively addressing this coordination problem requires agents to skillfully navigate these intricacies and strike the right balance between focusing on stability and adaptability.

\begin{itemize}
    \item Independent agents disregard the notion of coordination entirely. Moreover, their focus lies solely on converging to stable behaviors rather than adapting to the actions of other agents in their environment. In cooperative, adversarial, and mixed settings, such methods are referred to as coordination-free, opponent-independent, and agent-independent respectively, and each has demonstrated empirical success under restricted problem settings \cite{Littman2001, Lauer2000, Hernandez2017Survey}. However, these independent methods often result in sub-optimal outcomes or even failure to achieve desired goals as coordination becomes crucial in many scenarios to anticipate and respond strategically to the actions of other agents.
    \item Tracking agents prioritize adaptability over stability, placing a greater priority on coordination rather than learning a stable individual behavior. With a tracking approach, agents continuously adjust their strategies based on the observed behavior of other agents. Empirically, agent-tracking methods rely on agent modeling to guide the agents' action selection process \cite{Robinson1951FicitiousPlay, Weinberg2004}. However, the stability of the joint behavior may be compromised. The constant adaptation can lead to non-stationary behaviors, as the agents respond not only to changes in the environment but also to the changing strategies of other agents.
    \item Agent-aware agents strive to achieve a balance between stability and adaptability by being conscious of the other agents' strategies while preserving their individuality. Previous studies have explored approaches such as "Adapt When Everyone is Stationary, Otherwise Move to Equilibrium" (AWESOME) \cite{conitzer2003awesome} or "Win or Learn Fast" (WoLF) \cite{Bowling2004GigaWolf} to determine when to adapt or maintain their local strategy, mostly just by adjusting the learning rate or incorporating the other agent's anticipated learning to one another \cite{foerster2018learning}. These concepts primarily address handling the non-stationarity of the optimization problem, but this heightened awareness also establishes a well-rounded foundation for fostering social behaviors that lead to stable and successful coordination.
\end{itemize}

\paragraph{Learning with Awareness} When learning, agents can take into account the behaviors and information regarding other agents \cite{foerster2017lola}. One way this can be achieved is by extrapolating gradient updates for each agents, thereby performing a one-step look-ahead over the learning over all agents, which is known as extragradient \cite{Korpelevich1976TheEM}. To reduce complexity of extrapolation step of gradient update, we can instead use only a sample subset of agents in many-agent settings \cite{jelassi2020extragradient}. Similarly, LOLA \cite{foerster2017lola} takes a similar approach, however, extrapolates only the other agents using a second order correction term with a Taylor expansion approximation.

\subsection{Multi-agent Credit Assignment}
For many MARL applications, it may be intractable to define local rewards for each agent, hence necessitating the use of the Dec-MDP framework. In such settings, a global reward is instead provided, which represents the collective's utility. However, with this measure, it is unclear the direct contributions and local performances of each agent. This problem is known as multi-agent credit assignment (MACA) \cite{agogino2008analyzing}. We distinguish MACA from the traditional credit assignment problem associated with the casual aspects of sequential decision-making, where the actions themselves are evaluated on their impact \cite{Sutton2018RL}. In recent efforts, the challenge of MACA, as briefly mentioned in Section \ref{performance_eval}, is addressed in two approaches: difference rewards and value factorization.

\paragraph{Difference Rewards} Difference rewards aim to capture an agent’s contribution from a global performance measure by shaping a local reward signal that isolates the utility of individual agent's actions by removing the utility of other agents \cite{agogino2004unifying}. While in some applications, such as air traffic flow management \cite{tumer2007distributed}, this isolation is possible, generally, forming a theoretical setting that removes agents individually may be impossible. Hence, the marginalization of individual agents is often estimated by comparing them against the average actions, known as aristocrat utility \cite{wolpert2001optimal}. To extend to the realm of deep RL, COMA \cite{foerster2017counterfactual} leverages the concept of the advantage function to achieve the same effects.

\paragraph{Value Factorization} Value factorization decomposes a global value into local values for each agent \cite{guestrin2001}. A simple implementation of this is known as a value decomposition network (VDN) \cite{sunehag2017valuedecomposition}, where the sum of the learned local value functions can be treated as the global value. With the many extensions of VDN that have been proposed in the past decade, the standard constraints of Individual-Global-Max (IGM) \cite{rashid2018qmix} serve as the theoretical basis for guaranteeing and maintaining consistency between the global $Q$ and local $q_i$ value estimates.
\begin{equation}  
\argmax_a Q(s,a) = \begin{pmatrix} \argmax\limits_{a_0} q_0(\tau_0, a_0) \\ \vdots \\ \argmax\limits_{a_N} q_N(\tau_N, a_N) \end{pmatrix}
\end{equation}
where $\tau_i$ is the observation-action history of agent $i$. However, these constraints have been shown to restrict the expressiveness of the value function representations \cite{mahajan2020maven}, leading to sub-optimal value approximations and poor explorations. Hence, it remains a open research challenge to improve these limitations of value decomposition while trying to adhere to IGM.

\subsection{Communication}\label{subsection: communication}
Communication is a powerful capability of high interest within MARL literature that enables agents to exchange and propagate information between one another, leading them to behave as a collective rather than a collection of independent individuals \cite{zhu2024survey}. In many multi-agent tasks, communication proves vital to coordinate the behavior of multiple agents to achieve optimal performance \cite{foerster2016learning}, especially under settings with imperfect information and partial observability \cite{zaïem2019learning}.

However, an efficient and practical implementation of a communication mechanism presents several key challenges, necessitating consideration of not only what information to communicate \cite{sukhbaatar2016learning, foerster2016learning}, but also how \cite{shao2022self}, when \cite{singh2018learning}, and with whom \cite{jiang2018learning} to communicate. In \cite{zhu2024survey}, the topic of communication is broken down and categorized over $9$ dimensions on its implementation, so we recommend readers refer to this resource for more in-depth analysis.

\subsubsection{Communication Infrastructure}
A communication graph is introduced to define which agents each agent can communicate with, alongside the use of the networked stochastic game framework. The restriction placed on the existence of the graph's edges is bounded by the constraints of decentralization and requires solutions that address issues including limited range \cite{huh2023decentralized}, limited bandwidth \cite{foerster2016learning}, noisy communication channels \cite{Freed2020}, and contentions with shared communication mediums \cite{kim2019learning} such as a proxy. A consideration for each is understanding their practicality, which is dependent on the nature of the task, as well as their shortcomings.

\paragraph{Proxy} While decentralization prohibits centralized executors, a centralized communication medium is not prohibited. The role of a proxy is to serve as a coordinator and message aggregator. It gathers local observations or messages from agents in the environment, subsequently broadcasting messages to each of them \cite{kong2017revisiting}. Alternatively, it can connect nearby agents who opt to participate in a communication group, facilitating the sharing of coordinated messages with each group member \cite{jiang2018learning}. Canonically, a proxy solely acts as a communication medium, having no direct effect on the environment. A key challenge of a proxy is its design. Solutions that use proxies must consider their efficiency, ensuring that sufficient communication is achieved for the task at hand while also managing the computation load and expressiveness of the proxy.

\paragraph{Networked communication} In contrast to relying on a proxy for inter-agent communication, the networked communication protocol consists of agents that pass and receive messages directly to and from other agents \cite{zhang2018fully, chu2020multiagent}. While this form of communication may seem most fitting for a decentralized setting, its dynamic nature and lack of structure can lead to poor performance and scalability issues, especially when each agent has limited compute resources and is required to process many agents' messages.

\paragraph{Implicit Communication} Agents can also communicate with one another without explicit means, such as through stigmergy \cite{grasse1959reconstruction}. The concept of stigmery defines the influence agents have through their actions on one another, often through environmental changes or some other form of stimuli. More formally, stigmery describes the influence of the persisting environmental effects of prior behaviors on behaviors\cite{holland1999stigmergy}. We categorize the idea of stigmergy based on the intent/form of the stigmergic actions, the responses to the stigmergic behaviors, and the impact of stigmergic actions.
\begin{itemize}
    \item Sematectonic and marker-based stigmergy \cite{wilson2000sociobiology, marsh2008stigmergic} distinguish whether the stigmergic actions were directly aligned with true objectives of the agents or rather, to solely stigmergize. The intent and the actual actions are often considered when classifying the two forms of stigmergy.
    \item Quantitative and qualitative stigmergy \cite{theraulaz1999brief} differentiate whether the response to the stigmergic actions is an intensification of the resulting stimulus or triggering different stimuli, leading to a self-organization process. A self-organizing process refers to a set of dynamical local mechanisms, which through their applications and interactions, causes emergent global structures and behaviors.
    \item Active and passive stigmergy \cite{holland1999stigmergy} refer to the effects and outcomes of the stigmergic actions. Active stigmergy directly affects the agents, influencing the observations, actions, and parameters (e.g. frequency, latency, duration, intensity). However, passive stigmergy is more indirect and subtle, perhaps leading to no changes to any observations, actions, or their parameters, but only to changes to the outcome.
\end{itemize}
Historically, a central focus of stigmery in MARL has been deriving optimization algorithms, such as ant colony optimization \cite{dorigo2005ant}, that are inspired by these concepts. We propose that these patterns and behaviors of stigmergy should be further studied towards other forms of integration into our MAS, including, but not limited to, how to induce stigmergic behaviors and quantify and evaluate them.

\subsubsection{Communication Representation}
In this section, we look into the various forms of communication mechanisms that are used in practice, namely those that were realized using deep learning techniques such as graph neural networks (GNN).

\paragraph{Graph Neural Networks} Facilitating rich communication among agents requires a scalable framework that can naturally process information within the communication graph in an expressive manner. GNNs are a fundamental tool for handling non-Euclidean data, especially when dealing with information naturally occurring in graph structures. 

\begin{figure}
    \centering
    \includegraphics[width=\linewidth]{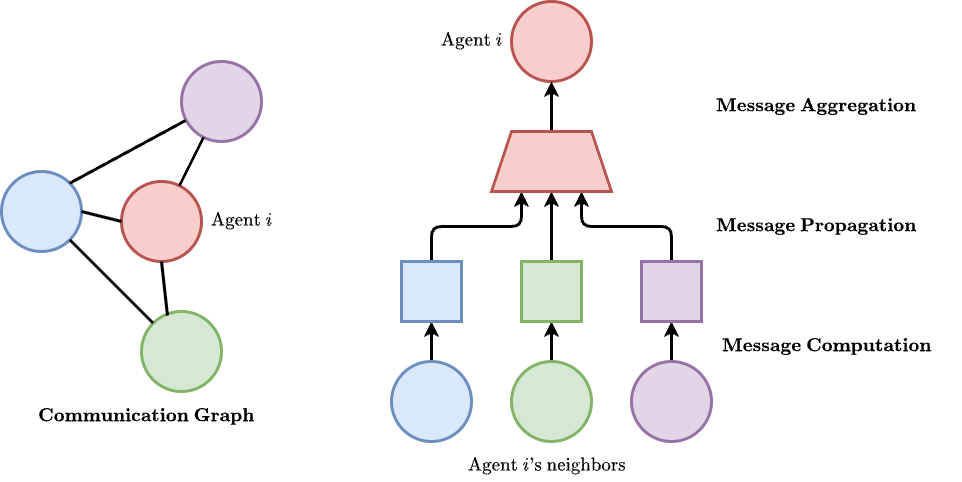}
    \caption{A visualization of the three operations of graph convolutions.}
    \label{fig: gnn}
\end{figure}

Concretely, GNNs learn to map input data to latent representations that can be used in subsequent tasks. GNNs generate these latent embeddings by iteratively performing the following operations: message computation, propagation, and aggregation. A visualization of these operations is provided in Figure \ref{fig: gnn}. Together, these three operations can be collectively described as a graph convolution \cite{kipf2017semisupervised}. Graph convolutions can be performed iteratively, increasing the receptive field and allowing GNNs to capture more global information as the number of message passing rounds increase, i.e. increase the number of GNN layers.

There have been several advancements with GNN algorithms, specifically to increase expressiveness, improve scalability, and importantly, compensate for the over-smoothing problem \cite{xu2018representation}, where learning on densely connected graphs often converged to redundant node embeddings. Current MARL research utilizes GNNs as the de-facto communication mechanism, such as graph convolution network (GCN) \cite{kipf2017semisupervised} in CommNet \cite{sukhbaatar2016learning} and BiCNet \cite{peng2017multiagent}, and graph attention network (GAT) \cite{brody2022attentive} in DGN \cite{jiang2020graph} and ATOC\cite{jiang2018learning}. While many of these GNN models were initially designed for prediction-based tasks such as supervised classification, there have been efforts to bridge these tools into the domain of control such that novel mechanisms are tailored to the intricacies of MAS (e.g. dynamic role assignments \cite{shao2022self} and limited communication \cite{kim2019learning}), although this remains an open problem in MARL.

\subsubsection{Learning to Communicate}
Learning communication involves considering the following three aspects of communication: the content of the outgoing messages, how the incoming messages are incorporated and the communication policy \cite{zhu2024survey}. The optimization itself can be devised to learn all these aspects together or individually and makes use of explicit and/or implicit feedback. This means the communication learning dynamics can use additional feedback signals, such as social influence \cite{jaques2019social}, which may be optimized in conjunction with the MARL training. A common practice is to seamlessly integrate the two learning dynamics of communication and control, often with differentiable modeling and backpropagation \cite{foerster2016learning, sukhbaatar2016learning, Freed2020}.

\paragraph{Message Computation} To initiate communication, each agent must compute messages to broadcast. The content of the message can vary from encoded or non-encoded information regarding the current and/or past local observations, actions, rewards, beliefs, objectives, previously received messages, or any other accessible information regarding the agents and the tasks. This can also include imagined information, such as future imagined trajectories/behaviors or intentions \cite{kim2020communication}. To embed the listed information, an explicit approach is often employed through an auto-encoding process. On the other hand, although not completely orthogonal to the explicit learning approach, an end-to-end learning process is an alternative trained using the MARL learning dynamics without any such grounding. The learning representation for the content can be either discrete symbols or continuous values \cite{foerster2016learning}.

In addition to the content itself, the concept of language is postulated to be an important aspect of generalizable coordinating behaviors, such as learning a lingua franca \cite{lin2021learning}. An interesting property of language often studied is compositionality \cite{lazaridou2018emergence}, which refers to the ability to produce complex meanings by combining simpler linguistic elements and symbols in systematic ways.

\paragraph{Communication Policy} A communication policy defines the properties of the edges in the communication graph, managing the following: the senders and recipients of all messages and the frequency of the message transmissions. This policy must thereby consider and adhere to the constraints of decentralized communication. If a proxy is used, the purpose of the proxy is to handle the operations of the communication policy.

The edges of the communication graph can be statically or dynamically defined. A static implementation can be fixed or make use of some heuristics \cite{huh2023decentralized}, whereas a dynamically defined solution is more involved. A natural approach is to make use of gating mechanisms \cite{jiang2018learning} and communication scheduling modules \cite{kim2019learning} to dictate the formation of the edges in a learnable fashion. This enables us to design communication paradigms dynamically dependent on the needs and constraints of the task. For example, we can regularize the communication overhead with penalty terms that directly impact the parameters of the gating mechanism \cite{hu2020event}.

Another consideration is the frequency of communication between agents. A common assumption is allowing all agents to communicate at every time step, however, this may not be necessary and in such cases, the over-communication can be detrimental in not only cost but also effectiveness due to the greater reliance and need for more expressive and consistent communication measures. To mitigate this issue, communication can be less frequent by setting a more sparse communication cycle \cite{shao2022self} or utilizing a mechanism that enables agents to be more selective, such as the gating mechanism or communication scheduling.

\paragraph{Message Integration} Lastly, we discuss message integration, which refers to how each agent processes and utilizes the messages they receive. A popular approach is to use message aggregation operations, as observed with GNN architectures, involving either a weighted or non-weighted summation over all incoming messages. While maintaining parameters for the weighting over the incoming messages is a great option to increase expressiveness in the aggregation process \cite{Zhang2021Coordinate, li2021deep}, learning the weighting may lead to challenges notably in generalization, as performance may suffer when dealing with ad-hoc team-play scenarios if there is a lack of adaptability.

\subsubsection{Evaluating Communication}
In recent efforts relating to communication, studies have explored evaluation metrics to grow our understanding and quantify the quality of the communication between agents \cite{lowe2019pitfalls}. These efforts range from taking a broader view by observing the changes in performance (i.e. agent's rewards or task success rate) under changes in communication methods to a more granular scope, where explicit metrics are defined that can account for the varying reasons for the impact achieved by communication. A broader view is often taken when communication is vital for the task at hand, often in the form of referential games, a common mode of game used to study the aspect of communication in MARL. Referential games can be thought of as a form of Lewis signaling game, where agents are each assigned the roles of speaker and listeners, and the speaker agent must communicate to the listener agents to complete their tasks \cite{lewis2008convention}. In many cases, the speaker agent has access to some private information not privied to the listener agents, making their communication significant. These specific roles designated to the agents result in a strong reliance on the communication system in place for the success of the task.

The metrics of communication can be divided into two classifications: positive signaling and positive listening \cite{lowe2019pitfalls}. Positive signaling quantifies some statistical dependence of the messages on the agents' observations or actions. Positive listening, on the other hand, is measured by the change in behavior of an agent if its incoming messages are obscured or omitted. Previous studies that focus on devising positive signaling metrics aim towards quantifying the alignment between the agent's messages and some inherent component relating to the agent, often defined using mutual information \cite{jaques2019social, bogin2019emergence}. For instance, speaker consistency (SC) measures the mutual information (MI) between an agent's messages and its actions \cite{jaques2019social}, whereas context independence (CI) instead measures the MI between the agent's messages and predefined task concepts. Intuitively, SC provides insights into how much uncertainty is reduced regarding an agent's action given its messages, and CI enforces the notion of language compositionality, where the content of the messages is induced to be related to inherent concepts within the environment.

On the other hand, the guiding principle for measuring positive listening has been to quantify the causal influence of an agent's message on another agent's behavior. Similarly, this computation can be achieved with MI, where in practice, we quantify the alignment between an agent's message to the actions of receiving agents \cite{lowe2019pitfalls, jaques2019social}. Usually, these metrics consider the ``one-step" behavior, meaning the causal influence of a message is referenced against the immediate response of an agent, such as its next action, where a multi-step behavior can lead to more accurate measures \cite{eccles2019biases}. In terms of future directions, it remains unclear the definitive relationship between these metrics, the concepts of positive listening and positive signaling, the true quality of communication, and their interactions as well as a more theoretical justification for the bias these approaches provide \cite{eccles2019biases}.

\subsection{Modeling Other Agents} \label{subsection: agent modeling}
An essential capability for agents is the ability to reason about the behaviors of other agents, which can be achieved by constructing models of other agents (MOA). This process is often referred to as agent modeling, or more traditionally known as opponent modeling. We divide our discussion of agent modeling into three parts: the representation of MOA, the optimization paradigm used with MOA, and how MOAs are utilized.

\paragraph{Representation of MOA} MOA can encompass a wide array of properties of other agents, including their observations, actions, goals, beliefs, and more intricate components such as intentions or agent types  \cite{hong2018deep, raileanu2018modeling}. MOA can also contain information regarding entire coalitions \cite{erdogan2011action}. Empirical applications of MOA can be accomplished using deep learning models tailored to represent the specific properties they embody and how the MOA is intended to be integrated \cite{he2016opponent}. Bayesian game is a common game mode used to represent MOA through a belief space, i.e. type \cite{harsanyi1967games}. The belief space of each agent contains any information which is not regarded as common knowledge including its private knowledge, which can hold local information regarding other agents. These beliefs, however, face the challenge of approximating uncertainty as agents must contend with incomplete information about the environment and the behaviors of other agents. Addressing this challenge often involves employing probabilistic techniques to estimate and reason about uncertainties within the belief space \cite{huh2024representation}. Another interesting concept with MOA is the theory of mind, where agents engage in recursive reasoning about the states of other agents \cite{Premack1978tom, yu2022model}. In practice, a nested reasoning approach is often approximated using belief nesting down to a fixed recursion depth, which can be implemented with game tree search techniques \cite{Carmel1996}. In practice, there exists a large inspiration from model-based single-agent RL (MBRL), and there remains much work to incorporate the unique aspects of MAS into such MBRL methods \cite{nashed2022survey}.

\paragraph{Learning MOA} Here, we discuss two approaches to optimize and learn information about other agents, in the form of discriminative and generative learning. The discriminative learning approach comprises training MOA to classify and predict explicit properties of other agents, such as their observations or policies, typically through methods of maximum likelihood estimation \cite{huh2024representation}. On the other hand, generative learning approaches rely more on maximum a posterior, where the goal is to model the joint distribution of observed data and latent variables, enabling the generation of realistic samples from the learned distribution. By capturing the underlying structure of the data, generative learning facilitates a deeper understanding of the relationships between different properties of other agents, allowing for more nuanced inference and decision-making in multi-agent environments \cite{erdogan2011action, nashed2022survey}. The learning process of MOA also must consider how it is integrated with the MARL training, which can be done separately or simultaneously, and what additional data or assumptions are required.

\paragraph{Using MOA}
MOA can be utilized in various manners, such as guiding the agent's decision-making process, i.e. planning and recursive reasoning, or helping construct a more accurate understanding of the environment the agent is presiding in \cite{huh2024representation}. An important consideration is the trust and robustness of utilizing these models, as agents may make incorrect or inaccurate predictions. An interesting application that remains an open research topic is an adversarial attempt to trick agents through deceptive actions to promote misleading synergies \cite{Albrecht_2018}.

\subsection{Ad-Hoc Team-Play} \label{subsection: adhoc teamplay}
Learning in multi-agent systems may be faced with a distributional mismatch resulting from unseen behaviors from other agents. The concept of ad-hoc team-play (AHTP) challenges the learned behaviors of agents to work with unknown partners who are capable of contributing to the task \cite{Stone2010adhoc}. For its evaluation, it is common to define a period of ad-hoc interactions between the agents to acclimate and devise their new joint strategy but assume that agents have no prior coordination before this ad-hoc interaction and also that agents have no direct control over other agents \cite{mirsky2022survey}. If we assume no ad-hoc interactions, we refer to this as zero-shot coordination (ZSC) \cite{treutlein2021new}. The notion of AHTP/ZSC often arises in settings of human-AI coordination \cite{nekoei2023towards, yu2023learning}, where the AI agents must interact with humans to achieve a task. Typically in such settings, the AI agents and the humans have not interacted with each other previously. In general, we note that despite the use of the term ``teammate" and ``team-play", these agents are not necessarily cooperative but can be adversarial or mixed.

A crucial aspect of achieving good AHTP behaviors is to avoid arbitrary conventions that often arise in traditional MARL training \cite{carroll2020utility}. This can be achieved by either training agents on a pool of diverse policies, known as population-based training (PBT) \cite{canaan2019diverse}, or removing grounded beliefs of the agents that rely on arbitrary social conventions by utilizing off-belief learning, which assumes the prior behaviors of others were derived from fixed random policies, but their future actions will be computed with actual behavioral policies \cite{hu2021offbelief}. Another option for addressing AHTP is for agents to learn to identify the behaviors of the other agents such that they can properly adjust their strategy to its current environment \cite{chen2020adhoc}. Similarly, agents can instead learn models of the other agents for fast and efficient teamwork in the absence of explicit prior coordination \cite{Barrett2017adhoc, Santos2021adhoc}. However, this approach is sample inefficient, as agents would need to learn these additional capabilities effectively.

To quantify the AHTP capability, there exist two popular metrics: cross-play and adaptation regret \cite{nekoei2023towards}. Cross-play is a static measure of the performance of agents with their new teammates and is often visualized through a cross-play matrix. Similarly, adaptation regret measures the cross-play performance but compares it to the performance achieved with the old teammates, i.e. the regret. The adaptation regret can be viewed under an adaptation curve, which helps view how quickly agents adapt to their new teammates. Similar to PBT,  how to define the pool of ``diverse" policies more optimally remains a question and often is generated using multiple independent runs of training with the same or different MARL algorithms.

AHTP remains an especially difficult challenge when dealing with heterogeneous agents, open environments with variation in the number of agents in the environment, imperfect information, unreliable nor robust communication mechanisms, highly adaptive, irrational or risk-averse agents, and diverse nature of interactions \cite{mirsky2022survey, guan2023efficient, nekoei2023towards}.

\subsection{Social Learning} \label{subsection: knowledge transfer}
The sharing of learned behaviors amongst agents is a powerful mechanism that can improve the efficiency and adaptability of a population's knowledge \cite{silva2018transfer}. This process is referred to as social learning \cite{ndousse2021emergent} and encompasses the concept of knowledge transfer, which manifests through two central approaches: intra-agent transfer, where knowledge is transferred and reused from different domains, and inter-agent transfer, where agents share knowledge within the same task and setting. While both forms of knowledge transfer play crucial roles in facilitating efficient social learning and adaptation within MAS, the focus will be on inter-agent transfer. Within the framework of inter-agent transfer, it is also important to consider the role assignments within the populations. In this work, we define three common role assignments: advisor/advisee, teacher/student, and mentor/observer  \cite{silva2018transfer}.

\begin{itemize}
    \item In the advisor/advisee relationship, the advisor receives requests from the advisee and observes their state, offering valuable information without presuming anything about the internal representation of agents.
    \item The interaction between a teacher and student is similar to that of an advisor/advisee, but in this case, certain assumptions are made, allowing for more informed designs for any information exchange.
    \item As for the mentor/observer relationship, the observer aims to emulate the behavior of the mentor, thereby learning from the mentor's expertise and experience.
\end{itemize}  

Concretely, we explore three forms of inter-agent transfer: action advising, reward shaping, and knowledge distillation. Each of these approaches offers unique ways for agents to leverage the knowledge of their peers and improve their performance in the collective endeavor.

\paragraph{Action Advising} The fundamental concept of action advising (AA) revolves around an experienced agent providing recommendations on the next best actions to take to a less experienced agent \cite{guo2023explainable, omidshafiei2018learning}. In many cases, one agent can offer action suggestions to another, even when the internal representation of the other agents remains unknown. In practice, its implementation must consider how this AA process is initiated, i.e. by the advisor and/or advisee. \cite{da2020uncertainty, fachantidis2017learning, amir2016interactive}, and how the advice is incorporated with some form of option learning, referring to whether or not the suggested policy should be followed \cite{sutton1999between, yang2021efficient}. However, an effective approach that generalizes this AA capability that benefits MARL training remains an open challenge.

\paragraph{Reward Shaping} Derived from the motivations of potential functions in single-agent RL \cite{ng1999policy}, reward shaping in MARL enables agents to influence the reward signals of other agents, typically in the form of an auxiliary signal that provides further learning guidance \cite{gupta2017learning}. Reward shaping approaches are particularly valuable in scenarios where rewards are sparse, as they allow for a learnable method to devise more informative learning signals \cite{wang2022individual}. However, similar to potential functions, it is important to take caution when using such methods, as the behaviors of agents are highly dependent and influenced by the reward function, thereby it may be pivotal in some settings to maintain some level of invariance to these auxiliary reward signals.

\paragraph{Knowledge Distillation} The process of knowledge distillation (KD) involves transferring knowledge from teacher agents to student agents \cite{buciluǎ2006model, hinton2015distilling}. Typically, KD is primarily used for model compression, where the capabilities of a larger model or multiple models are distilled into a single/smaller model for parameter efficiency and potential performance benefits, or adaptation to a new state and/or action space, where a teacher model is initially trained on a more complete state-action space and a student model must make use of a more restricted state-action space \cite{czarnecki2019distilling, lai2020dual}. In MARL applications, KD can further leverage the multi-agent nature, through structural relations distillation, where the relations between multi-agents' features are preserved \cite{tseng2022offline}.

\section{Concluding Remarks}\label{section: open challenges}
While the challenges within MARL have been extensively studied and assessed, the existing methodologies for acquiring multi-agent behaviors fall short of fully harnessing the myriad opportunities within a MAS. Despite substantial progress, particularly in unique areas of learning in a MAS, i.e. its prospects, and the fundamental challenges of MARL, there remain open research challenges that demand further exploration and refinement.

The intricacies of MARL extend beyond individual agent behavior to encompass the dynamic interactions unfolding within complex environments. Other properties of certain applications, such as open environments, human-robot interactions, and heterogeneous agents, merit additional considerations, for instance, how to handle the unbounded and evolving nature of open environments, safety and proper coordination with human-robot interactions, and manage the diverse capabilities, behaviors, and learning speeds within heterogeneous populations.

To conclude our discussion of MARL, our exploration of MARL discussed both the progress made and the avenues yet to be fully explored. The multifaceted nature of multi-agent interactions within dynamic environments demands ongoing research and refinement of methodologies to unlock the full potential of MARL in harnessing the complexities inherent in a MAS.

\bibliography{citation}

\begin{thebibliography}{361}
\providecommand{\natexlab}[1]{#1}
\providecommand{\url}[1]{\texttt{#1}}
\expandafter\ifx\csname urlstyle\endcsname\relax
  \providecommand{\doi}[1]{doi: #1}\else
  \providecommand{\doi}{doi: \begingroup \urlstyle{rm}\Url}\fi

\bibitem[Adam et~al.(2021)Adam, Hor{\v{c}}{\'\i}k, Kasl, and Kroupa]{adam2021double}
L.~Adam, R.~Hor{\v{c}}{\'\i}k, T.~Kasl, and T.~Kroupa.
\newblock Double oracle algorithm for computing equilibria in continuous games.
\newblock In \emph{Proceedings of the AAAI Conference on Artificial Intelligence}, volume~35, pages 5070--5077, 2021.

\bibitem[Agogino and Tumer(2004)]{agogino2004unifying}
A.~K. Agogino and K.~Tumer.
\newblock Unifying temporal and structural credit assignment problems.
\newblock In \emph{Autonomous agents and multi-agent systems conference}, 2004.

\bibitem[Agogino and Tumer(2008)]{agogino2008analyzing}
A.~K. Agogino and K.~Tumer.
\newblock Analyzing and visualizing multiagent rewards in dynamic and stochastic domains.
\newblock \emph{Autonomous Agents and Multi-Agent Systems}, 17:\penalty0 320--338, 2008.

\bibitem[Albrecht and Stone(2018)]{Albrecht_2018}
S.~V. Albrecht and P.~Stone.
\newblock Autonomous agents modelling other agents: A comprehensive survey and open problems.
\newblock \emph{Artificial Intelligence}, 258:\penalty0 66--95, may 2018.
\newblock \doi{10.1016/j.artint.2018.01.002}.
\newblock URL \url{https://doi.org/10.1016%2Fj.artint.2018.01.002}.

\bibitem[Albrecht et~al.(2024)Albrecht, Christianos, and Sch\"afer]{Albrecht2024Book}
S.~V. Albrecht, F.~Christianos, and L.~Sch\"afer.
\newblock Multi-agent reinforcement learning: Foundations and modern approaches.
\newblock 2024.
\newblock URL \url{https://www.marl-article.com}.

\bibitem[Amato and Oliehoek(2014)]{amato2014scalable}
C.~Amato and F.~A. Oliehoek.
\newblock Scalable planning and learning for multiagent pomdps: Extended version, 2014.

\bibitem[Amato et~al.(2019)Amato, Konidaris, Kaelbling, and How]{amato2019modeling}
C.~Amato, G.~Konidaris, L.~P. Kaelbling, and J.~P. How.
\newblock Modeling and planning with macro-actions in decentralized pomdps.
\newblock \emph{Journal of Artificial Intelligence Research}, 64:\penalty0 817--859, 2019.

\bibitem[Amir et~al.(2016)Amir, Kamar, Kolobov, and Grosz]{amir2016interactive}
O.~Amir, E.~Kamar, A.~Kolobov, and B.~Grosz.
\newblock Interactive teaching strategies for agent training.
\newblock In \emph{In Proceedings of IJCAI 2016}, 2016.

\bibitem[Andrychowicz et~al.(2017)Andrychowicz, Wolski, Ray, Schneider, Fong, Welinder, McGrew, Tobin, Abbeel, and Zaremba]{andrychowicz2017hindsight}
M.~Andrychowicz, F.~Wolski, A.~Ray, J.~Schneider, R.~Fong, P.~Welinder, B.~McGrew, J.~Tobin, P.~Abbeel, and W.~Zaremba.
\newblock Hindsight experience replay.
\newblock \emph{arXiv preprint arXiv:1707.01495}, 2017.

\bibitem[Anshelevich et~al.(2008)Anshelevich, Dasgupta, Kleinberg, Tardos, Wexler, and Roughgarden]{anshelevich2008price}
E.~Anshelevich, A.~Dasgupta, J.~Kleinberg, {\'E}.~Tardos, T.~Wexler, and T.~Roughgarden.
\newblock The price of stability for network design with fair cost allocation.
\newblock \emph{SIAM Journal on Computing}, 38\penalty0 (4):\penalty0 1602--1623, 2008.

\bibitem[Arora et~al.(2012)Arora, Hazan, and Kale]{Arora2012}
S.~Arora, E.~Hazan, and S.~Kale.
\newblock The multiplicative weights update method: a meta-algorithm and applications.
\newblock \emph{Theory of Computing}, 8\penalty0 (6):\penalty0 121--164, 2012.
\newblock \doi{10.4086/toc.2012.v008a006}.
\newblock URL \url{https://theoryofcomputing.org/articles/v008a006}.

\bibitem[Aumann(1974)]{Aumann1974CE}
R.~J. Aumann.
\newblock Subjectivity and correlation in randomized strategies.
\newblock \emph{Journal of Mathematical Economics}, 1\penalty0 (1):\penalty0 67--96, 1974.
\newblock ISSN 0304-4068.
\newblock \doi{https://doi.org/10.1016/0304-4068(74)90037-8}.
\newblock URL \url{https://www.sciencedirect.com/science/article/pii/0304406874900378}.

\bibitem[Baarslag et~al.(2016)Baarslag, Hendrikx, Hindriks, and Jonker]{Baarslag2016}
T.~Baarslag, M.~Hendrikx, K.~Hindriks, and C.~Jonker.
\newblock Learning about the opponent in automated bilateral negotiation: a comprehensive survey of opponent modeling techniques.
\newblock \emph{Autonomous Agents and Multi-Agent Systems}, 30, 09 2016.
\newblock \doi{10.1007/s10458-015-9309-1}.

\bibitem[Badia et~al.(2020{\natexlab{a}})Badia, Piot, Kapturowski, Sprechmann, Vitvitskyi, Guo, and Blundell]{badia2020agent57}
A.~P. Badia, B.~Piot, S.~Kapturowski, P.~Sprechmann, A.~Vitvitskyi, D.~Guo, and C.~Blundell.
\newblock Agent57: Outperforming the atari human benchmark, 2020{\natexlab{a}}.

\bibitem[Badia et~al.(2020{\natexlab{b}})Badia, Sprechmann, Vitvitskyi, Guo, Piot, Kapturowski, Tieleman, Arjovsky, Pritzel, Bolt, and Blundell]{badia2020up}
A.~P. Badia, P.~Sprechmann, A.~Vitvitskyi, D.~Guo, B.~Piot, S.~Kapturowski, O.~Tieleman, M.~Arjovsky, A.~Pritzel, A.~Bolt, and C.~Blundell.
\newblock Never give up: Learning directed exploration strategies, 2020{\natexlab{b}}.

\bibitem[Baker et~al.(2022)Baker, Akkaya, Zhokhov, Huizinga, Tang, Ecoffet, Houghton, Sampedro, and Clune]{baker2022video}
B.~Baker, I.~Akkaya, P.~Zhokhov, J.~Huizinga, J.~Tang, A.~Ecoffet, B.~Houghton, R.~Sampedro, and J.~Clune.
\newblock Video pretraining (vpt): Learning to act by watching unlabeled online videos, 2022.

\bibitem[Barrett et~al.(2017)Barrett, Rosenfeld, Kraus, and Stone]{Barrett2017adhoc}
S.~Barrett, A.~Rosenfeld, S.~Kraus, and P.~Stone.
\newblock Making friends on the fly: Cooperating with new teammates.
\newblock \emph{Artificial Intelligence}, 242:\penalty0 132--171, 2017.
\newblock ISSN 0004-3702.
\newblock \doi{https://doi.org/10.1016/j.artint.2016.10.005}.
\newblock URL \url{https://www.sciencedirect.com/science/article/pii/S0004370216301266}.

\bibitem[Ba{\c{s}}ar and Olsder(1998)]{bacsar1998dynamic}
T.~Ba{\c{s}}ar and G.~J. Olsder.
\newblock \emph{Dynamic noncooperative game theory}.
\newblock SIAM, 1998.

\bibitem[Bellemare et~al.(2017)Bellemare, Dabney, and Munos]{bellemare17c51}
M.~G. Bellemare, W.~Dabney, and R.~Munos.
\newblock A distributional perspective on reinforcement learning.
\newblock 70:\penalty0 449--458, 06--11 Aug 2017.
\newblock URL \url{https://proceedings.mlr.press/v70/bellemare17a.html}.

\bibitem[Bellman(1957)]{bellman1957dynamic}
R.~Bellman.
\newblock Dynamic programming princeton university press princeton.
\newblock \emph{New Jersey Google Scholar}, pages 24--73, 1957.

\bibitem[Bergerson(2021)]{bergerson2021multiagent}
S.~Bergerson.
\newblock Multi-agent inverse reinforcement learning: Suboptimal demonstrations and alternative solution concepts, 2021.

\bibitem[Bernstein et~al.(2013)Bernstein, Zilberstein, and Immerman]{Bernstein2013Complexity}
D.~S. Bernstein, S.~Zilberstein, and N.~Immerman.
\newblock The complexity of decentralized control of markov decision processes.
\newblock \emph{CoRR}, abs/1301.3836, 2013.
\newblock URL \url{http://arxiv.org/abs/1301.3836}.

\bibitem[Bettini et~al.(2022)Bettini, Kortvelesy, Blumenkamp, and Prorok]{bettini2022vmas}
M.~Bettini, R.~Kortvelesy, J.~Blumenkamp, and A.~Prorok.
\newblock Vmas: A vectorized multi-agent simulator for collective robot learning.
\newblock \emph{The 16th International Symposium on Distributed Autonomous Robotic Systems}, 2022.

\bibitem[Bielefeld(1988)]{Selten1988PE}
R.~S. Bielefeld.
\newblock Reexamination of the perfectness concept for equilibrium points in extensive games, 1988.
\newblock URL \url{https://doi.org/10.1007/978-94-015-7774-8_1}.

\bibitem[Bighashdel et~al.(2024)Bighashdel, Wang, McAleer, Savani, and Oliehoek]{bighashdel2024policy}
A.~Bighashdel, Y.~Wang, S.~McAleer, R.~Savani, and F.~A. Oliehoek.
\newblock Policy space response oracles: A survey, 2024.

\bibitem[Blackwell(1956)]{blackwell1956analog}
D.~Blackwell.
\newblock An analog of the minimax theorem for vector payoffs.
\newblock 1956.

\bibitem[Blier et~al.(2021)Blier, Tallec, and Ollivier]{blier2021learning}
L.~Blier, C.~Tallec, and Y.~Ollivier.
\newblock Learning successor states and goal-dependent values: A mathematical viewpoint, 2021.

\bibitem[Bloembergen et~al.(2015)Bloembergen, Tuyls, Hennes, and Kaisers]{Bloembergen2015}
D.~Bloembergen, K.~Tuyls, D.~Hennes, and M.~Kaisers.
\newblock Evolutionary dynamics of multi-agent learning: A survey.
\newblock \emph{Journal of Artificial Intelligence Research}, 53:\penalty0 659--697, 08 2015.
\newblock \doi{10.1613/jair.4818}.

\bibitem[Blundell et~al.(2015)Blundell, Cornebise, Kavukcuoglu, and Wierstra]{blundell2015weight}
C.~Blundell, J.~Cornebise, K.~Kavukcuoglu, and D.~Wierstra.
\newblock Weight uncertainty in neural networks, 2015.

\bibitem[Bogin et~al.(2019)Bogin, Geva, and Berant]{bogin2019emergence}
B.~Bogin, M.~Geva, and J.~Berant.
\newblock Emergence of communication in an interactive world with consistent speakers, 2019.

\bibitem[Bommasani et~al.(2022)Bommasani, Hudson, Adeli, Altman, Arora, von Arx, Bernstein, Bohg, Bosselut, Brunskill, Brynjolfsson, Buch, Card, Castellon, Chatterji, Chen, Creel, Davis, Demszky, Donahue, Doumbouya, Durmus, Ermon, Etchemendy, Ethayarajh, Fei-Fei, Finn, Gale, Gillespie, Goel, Goodman, Grossman, Guha, Hashimoto, Henderson, Hewitt, Ho, Hong, Hsu, Huang, Icard, Jain, Jurafsky, Kalluri, Karamcheti, Keeling, Khani, Khattab, Koh, Krass, Krishna, Kuditipudi, Kumar, Ladhak, Lee, Lee, Leskovec, Levent, Li, Li, Ma, Malik, Manning, Mirchandani, Mitchell, Munyikwa, Nair, Narayan, Narayanan, Newman, Nie, Niebles, Nilforoshan, Nyarko, Ogut, Orr, Papadimitriou, Park, Piech, Portelance, Potts, Raghunathan, Reich, Ren, Rong, Roohani, Ruiz, Ryan, Ré, Sadigh, Sagawa, Santhanam, Shih, Srinivasan, Tamkin, Taori, Thomas, Tramèr, Wang, Wang, Wu, Wu, Wu, Xie, Yasunaga, You, Zaharia, Zhang, Zhang, Zhang, Zhang, Zheng, Zhou, and Liang]{bommasani2022opportunities}
R.~Bommasani, D.~A. Hudson, E.~Adeli, R.~Altman, S.~Arora, S.~von Arx, M.~S. Bernstein, J.~Bohg, A.~Bosselut, E.~Brunskill, E.~Brynjolfsson, S.~Buch, D.~Card, R.~Castellon, N.~Chatterji, A.~Chen, K.~Creel, J.~Q. Davis, D.~Demszky, C.~Donahue, M.~Doumbouya, E.~Durmus, S.~Ermon, J.~Etchemendy, K.~Ethayarajh, L.~Fei-Fei, C.~Finn, T.~Gale, L.~Gillespie, K.~Goel, N.~Goodman, S.~Grossman, N.~Guha, T.~Hashimoto, P.~Henderson, J.~Hewitt, D.~E. Ho, J.~Hong, K.~Hsu, J.~Huang, T.~Icard, S.~Jain, D.~Jurafsky, P.~Kalluri, S.~Karamcheti, G.~Keeling, F.~Khani, O.~Khattab, P.~W. Koh, M.~Krass, R.~Krishna, R.~Kuditipudi, A.~Kumar, F.~Ladhak, M.~Lee, T.~Lee, J.~Leskovec, I.~Levent, X.~L. Li, X.~Li, T.~Ma, A.~Malik, C.~D. Manning, S.~Mirchandani, E.~Mitchell, Z.~Munyikwa, S.~Nair, A.~Narayan, D.~Narayanan, B.~Newman, A.~Nie, J.~C. Niebles, H.~Nilforoshan, J.~Nyarko, G.~Ogut, L.~Orr, I.~Papadimitriou, J.~S. Park, C.~Piech, E.~Portelance, C.~Potts, A.~Raghunathan, R.~Reich, H.~Ren, F.~Rong, Y.~Roohani, C.~Ruiz, J.~Ryan, C.~Ré,
  D.~Sadigh, S.~Sagawa, K.~Santhanam, A.~Shih, K.~Srinivasan, A.~Tamkin, R.~Taori, A.~W. Thomas, F.~Tramèr, R.~E. Wang, W.~Wang, B.~Wu, J.~Wu, Y.~Wu, S.~M. Xie, M.~Yasunaga, J.~You, M.~Zaharia, M.~Zhang, T.~Zhang, X.~Zhang, Y.~Zhang, L.~Zheng, K.~Zhou, and P.~Liang.
\newblock On the opportunities and risks of foundation models, 2022.

\bibitem[Bowling(2004)]{Bowling2004GigaWolf}
M.~Bowling.
\newblock Convergence and no-regret in multiagent learning.
\newblock \emph{Advances in neural information processing systems}, 17, 2004.

\bibitem[Bowling and Veloso(2001)]{Bowling2001Analysis}
M.~Bowling and M.~Veloso.
\newblock An analysis of stochastic game theory for multiagent reinforcement learning.
\newblock 08 2001.

\bibitem[Bowling and Veloso(2002)]{Bowling2002Wolf}
M.~Bowling and M.~Veloso.
\newblock Multiagent learning using a variable learning rate.
\newblock \emph{Artificial Intelligence}, 136\penalty0 (2):\penalty0 215--250, 2002.
\newblock ISSN 0004-3702.
\newblock \doi{https://doi.org/10.1016/S0004-3702(02)00121-2}.
\newblock URL \url{https://www.sciencedirect.com/science/article/pii/S0004370202001212}.

\bibitem[Bowling et~al.(2015)Bowling, Burch, Johanson, and Tammelin]{bowling2015heads}
M.~Bowling, N.~Burch, M.~Johanson, and O.~Tammelin.
\newblock Heads-up limit hold’em poker is solved.
\newblock \emph{Science}, 347\penalty0 (6218):\penalty0 145--149, 2015.

\bibitem[Brafman and Tennenholtz(2003)]{Brafman_2003}
R.~I. Brafman and M.~Tennenholtz.
\newblock Learning to coordinate efficiently: A model-based approach.
\newblock \emph{Journal of Artificial Intelligence Research}, 19:\penalty0 11--23, jul 2003.
\newblock \doi{10.1613/jair.1154}.
\newblock URL \url{https://doi.org/10.1613%2Fjair.1154}.

\bibitem[Brody et~al.(2022)Brody, Alon, and Yahav]{brody2022attentive}
S.~Brody, U.~Alon, and E.~Yahav.
\newblock How attentive are graph attention networks?, 2022.

\bibitem[Buciluǎ et~al.(2006)Buciluǎ, Caruana, and Niculescu-Mizil]{buciluǎ2006model}
C.~Buciluǎ, R.~Caruana, and A.~Niculescu-Mizil.
\newblock Model compression.
\newblock In \emph{Proceedings of the 12th ACM SIGKDD international conference on Knowledge discovery and data mining}, pages 535--541, 2006.

\bibitem[Buckman et~al.(2018)Buckman, Hafner, Tucker, Brevdo, and Lee]{Buckman2018smve}
J.~Buckman, D.~Hafner, G.~Tucker, E.~Brevdo, and H.~Lee.
\newblock Sample-efficient reinforcement learning with stochastic ensemble value expansion.
\newblock \emph{CoRR}, abs/1807.01675, 2018.
\newblock URL \url{http://arxiv.org/abs/1807.01675}.

\bibitem[Busoniu et~al.(2008)Busoniu, Babuska, and De~Schutter]{Busoniu2008ComprehensiveSurvey}
L.~Busoniu, R.~Babuska, and B.~De~Schutter.
\newblock A comprehensive survey of multiagent reinforcement learning.
\newblock \emph{IEEE Transactions on Systems, Man, and Cybernetics, Part C (Applications and Reviews)}, 38\penalty0 (2):\penalty0 156--172, 2008.
\newblock \doi{10.1109/TSMCC.2007.913919}.

\bibitem[Busoniu et~al.(2010)Busoniu, Babuvska, and Schutter]{Buoniu2010MultiagentRL}
L.~Busoniu, R.~Babuvska, and B.~D. Schutter.
\newblock Multi-agent reinforcement learning: An overview.
\newblock 2010.
\newblock URL \url{https://api.semanticscholar.org/CorpusID:17136625}.

\bibitem[Böhmer et~al.(2020)Böhmer, Kurin, and Whiteson]{böhmer2020deep}
W.~Böhmer, V.~Kurin, and S.~Whiteson.
\newblock Deep coordination graphs, 2020.

\bibitem[Cai et~al.(2007)Cai, Yang, Zhu, and Liang]{Cai2007collision}
C.~Cai, C.~Yang, Q.~Zhu, and Y.~Liang.
\newblock Collision avoidance in multi-robot systems.
\newblock pages 2795--2800, 2007.
\newblock \doi{10.1109/ICMA.2007.4304002}.

\bibitem[Calvo and Dusparic(2018)]{Calvo2018HeterogeneousMD}
J.~A. Calvo and I.~Dusparic.
\newblock Heterogeneous multi-agent deep reinforcement learning for traffic lights control.
\newblock 2018.
\newblock URL \url{https://api.semanticscholar.org/CorpusID:57661298}.

\bibitem[Camerer(1997)]{camerer1997progress}
C.~F. Camerer.
\newblock Progress in behavioral game theory.
\newblock \emph{Journal of economic perspectives}, 11\penalty0 (4):\penalty0 167--188, 1997.

\bibitem[Canaan et~al.(2019)Canaan, Togelius, Nealen, and Menzel]{canaan2019diverse}
R.~Canaan, J.~Togelius, A.~Nealen, and S.~Menzel.
\newblock Diverse agents for ad-hoc cooperation in hanabi.
\newblock In \emph{2019 IEEE Conference on Games (CoG)}, pages 1--8. IEEE, 2019.

\bibitem[Carmel and Markovitch(1996)]{Carmel1996}
D.~Carmel and S.~Markovitch.
\newblock Incorporating opponent models into adversary search.
\newblock pages 120--125, 01 1996.

\bibitem[Carroll et~al.(2020)Carroll, Shah, Ho, Griffiths, Seshia, Abbeel, and Dragan]{carroll2020utility}
M.~Carroll, R.~Shah, M.~K. Ho, T.~L. Griffiths, S.~A. Seshia, P.~Abbeel, and A.~Dragan.
\newblock On the utility of learning about humans for human-ai coordination, 2020.

\bibitem[Chalkiadakis and Boutilier(2003)]{chalkiadakis2003coordination}
G.~Chalkiadakis and C.~Boutilier.
\newblock Coordination in multiagent reinforcement learning: A bayesian approach.
\newblock In \emph{Proceedings of the second international joint conference on Autonomous agents and multiagent systems}, pages 709--716, 2003.

\bibitem[Chebotar et~al.(2023)Chebotar, Vuong, Irpan, Hausman, Xia, Lu, Kumar, Yu, Herzog, Pertsch, Gopalakrishnan, Ibarz, Nachum, Sontakke, Salazar, Tran, Peralta, Tan, Manjunath, Singht, Zitkovich, Jackson, Rao, Finn, and Levine]{chebotar2023qtransformer}
Y.~Chebotar, Q.~Vuong, A.~Irpan, K.~Hausman, F.~Xia, Y.~Lu, A.~Kumar, T.~Yu, A.~Herzog, K.~Pertsch, K.~Gopalakrishnan, J.~Ibarz, O.~Nachum, S.~Sontakke, G.~Salazar, H.~T. Tran, J.~Peralta, C.~Tan, D.~Manjunath, J.~Singht, B.~Zitkovich, T.~Jackson, K.~Rao, C.~Finn, and S.~Levine.
\newblock Q-transformer: Scalable offline reinforcement learning via autoregressive q-functions, 2023.

\bibitem[Chen et~al.(2020)Chen, Andrejczuk, Cao, and Zhang]{chen2020adhoc}
S.~Chen, E.~Andrejczuk, Z.~Cao, and J.~Zhang.
\newblock Aateam: Achieving the ad hoc teamwork by employing the attention mechanism.
\newblock \emph{Proceedings of the AAAI Conference on Artificial Intelligence}, 34\penalty0 (05):\penalty0 7095--7102, Apr. 2020.
\newblock \doi{10.1609/aaai.v34i05.6196}.
\newblock URL \url{https://ojs.aaai.org/index.php/AAAI/article/view/6196}.

\bibitem[Chen et~al.(2007)Chen, Deng, and Teng]{chen2007settling}
X.~Chen, X.~Deng, and S.-H. Teng.
\newblock Settling the complexity of computing two-player nash equilibria, 2007.

\bibitem[Chen et~al.(2022)Chen, Yang, Wu, Wang, Feng, Jiang, Lu, McAleer, Dong, and Zhu]{chen2022towards}
Y.~Chen, Y.~Yang, T.~Wu, S.~Wang, X.~Feng, J.~Jiang, Z.~Lu, S.~M. McAleer, H.~Dong, and S.-C. Zhu.
\newblock Towards human-level bimanual dexterous manipulation with reinforcement learning.
\newblock 2022.
\newblock URL \url{https://openreview.net/forum?id=D29JbExncTP}.

\bibitem[Choi et~al.(1999)Choi, Yeung, and Zhang]{Choi1999Nonstationarity}
S.~Choi, D.-Y. Yeung, and N.~Zhang.
\newblock An environment model for nonstationary reinforcement learning.
\newblock 12, 1999.
\newblock URL \url{https://proceedings.neurips.cc/paper_files/paper/1999/file/e8d92f99edd25e2cef48eca48320a1a5-Paper.pdf}.

\bibitem[Christianos et~al.(2020)Christianos, Schäfer, and Albrecht]{christianos2020shared}
F.~Christianos, L.~Schäfer, and S.~V. Albrecht.
\newblock Shared experience actor-critic for multi-agent reinforcement learning.
\newblock 2020.

\bibitem[Chu et~al.(2019)Chu, Wang, Codecà, and Li]{chu2019multiagent}
T.~Chu, J.~Wang, L.~Codecà, and Z.~Li.
\newblock Multi-agent deep reinforcement learning for large-scale traffic signal control, 2019.

\bibitem[Chu et~al.(2020)Chu, Chinchali, and Katti]{chu2020multiagent}
T.~Chu, S.~Chinchali, and S.~Katti.
\newblock Multi-agent reinforcement learning for networked system control, 2020.

\bibitem[Chua et~al.(2018)Chua, Calandra, McAllister, and Levine]{chua2018deep}
K.~Chua, R.~Calandra, R.~McAllister, and S.~Levine.
\newblock Deep reinforcement learning in a handful of trials using probabilistic dynamics models, 2018.

\bibitem[Claus and Boutilier(1998)]{Claus1998dynamics}
C.~Claus and C.~Boutilier.
\newblock The dynamics of reinforcement learning in cooperative multiagent systems.
\newblock page 746–752, 1998.

\bibitem[Clavera et~al.(2018)Clavera, Rothfuss, Schulman, Fujita, Asfour, and Abbeel]{clavera2018modelbased}
I.~Clavera, J.~Rothfuss, J.~Schulman, Y.~Fujita, T.~Asfour, and P.~Abbeel.
\newblock Model-based reinforcement learning via meta-policy optimization, 2018.

\bibitem[Cobbe et~al.(2020)Cobbe, Hilton, Klimov, and Schulman]{cobbe2020phasic}
K.~Cobbe, J.~Hilton, O.~Klimov, and J.~Schulman.
\newblock Phasic policy gradient, 2020.

\bibitem[Collaboration et~al.(2023)Collaboration, Padalkar, Pooley, Mandlekar, Jain, Tung, Bewley, Herzog, Irpan, Khazatsky, Rai, Singh, Garg, Brohan, Raffin, Wahid, Burgess-Limerick, Kim, Schölkopf, Ichter, Lu, Xu, Finn, Xu, Chi, Huang, Chan, Pan, Fu, Devin, Driess, Pathak, Shah, Büchler, Kalashnikov, Sadigh, Johns, Ceola, Xia, Stulp, Zhou, Sukhatme, Salhotra, Yan, Schiavi, Kahn, Su, Fang, Shi, Amor, Christensen, Furuta, Walke, Fang, Mordatch, Radosavovic, Leal, Liang, Abou-Chakra, Kim, Peters, Schneider, Hsu, Bohg, Bingham, Wu, Wu, Luo, Gu, Tan, Oh, Malik, Booher, Tompson, Yang, Lim, Silvério, Han, Rao, Pertsch, Hausman, Go, Gopalakrishnan, Goldberg, Byrne, Oslund, Kawaharazuka, Zhang, Rana, Srinivasan, Chen, Pinto, Fei-Fei, Tan, Ott, Lee, Tomizuka, Spero, Du, Ahn, Zhang, Ding, Srirama, Sharma, Kim, Kanazawa, Hansen, Heess, Joshi, Suenderhauf, Palo, Shafiullah, Mees, Kroemer, Sanketi, Wohlhart, Xu, Sermanet, Sundaresan, Vuong, Rafailov, Tian, Doshi, Martín-Martín, Mendonca, Shah, Hoque, Julian,
  Bustamante, Kirmani, Levine, Moore, Bahl, Dass, Sonawani, Song, Xu, Haldar, Adebola, Guist, Nasiriany, Schaal, Welker, Tian, Dasari, Belkhale, Osa, Harada, Matsushima, Xiao, Yu, Ding, Davchev, Zhao, Armstrong, Darrell, Jain, Vanhoucke, Zhan, Zhou, Burgard, Chen, Wang, Zhu, Li, Lu, Chebotar, Zhou, Zhu, Xu, Wang, Bisk, Cho, Lee, Cui, Wu, Tang, Zhu, Li, Iwasawa, Matsuo, Xu, and Cui]{embodimentcollaboration2023open}
E.~Collaboration, A.~Padalkar, A.~Pooley, A.~Mandlekar, A.~Jain, A.~Tung, A.~Bewley, A.~Herzog, A.~Irpan, A.~Khazatsky, A.~Rai, A.~Singh, A.~Garg, A.~Brohan, A.~Raffin, A.~Wahid, B.~Burgess-Limerick, B.~Kim, B.~Schölkopf, B.~Ichter, C.~Lu, C.~Xu, C.~Finn, C.~Xu, C.~Chi, C.~Huang, C.~Chan, C.~Pan, C.~Fu, C.~Devin, D.~Driess, D.~Pathak, D.~Shah, D.~Büchler, D.~Kalashnikov, D.~Sadigh, E.~Johns, F.~Ceola, F.~Xia, F.~Stulp, G.~Zhou, G.~S. Sukhatme, G.~Salhotra, G.~Yan, G.~Schiavi, G.~Kahn, H.~Su, H.-S. Fang, H.~Shi, H.~B. Amor, H.~I. Christensen, H.~Furuta, H.~Walke, H.~Fang, I.~Mordatch, I.~Radosavovic, I.~Leal, J.~Liang, J.~Abou-Chakra, J.~Kim, J.~Peters, J.~Schneider, J.~Hsu, J.~Bohg, J.~Bingham, J.~Wu, J.~Wu, J.~Luo, J.~Gu, J.~Tan, J.~Oh, J.~Malik, J.~Booher, J.~Tompson, J.~Yang, J.~J. Lim, J.~Silvério, J.~Han, K.~Rao, K.~Pertsch, K.~Hausman, K.~Go, K.~Gopalakrishnan, K.~Goldberg, K.~Byrne, K.~Oslund, K.~Kawaharazuka, K.~Zhang, K.~Rana, K.~Srinivasan, L.~Y. Chen, L.~Pinto, L.~Fei-Fei, L.~Tan, L.~Ott,
  L.~Lee, M.~Tomizuka, M.~Spero, M.~Du, M.~Ahn, M.~Zhang, M.~Ding, M.~K. Srirama, M.~Sharma, M.~J. Kim, N.~Kanazawa, N.~Hansen, N.~Heess, N.~J. Joshi, N.~Suenderhauf, N.~D. Palo, N.~M.~M. Shafiullah, O.~Mees, O.~Kroemer, P.~R. Sanketi, P.~Wohlhart, P.~Xu, P.~Sermanet, P.~Sundaresan, Q.~Vuong, R.~Rafailov, R.~Tian, R.~Doshi, R.~Martín-Martín, R.~Mendonca, R.~Shah, R.~Hoque, R.~Julian, S.~Bustamante, S.~Kirmani, S.~Levine, S.~Moore, S.~Bahl, S.~Dass, S.~Sonawani, S.~Song, S.~Xu, S.~Haldar, S.~Adebola, S.~Guist, S.~Nasiriany, S.~Schaal, S.~Welker, S.~Tian, S.~Dasari, S.~Belkhale, T.~Osa, T.~Harada, T.~Matsushima, T.~Xiao, T.~Yu, T.~Ding, T.~Davchev, T.~Z. Zhao, T.~Armstrong, T.~Darrell, V.~Jain, V.~Vanhoucke, W.~Zhan, W.~Zhou, W.~Burgard, X.~Chen, X.~Wang, X.~Zhu, X.~Li, Y.~Lu, Y.~Chebotar, Y.~Zhou, Y.~Zhu, Y.~Xu, Y.~Wang, Y.~Bisk, Y.~Cho, Y.~Lee, Y.~Cui, Y.-H. Wu, Y.~Tang, Y.~Zhu, Y.~Li, Y.~Iwasawa, Y.~Matsuo, Z.~Xu, and Z.~J. Cui.
\newblock Open x-embodiment: Robotic learning datasets and rt-x models, 2023.

\bibitem[Conitzer and Sandholm(2003)]{conitzer2003awesome}
V.~Conitzer and T.~Sandholm.
\newblock Awesome: A general multiagent learning algorithm that converges in self-play and learns a best response against stationary opponents, 2003.

\bibitem[Conitzer and Sandholm(2008)]{Conitzer2008Complexity}
V.~Conitzer and T.~Sandholm.
\newblock New complexity results about nash equilibria.
\newblock \emph{Games and Economic Behavior}, 63\penalty0 (2):\penalty0 621--641, 2008.
\newblock ISSN 0899-8256.
\newblock \doi{https://doi.org/10.1016/j.geb.2008.02.015}.
\newblock URL \url{https://www.sciencedirect.com/science/article/pii/S0899825608000936}.
\newblock Second World Congress of the Game Theory Society.

\bibitem[Cutkosky(2019)]{pmlr-v99-cutkosky19a}
A.~Cutkosky.
\newblock Artificial constraints and hints for unbounded online learning.
\newblock In A.~Beygelzimer and D.~Hsu, editors, \emph{Proceedings of the Thirty-Second Conference on Learning Theory}, volume~99 of \emph{Proceedings of Machine Learning Research}, pages 874--894. PMLR, 25--28 Jun 2019.
\newblock URL \url{https://proceedings.mlr.press/v99/cutkosky19a.html}.

\bibitem[Czarnecki et~al.(2019)Czarnecki, Pascanu, Osindero, Jayakumar, Swirszcz, and Jaderberg]{czarnecki2019distilling}
W.~M. Czarnecki, R.~Pascanu, S.~Osindero, S.~Jayakumar, G.~Swirszcz, and M.~Jaderberg.
\newblock Distilling policy distillation.
\newblock In \emph{The 22nd international conference on artificial intelligence and statistics}, pages 1331--1340. PMLR, 2019.

\bibitem[Da~Silva et~al.(2020)Da~Silva, Hernandez-Leal, Kartal, and Taylor]{da2020uncertainty}
F.~L. Da~Silva, P.~Hernandez-Leal, B.~Kartal, and M.~E. Taylor.
\newblock Uncertainty-aware action advising for deep reinforcement learning agents.
\newblock In \emph{Proceedings of the AAAI conference on artificial intelligence}, volume~34, pages 5792--5799, 2020.

\bibitem[Dabney et~al.(2018{\natexlab{a}})Dabney, Ostrovski, Silver, and Munos]{dabney2018implicit}
W.~Dabney, G.~Ostrovski, D.~Silver, and R.~Munos.
\newblock Implicit quantile networks for distributional reinforcement learning.
\newblock pages 1096--1105, 2018{\natexlab{a}}.

\bibitem[Dabney et~al.(2018{\natexlab{b}})Dabney, Rowland, Bellemare, and Munos]{dabney2018distributional}
W.~Dabney, M.~Rowland, M.~G. Bellemare, and R.~Munos.
\newblock Distributional reinforcement learning with quantile regression.
\newblock 2018{\natexlab{b}}.

\bibitem[Dalton et~al.(2019)Dalton, Frosio, and Garland]{Dalton2019atari}
S.~Dalton, I.~Frosio, and M.~Garland.
\newblock Gpu-accelerated atari emulation for reinforcement learning.
\newblock \emph{CoRR}, abs/1907.08467, 2019.
\newblock URL \url{http://arxiv.org/abs/1907.08467}.

\bibitem[Daskalakis(2022)]{Daskalakis2022EC}
C.~Daskalakis.
\newblock Equilibrium computation and machine learning, 2022.

\bibitem[Daskalakis and Panageas(2018)]{daskalakis2018limit}
C.~Daskalakis and I.~Panageas.
\newblock The limit points of (optimistic) gradient descent in min-max optimization, 2018.

\bibitem[Daskalakis et~al.(2009)Daskalakis, Goldberg, and Papadimitriou]{Daskalakis2009Complexity}
C.~Daskalakis, P.~W. Goldberg, and C.~H. Papadimitriou.
\newblock The complexity of computing a nash equilibrium.
\newblock \emph{SIAM Journal on Computing}, 39\penalty0 (1):\penalty0 195--259, 2009.
\newblock \doi{10.1137/070699652}.
\newblock URL \url{https://doi.org/10.1137/070699652}.

\bibitem[Daskalakis et~al.(2018)Daskalakis, Ilyas, Syrgkanis, and Zeng]{daskalakis2018traininggansoptimism}
C.~Daskalakis, A.~Ilyas, V.~Syrgkanis, and H.~Zeng.
\newblock Training gans with optimism, 2018.
\newblock URL \url{https://arxiv.org/abs/1711.00141}.

\bibitem[Daskalakis et~al.(2022)Daskalakis, Golowich, and Zhang]{daskalakis2022complexity}
C.~Daskalakis, N.~Golowich, and K.~Zhang.
\newblock The complexity of markov equilibrium in stochastic games, 2022.

\bibitem[Dayan(1993)]{Dayan1993SR}
P.~Dayan.
\newblock Improving generalization for temporal difference learning: The successor representation.
\newblock \emph{Neural Computation}, 5\penalty0 (4):\penalty0 613--624, 1993.
\newblock \doi{10.1162/neco.1993.5.4.613}.

\bibitem[De~Hauwere et~al.(2010)De~Hauwere, Vrancx, and Now{\'e}]{de2010learning}
Y.-M. De~Hauwere, P.~Vrancx, and A.~Now{\'e}.
\newblock Learning multi-agent state space representations.
\newblock In \emph{Proceedings of the 9th International Conference on Autonomous Agents and Multiagent Systems: volume 1-Volume 1}, pages 715--722, 2010.

\bibitem[De~Rooij et~al.(2014)De~Rooij, Van~Erven, Gr{\"u}nwald, and Koolen]{de2014follow}
S.~De~Rooij, T.~Van~Erven, P.~D. Gr{\"u}nwald, and W.~M. Koolen.
\newblock Follow the leader if you can, hedge if you must.
\newblock \emph{The Journal of Machine Learning Research}, 15\penalty0 (1):\penalty0 1281--1316, 2014.

\bibitem[Degris et~al.(2012)Degris, White, and Sutton]{degris2012off}
T.~Degris, M.~White, and R.~S. Sutton.
\newblock Off-policy actor-critic.
\newblock \emph{arXiv preprint arXiv:1205.4839}, 2012.

\bibitem[Deisenroth and Rasmussen(2011)]{deisenroth2011pilco}
M.~Deisenroth and C.~E. Rasmussen.
\newblock Pilco: A model-based and data-efficient approach to policy search.
\newblock pages 465--472, 2011.

\bibitem[Dorigo and Blum(2005)]{dorigo2005ant}
M.~Dorigo and C.~Blum.
\newblock Ant colony optimization theory: A survey.
\newblock \emph{Theoretical computer science}, 344\penalty0 (2-3):\penalty0 243--278, 2005.

\bibitem[Du et~al.(2019)Du, Han, Fang, Dai, Liu, and Tao]{du2019learning}
Y.~Du, L.~Han, M.~Fang, T.~Dai, J.~Liu, and D.~Tao.
\newblock Liir: Learning individual intrinsic reward in multi-agent reinforcement learning.
\newblock 2019.

\bibitem[Eccles et~al.(2019)Eccles, Bachrach, Lever, Lazaridou, and Graepel]{eccles2019biases}
T.~Eccles, Y.~Bachrach, G.~Lever, A.~Lazaridou, and T.~Graepel.
\newblock Biases for emergent communication in multi-agent reinforcement learning.
\newblock \emph{Advances in neural information processing systems}, 32, 2019.

\bibitem[Ellis et~al.(2022)Ellis, Moalla, Samvelyan, Sun, Mahajan, Foerster, and Whiteson]{ellis2022smacv2}
B.~Ellis, S.~Moalla, M.~Samvelyan, M.~Sun, A.~Mahajan, J.~N. Foerster, and S.~Whiteson.
\newblock Smacv2: An improved benchmark for cooperative multi-agent reinforcement learning, 2022.
\newblock URL \url{https://arxiv.org/abs/2212.07489}.

\bibitem[Erdogan and Veloso(2011)]{erdogan2011action}
C.~Erdogan and M.~Veloso.
\newblock Action selection via learning behavior patterns in multi-robot domains.
\newblock In \emph{Proc. International Joint Conference on Artificial Intelligence}, pages 192--197. Citeseer, 2011.

\bibitem[Espeholt et~al.(2018)Espeholt, Soyer, Munos, Simonyan, Mnih, Ward, Doron, Firoiu, Harley, Dunning, Legg, and Kavukcuoglu]{espeholt2018impala}
L.~Espeholt, H.~Soyer, R.~Munos, K.~Simonyan, V.~Mnih, T.~Ward, Y.~Doron, V.~Firoiu, T.~Harley, I.~Dunning, S.~Legg, and K.~Kavukcuoglu.
\newblock Impala: Scalable distributed deep-rl with importance weighted actor-learner architectures, 2018.

\bibitem[Fachantidis et~al.(2017)Fachantidis, Taylor, and Vlahavas]{fachantidis2017learning}
A.~Fachantidis, M.~E. Taylor, and I.~Vlahavas.
\newblock Learning to teach reinforcement learning agents.
\newblock \emph{Machine Learning and Knowledge Extraction}, 1\penalty0 (1):\penalty0 21--42, 2017.

\bibitem[Feinberg et~al.(2018)Feinberg, Wan, Stoica, Jordan, Gonzalez, and Levine]{feinberg2018modelbased}
V.~Feinberg, A.~Wan, I.~Stoica, M.~I. Jordan, J.~E. Gonzalez, and S.~Levine.
\newblock Model-based value estimation for efficient model-free reinforcement learning, 2018.

\bibitem[Fishburn et~al.(1979)Fishburn, Fishburn, et~al.]{fishburn1979utility}
P.~C. Fishburn, P.~C. Fishburn, et~al.
\newblock Utility theory for decision making.
\newblock 1979.

\bibitem[Foerster et~al.(2017{\natexlab{a}})Foerster, Farquhar, Afouras, Nardelli, and Whiteson]{foerster2017counterfactual}
J.~Foerster, G.~Farquhar, T.~Afouras, N.~Nardelli, and S.~Whiteson.
\newblock Counterfactual multi-agent policy gradients, 2017{\natexlab{a}}.

\bibitem[Foerster et~al.(2017{\natexlab{b}})Foerster, Nardelli, Farquhar, Afouras, Torr, Kohli, and Whiteson]{foerster17er}
J.~Foerster, N.~Nardelli, G.~Farquhar, T.~Afouras, P.~H.~S. Torr, P.~Kohli, and S.~Whiteson.
\newblock Stabilising experience replay for deep multi-agent reinforcement learning.
\newblock 70:\penalty0 1146--1155, 06--11 Aug 2017{\natexlab{b}}.
\newblock URL \url{https://proceedings.mlr.press/v70/foerster17b.html}.

\bibitem[Foerster et~al.(2016)Foerster, Assael, de~Freitas, and Whiteson]{foerster2016learning}
J.~N. Foerster, Y.~M. Assael, N.~de~Freitas, and S.~Whiteson.
\newblock Learning to communicate with deep multi-agent reinforcement learning, 2016.

\bibitem[Foerster et~al.(2017{\natexlab{c}})Foerster, Chen, Al-Shedivat, Whiteson, Abbeel, and Mordatch]{foerster2017lola}
J.~N. Foerster, R.~Y. Chen, M.~Al-Shedivat, S.~Whiteson, P.~Abbeel, and I.~Mordatch.
\newblock Learning with opponent-learning awareness.
\newblock \emph{arXiv preprint arXiv:1709.04326}, 2017{\natexlab{c}}.

\bibitem[Foerster et~al.(2018)Foerster, Chen, Al-Shedivat, Whiteson, Abbeel, and Mordatch]{foerster2018learning}
J.~N. Foerster, R.~Y. Chen, M.~Al-Shedivat, S.~Whiteson, P.~Abbeel, and I.~Mordatch.
\newblock Learning with opponent-learning awareness, 2018.

\bibitem[Fortunato et~al.(2017)Fortunato, Azar, Piot, Menick, Osband, Graves, Mnih, Munos, Hassabis, Pietquin, et~al.]{fortunato2017noisy}
M.~Fortunato, M.~G. Azar, B.~Piot, J.~Menick, I.~Osband, A.~Graves, V.~Mnih, R.~Munos, D.~Hassabis, O.~Pietquin, et~al.
\newblock Noisy networks for exploration.
\newblock \emph{arXiv preprint arXiv:1706.10295}, 2017.

\bibitem[Freed et~al.(2020)Freed, Sartoretti, Hu, and Choset]{Freed2020}
B.~Freed, G.~Sartoretti, J.~Hu, and H.~Choset.
\newblock Communication learning via backpropagation in discrete channels with unknown noise.
\newblock \emph{Proceedings of the AAAI Conference on Artificial Intelligence}, 34\penalty0 (05):\penalty0 7160--7168, Apr. 2020.
\newblock \doi{10.1609/aaai.v34i05.6205}.
\newblock URL \url{https://ojs.aaai.org/index.php/AAAI/article/view/6205}.

\bibitem[Freeman et~al.(2021)Freeman, Frey, Raichuk, Girgin, Mordatch, and Bachem]{freeman2021brax}
C.~D. Freeman, E.~Frey, A.~Raichuk, S.~Girgin, I.~Mordatch, and O.~Bachem.
\newblock Brax -- a differentiable physics engine for large scale rigid body simulation, 2021.

\bibitem[Freund and Schapire(1999)]{freund1999adaptive}
Y.~Freund and R.~E. Schapire.
\newblock Adaptive game playing using multiplicative weights.
\newblock \emph{Games and Economic Behavior}, 29\penalty0 (1-2):\penalty0 79--103, 1999.

\bibitem[Fu et~al.(2022)Fu, Yu, Xu, Yang, and Wu]{fu2022revisiting}
W.~Fu, C.~Yu, Z.~Xu, J.~Yang, and Y.~Wu.
\newblock Revisiting some common practices in cooperative multi-agent reinforcement learning, 2022.

\bibitem[Fudenberg and Kreps(1993)]{fudenberg1993learning}
D.~Fudenberg and D.~M. Kreps.
\newblock Learning mixed equilibria.
\newblock \emph{Games and economic behavior}, 5\penalty0 (3):\penalty0 320--367, 1993.

\bibitem[Fujimoto and Gu(2021)]{fujimoto2021minimalist}
S.~Fujimoto and S.~S. Gu.
\newblock A minimalist approach to offline reinforcement learning, 2021.

\bibitem[Fujimoto et~al.(2018{\natexlab{a}})Fujimoto, van Hoof, and Meger]{fujimoto2018addressing}
S.~Fujimoto, H.~van Hoof, and D.~Meger.
\newblock Addressing function approximation error in actor-critic methods, 2018{\natexlab{a}}.

\bibitem[Fujimoto et~al.(2018{\natexlab{b}})Fujimoto, van Hoof, and Meger]{fujimoto2018td3}
S.~Fujimoto, H.~van Hoof, and D.~Meger.
\newblock Addressing function approximation error in actor-critic methods, 2018{\natexlab{b}}.

\bibitem[Fulda and Ventura(2007)]{fulda2007shadowing}
N.~Fulda and D.~Ventura.
\newblock Predicting and preventing coordination problems in cooperative q-learning systems.
\newblock page 780–785, 2007.

\bibitem[Gal et~al.(2017)Gal, Hron, and Kendall]{gal2017concrete}
Y.~Gal, J.~Hron, and A.~Kendall.
\newblock Concrete dropout, 2017.

\bibitem[Ghosh et~al.(2023)Ghosh, Bhateja, and Levine]{ghosh2023reinforcement}
D.~Ghosh, C.~Bhateja, and S.~Levine.
\newblock Reinforcement learning from passive data via latent intentions, 2023.

\bibitem[Ghosh(1986)]{ghosh1986concept}
S.~Ghosh.
\newblock On the concept of dynamic multi-level simulation.
\newblock In \emph{Proceedings of the 19th annual symposium on Simulation}, pages 201--205, 1986.

\bibitem[Gidel et~al.(2019)Gidel, Hemmat, Pezeshki, Le~Priol, Huang, Lacoste-Julien, and Mitliagkas]{gidel2019negative}
G.~Gidel, R.~A. Hemmat, M.~Pezeshki, R.~Le~Priol, G.~Huang, S.~Lacoste-Julien, and I.~Mitliagkas.
\newblock Negative momentum for improved game dynamics.
\newblock In \emph{The 22nd International Conference on Artificial Intelligence and Statistics}, pages 1802--1811. PMLR, 2019.

\bibitem[Gilboa and Zemel(1989)]{Gilboa1989Complexity}
I.~Gilboa and E.~Zemel.
\newblock Nash and correlated equilibria: Some complexity considerations.
\newblock \emph{Games and Economic Behavior}, 1\penalty0 (1):\penalty0 80--93, 1989.
\newblock ISSN 0899-8256.
\newblock \doi{https://doi.org/10.1016/0899-8256(89)90006-7}.
\newblock URL \url{https://www.sciencedirect.com/science/article/pii/0899825689900067}.

\bibitem[Goodfellow et~al.(2016)Goodfellow, Bengio, and Courville]{Goodfellow2016DL}
I.~Goodfellow, Y.~Bengio, and A.~Courville.
\newblock Deep learning.
\newblock 2016.
\newblock \url{http://www.deeplearningarticle.org}.

\bibitem[Grass{\'e}(1959)]{grasse1959reconstruction}
P.-P. Grass{\'e}.
\newblock La reconstruction du nid et les coordinations interindividuelles chez bellicositermes natalensis et cubitermes sp. la th{\'e}orie de la stigmergie: Essai d'interpr{\'e}tation du comportement des termites constructeurs.
\newblock \emph{Insectes sociaux}, 6:\penalty0 41--80, 1959.

\bibitem[Greenwald and Hall(2003)]{Greenwald2003CEQ}
A.~Greenwald and K.~Hall.
\newblock Correlated-q learning.
\newblock page 242–249, 2003.

\bibitem[Gronauer and Diepold(2022)]{Gronauer2022survey}
S.~Gronauer and K.~Diepold.
\newblock Multi-agent deep reinforcement learning: A survey.
\newblock \emph{Artif. Intell. Rev.}, 55\penalty0 (2):\penalty0 895–943, feb 2022.
\newblock ISSN 0269-2821.
\newblock \doi{10.1007/s10462-021-09996-w}.
\newblock URL \url{https://doi.org/10.1007/s10462-021-09996-w}.

\bibitem[Gu et~al.(2016)Gu, Lillicrap, Sutskever, and Levine]{gu2016continuous}
S.~Gu, T.~Lillicrap, I.~Sutskever, and S.~Levine.
\newblock Continuous deep q-learning with model-based acceleration.
\newblock pages 2829--2838, 2016.

\bibitem[Gu et~al.(2017)Gu, Lillicrap, Ghahramani, Turner, and Levine]{gu2017qprop}
S.~Gu, T.~Lillicrap, Z.~Ghahramani, R.~E. Turner, and S.~Levine.
\newblock Q-prop: Sample-efficient policy gradient with an off-policy critic, 2017.

\bibitem[Gu et~al.(2023)Gu, Kuba, Chen, Du, Yang, Knoll, and Yang]{gu2023safe}
S.~Gu, J.~G. Kuba, Y.~Chen, Y.~Du, L.~Yang, A.~Knoll, and Y.~Yang.
\newblock Safe multi-agent reinforcement learning for multi-robot control.
\newblock \emph{Artificial Intelligence}, page 103905, 2023.

\bibitem[Guan et~al.(2023)Guan, Zhang, Fan, Li, Chen, Li, Tian, Yuan, and Yu]{guan2023efficient}
C.~Guan, L.~Zhang, C.~Fan, Y.~Li, F.~Chen, L.~Li, Y.~Tian, L.~Yuan, and Y.~Yu.
\newblock Efficient human-ai coordination via preparatory language-based convention.
\newblock \emph{arXiv preprint arXiv:2311.00416}, 2023.

\bibitem[Guestrin et~al.(2001)Guestrin, Koller, and Parr]{guestrin2001}
C.~Guestrin, D.~Koller, and R.~Parr.
\newblock Multiagent planning with factored mdps.
\newblock 14, 2001.
\newblock URL \url{https://proceedings.neurips.cc/paper_files/paper/2001/file/7af6266cc52234b5aa339b16695f7fc4-Paper.pdf}.

\bibitem[Guo et~al.(2023)Guo, Campbell, Stepputtis, Li, Hughes, Fang, and Sycara]{guo2023explainable}
Y.~Guo, J.~Campbell, S.~Stepputtis, R.~Li, D.~Hughes, F.~Fang, and K.~Sycara.
\newblock Explainable action advising for multi-agent reinforcement learning.
\newblock In \emph{2023 IEEE International Conference on Robotics and Automation (ICRA)}, pages 5515--5521. IEEE, 2023.

\bibitem[Gupta et~al.(2017{\natexlab{a}})Gupta, Devin, Liu, Abbeel, and Levine]{gupta2017learning}
A.~Gupta, C.~Devin, Y.~Liu, P.~Abbeel, and S.~Levine.
\newblock Learning invariant feature spaces to transfer skills with reinforcement learning, 2017{\natexlab{a}}.

\bibitem[Gupta et~al.(2017{\natexlab{b}})Gupta, Egorov, and Kochenderfer]{Gupta2017CooperativeMC}
J.~K. Gupta, M.~Egorov, and M.~J. Kochenderfer.
\newblock Cooperative multi-agent control using deep reinforcement learning.
\newblock 2017{\natexlab{b}}.

\bibitem[Ha and Schmidhuber(2018)]{ha2018worldmodels}
D.~Ha and J.~Schmidhuber.
\newblock Recurrent world models facilitate policy evolution.
\newblock pages 2451--2463. Curran Associates, Inc., 2018.
\newblock URL \url{https://papers.nips.cc/paper/7512-recurrent-world-models-facilitate-policy-evolution}.
\newblock \url{https://worldmodels.github.io}.

\bibitem[Haarnoja et~al.(2017)Haarnoja, Tang, Abbeel, and Levine]{haarnoja2017reinforcement}
T.~Haarnoja, H.~Tang, P.~Abbeel, and S.~Levine.
\newblock Reinforcement learning with deep energy-based policies.
\newblock pages 1352--1361, 2017.

\bibitem[Haarnoja et~al.(2018)Haarnoja, Zhou, Abbeel, and Levine]{haarnoja2018soft}
T.~Haarnoja, A.~Zhou, P.~Abbeel, and S.~Levine.
\newblock Soft actor-critic: Off-policy maximum entropy deep reinforcement learning with a stochastic actor, 2018.

\bibitem[Haarnoja et~al.(2019)Haarnoja, Zhou, Hartikainen, Tucker, Ha, Tan, Kumar, Zhu, Gupta, Abbeel, and Levine]{haarnoja2019soft}
T.~Haarnoja, A.~Zhou, K.~Hartikainen, G.~Tucker, S.~Ha, J.~Tan, V.~Kumar, H.~Zhu, A.~Gupta, P.~Abbeel, and S.~Levine.
\newblock Soft actor-critic algorithms and applications, 2019.

\bibitem[Hafner et~al.(2020)Hafner, Lillicrap, Ba, and Norouzi]{hafner2020dream}
D.~Hafner, T.~Lillicrap, J.~Ba, and M.~Norouzi.
\newblock Dream to control: Learning behaviors by latent imagination, 2020.

\bibitem[Hafner et~al.(2022)Hafner, Lillicrap, Norouzi, and Ba]{hafner2022mastering}
D.~Hafner, T.~Lillicrap, M.~Norouzi, and J.~Ba.
\newblock Mastering atari with discrete world models, 2022.

\bibitem[Hafner et~al.(2023)Hafner, Pasukonis, Ba, and Lillicrap]{hafner2023mastering}
D.~Hafner, J.~Pasukonis, J.~Ba, and T.~Lillicrap.
\newblock Mastering diverse domains through world models, 2023.

\bibitem[Haman et~al.(2017)Haman, Kamla, Galland, and Kamgang]{haman2017towards}
I.~T. Haman, V.~C. Kamla, S.~Galland, and J.~C. Kamgang.
\newblock Towards an multilevel agent-based model for traffic simulation.
\newblock \emph{Procedia Computer Science}, 109:\penalty0 887--892, 2017.

\bibitem[Hao et~al.(2023)Hao, Yang, Tang, Bai, Liu, Meng, Liu, and Wang]{Hao_2023}
J.~Hao, T.~Yang, H.~Tang, C.~Bai, J.~Liu, Z.~Meng, P.~Liu, and Z.~Wang.
\newblock Exploration in deep reinforcement learning: From single-agent to multiagent domain.
\newblock \emph{{IEEE} Transactions on Neural Networks and Learning Systems}, pages 1--21, 2023.
\newblock \doi{10.1109/tnnls.2023.3236361}.
\newblock URL \url{https://doi.org/10.1109%2Ftnnls.2023.3236361}.

\bibitem[Harker and Pang(1990)]{harker1990finite}
P.~T. Harker and J.-S. Pang.
\newblock Finite-dimensional variational inequality and nonlinear complementarity problems: a survey of theory, algorithms and applications.
\newblock \emph{Mathematical programming}, 48\penalty0 (1):\penalty0 161--220, 1990.

\bibitem[Harsanyi(1967)]{harsanyi1967games}
J.~C. Harsanyi.
\newblock Games with incomplete information played by “bayesian” players, i--iii part i. the basic model.
\newblock \emph{Management science}, 14\penalty0 (3):\penalty0 159--182, 1967.

\bibitem[Hart and Mas-Colell(2000)]{hart2000simple}
S.~Hart and A.~Mas-Colell.
\newblock A simple adaptive procedure leading to correlated equilibrium.
\newblock \emph{Econometrica}, 68\penalty0 (5):\penalty0 1127--1150, 2000.

\bibitem[Hausknecht and Stone(2015)]{hausknecht2015deep}
M.~Hausknecht and P.~Stone.
\newblock Deep recurrent q-learning for partially observable mdps.
\newblock 2015.

\bibitem[Hausknecht(2016)]{Hausknecht2016thesis}
M.~J. Hausknecht.
\newblock \emph{Cooperation and communication in multiagent deep reinforcement learning}.
\newblock PhD thesis, The University of Texas at Austin, 2016.

\bibitem[Havrylov and Titov(2017)]{havrylov2017emergence}
S.~Havrylov and I.~Titov.
\newblock Emergence of language with multi-agent games: Learning to communicate with sequences of symbols, 2017.

\bibitem[He et~al.(2016)He, Boyd-Graber, Kwok, and Daum{\'e}~III]{he2016opponent}
H.~He, J.~Boyd-Graber, K.~Kwok, and H.~Daum{\'e}~III.
\newblock Opponent modeling in deep reinforcement learning.
\newblock In \emph{International conference on machine learning}, pages 1804--1813. PMLR, 2016.

\bibitem[Heinrich et~al.(2015)Heinrich, Lanctot, and Silver]{heinrich2015fictitious}
J.~Heinrich, M.~Lanctot, and D.~Silver.
\newblock Fictitious self-play in extensive-form games.
\newblock In \emph{International conference on machine learning}, pages 805--813. PMLR, 2015.

\bibitem[Henderson et~al.(2019)Henderson, Islam, Bachman, Pineau, Precup, and Meger]{henderson2019deep}
P.~Henderson, R.~Islam, P.~Bachman, J.~Pineau, D.~Precup, and D.~Meger.
\newblock Deep reinforcement learning that matters, 2019.

\bibitem[Hernandez{-}Leal et~al.(2017)Hernandez{-}Leal, Kaisers, Baarslag, and de~Cote]{Hernandez2017Survey}
P.~Hernandez{-}Leal, M.~Kaisers, T.~Baarslag, and E.~M. de~Cote.
\newblock A survey of learning in multiagent environments: Dealing with non-stationarity.
\newblock \emph{CoRR}, abs/1707.09183, 2017.
\newblock URL \url{http://arxiv.org/abs/1707.09183}.

\bibitem[Hessel et~al.(2017)Hessel, Modayil, van Hasselt, Schaul, Ostrovski, Dabney, Horgan, Piot, Azar, and Silver]{hessel2017rainbow}
M.~Hessel, J.~Modayil, H.~van Hasselt, T.~Schaul, G.~Ostrovski, W.~Dabney, D.~Horgan, B.~Piot, M.~Azar, and D.~Silver.
\newblock Rainbow: Combining improvements in deep reinforcement learning, 2017.

\bibitem[Hinton et~al.(2015)Hinton, Vinyals, and Dean]{hinton2015distilling}
G.~Hinton, O.~Vinyals, and J.~Dean.
\newblock Distilling the knowledge in a neural network.
\newblock \emph{arXiv preprint arXiv:1503.02531}, 2015.

\bibitem[Holland and Melhuish(1999)]{holland1999stigmergy}
O.~Holland and C.~Melhuish.
\newblock Stigmergy, self-organization, and sorting in collective robotics.
\newblock \emph{Artificial life}, 5\penalty0 (2):\penalty0 173--202, 1999.

\bibitem[Hong et~al.(2018)Hong, Su, Shann, Chang, and Lee]{hong2018deep}
Z.-W. Hong, S.-Y. Su, T.-Y. Shann, Y.-H. Chang, and C.-Y. Lee.
\newblock A deep policy inference q-network for multi-agent systems, 2018.

\bibitem[Houthooft et~al.(2017)Houthooft, Chen, Duan, Schulman, Turck, and Abbeel]{houthooft2017vime}
R.~Houthooft, X.~Chen, Y.~Duan, J.~Schulman, F.~D. Turck, and P.~Abbeel.
\newblock Vime: Variational information maximizing exploration, 2017.

\bibitem[Hu et~al.(2020)Hu, Zhu, Zhao, Zhao, and Hao]{hu2020event}
G.~Hu, Y.~Zhu, D.~Zhao, M.~Zhao, and J.~Hao.
\newblock Event-triggered multi-agent reinforcement learning with communication under limited-bandwidth constraint.
\newblock \emph{arXiv preprint arXiv:2010.04978}, 2020.

\bibitem[Hu et~al.(2021)Hu, Lerer, Cui, Wu, Pineda, Brown, and Foerster]{hu2021offbelief}
H.~Hu, A.~Lerer, B.~Cui, D.~Wu, L.~Pineda, N.~Brown, and J.~Foerster.
\newblock Off-belief learning, 2021.

\bibitem[Hu and Wellman(2003)]{Hu2003NashQ}
J.~Hu and M.~P. Wellman.
\newblock Nash q-learning for general-sum stochastic games.
\newblock \emph{J. Mach. Learn. Res.}, 4\penalty0 (null):\penalty0 1039–1069, dec 2003.
\newblock ISSN 1532-4435.

\bibitem[Huh(2021)]{Huh2021MixAM}
D.~Huh.
\newblock Mix and mask actor-critic methods.
\newblock \emph{ArXiv}, abs/2106.13037, 2021.
\newblock URL \url{https://api.semanticscholar.org/CorpusID:235691945}.

\bibitem[Huh and Mohapatra(2023)]{huh2023decentralized}
D.~Huh and P.~Mohapatra.
\newblock Decentralized multi-agent filtering, 2023.

\bibitem[Huh and Mohapatra(2024{\natexlab{a}})]{Huh2023isaacteams}
D.~Huh and P.~Mohapatra.
\newblock Isaacteams: Extending gpu-based physics simulator for multi-agent learning.
\newblock 1 2024{\natexlab{a}}.

\bibitem[Huh and Mohapatra(2024{\natexlab{b}})]{huh2024representation}
D.~Huh and P.~Mohapatra.
\newblock Representation learning for efficient deep multi-agent reinforcement learning, 2024{\natexlab{b}}.

\bibitem[Hurwicz and Reiter(2006)]{hurwicz2006designing}
L.~Hurwicz and S.~Reiter.
\newblock \emph{Designing economic mechanisms}.
\newblock Cambridge university press, 2006.

\bibitem[Iqbal and Sha(2021)]{iqbal2021coordinated}
S.~Iqbal and F.~Sha.
\newblock Coordinated exploration via intrinsic rewards for multi-agent reinforcement learning, 2021.

\bibitem[Jaderberg et~al.(2019)Jaderberg, Czarnecki, Dunning, Marris, Lever, Casta{\~{n} }eda, Beattie, Rabinowitz, Morcos, Ruderman, Sonnerat, Green, Deason, Leibo, Silver, Hassabis, Kavukcuoglu, and Graepel]{Jaderberg_2019}
M.~Jaderberg, W.~M. Czarnecki, I.~Dunning, L.~Marris, G.~Lever, A.~G. Casta{\~{n} }eda, C.~Beattie, N.~C. Rabinowitz, A.~S. Morcos, A.~Ruderman, N.~Sonnerat, T.~Green, L.~Deason, J.~Z. Leibo, D.~Silver, D.~Hassabis, K.~Kavukcuoglu, and T.~Graepel.
\newblock Human-level performance in 3d multiplayer games with population-based reinforcement learning.
\newblock \emph{Science}, 364\penalty0 (6443):\penalty0 859--865, may 2019.
\newblock \doi{10.1126/science.aau6249}.
\newblock URL \url{https://doi.org/10.1126%2Fscience.aau6249}.

\bibitem[James et~al.(2013)James, Witten, Hastie, Tibshirani, et~al.]{james2013introduction}
G.~James, D.~Witten, T.~Hastie, R.~Tibshirani, et~al.
\newblock \emph{An introduction to statistical learning}, volume 112.
\newblock Springer, 2013.

\bibitem[Jaques et~al.(2019)Jaques, Lazaridou, Hughes, Gulcehre, Ortega, Strouse, Leibo, and de~Freitas]{jaques2019social}
N.~Jaques, A.~Lazaridou, E.~Hughes, C.~Gulcehre, P.~A. Ortega, D.~Strouse, J.~Z. Leibo, and N.~de~Freitas.
\newblock Social influence as intrinsic motivation for multi-agent deep reinforcement learning, 2019.

\bibitem[Jelassi et~al.(2020)Jelassi, Domingo-Enrich, Scieur, Mensch, and Bruna]{jelassi2020extragradient}
S.~Jelassi, C.~Domingo-Enrich, D.~Scieur, A.~Mensch, and J.~Bruna.
\newblock Extragradient with player sampling for faster nash equilibrium finding.
\newblock In \emph{Proceedings of the International Conference on Machine Learning}, 2020.

\bibitem[Jiang et~al.(2011)Jiang, Leyton-Brown, and Bhat]{jiang2011action}
A.~X. Jiang, K.~Leyton-Brown, and N.~A. Bhat.
\newblock Action-graph games.
\newblock \emph{Games and Economic Behavior}, 71\penalty0 (1):\penalty0 141--173, 2011.

\bibitem[Jiang and Lu(2018)]{jiang2018learning}
J.~Jiang and Z.~Lu.
\newblock Learning attentional communication for multi-agent cooperation, 2018.

\bibitem[Jiang et~al.(2020)Jiang, Dun, Huang, and Lu]{jiang2020graph}
J.~Jiang, C.~Dun, T.~Huang, and Z.~Lu.
\newblock Graph convolutional reinforcement learning, 2020.

\bibitem[Jie and Abbeel(2010)]{Jie2010offpolicypg}
T.~Jie and P.~Abbeel.
\newblock On a connection between importance sampling and the likelihood ratio policy gradient.
\newblock 23, 2010.
\newblock URL \url{https://proceedings.neurips.cc/paper/2010/file/35cf8659cfcb13224cbd47863a34fc58-Paper.pdf}.

\bibitem[Jin et~al.(2018{\natexlab{a}})Jin, Song, Li, Gai, Wang, and Zhang]{Jin2018bidding}
J.~Jin, C.~Song, H.~Li, K.~Gai, J.~Wang, and W.~Zhang.
\newblock Real-time bidding with multi-agent reinforcement learning in display advertising.
\newblock oct 2018{\natexlab{a}}.
\newblock \doi{10.1145/3269206.3272021}.
\newblock URL \url{https://doi.org/10.1145%2F3269206.3272021}.

\bibitem[Jin et~al.(2018{\natexlab{b}})Jin, Keutzer, and Levine]{jin2018regret}
P.~Jin, K.~Keutzer, and S.~Levine.
\newblock Regret minimization for partially observable deep reinforcement learning.
\newblock In \emph{International conference on machine learning}, pages 2342--2351. PMLR, 2018{\natexlab{b}}.

\bibitem[Jordan(1991)]{Jordan1991}
J.~Jordan.
\newblock Bayesian learning in normal form games.
\newblock \emph{Games and Economic Behavior}, 3\penalty0 (1):\penalty0 60--81, 1991.
\newblock ISSN 0899-8256.
\newblock \doi{https://doi.org/10.1016/0899-8256(91)90005-Y}.
\newblock URL \url{https://www.sciencedirect.com/science/article/pii/089982569190005Y}.

\bibitem[Kahneman and Tversky(2013)]{kahneman2013prospect}
D.~Kahneman and A.~Tversky.
\newblock Prospect theory: An analysis of decision under risk.
\newblock In \emph{Handbook of the fundamentals of financial decision making: Part I}, pages 99--127. World Scientific, 2013.

\bibitem[Kakade(2001)]{kakade2001natural}
S.~M. Kakade.
\newblock A natural policy gradient.
\newblock \emph{Advances in neural information processing systems}, 14, 2001.

\bibitem[Kakade and Langford(2002)]{Kakade2002ApproximatelyOA}
S.~M. Kakade and J.~Langford.
\newblock Approximately optimal approximate reinforcement learning.
\newblock 2002.
\newblock URL \url{https://api.semanticscholar.org/CorpusID:31442909}.

\bibitem[Kalai and Lehrer(1993)]{kalai1993rational}
E.~Kalai and E.~Lehrer.
\newblock Rational learning leads to nash equilibrium.
\newblock \emph{Econometrica: Journal of the Econometric Society}, pages 1019--1045, 1993.

\bibitem[Kalashnikov et~al.(2018)Kalashnikov, Irpan, Pastor, Ibarz, Herzog, Jang, Quillen, Holly, Kalakrishnan, Vanhoucke, et~al.]{kalashnikov2018qt}
D.~Kalashnikov, A.~Irpan, P.~Pastor, J.~Ibarz, A.~Herzog, E.~Jang, D.~Quillen, E.~Holly, M.~Kalakrishnan, V.~Vanhoucke, et~al.
\newblock Qt-opt: Scalable deep reinforcement learning for vision-based robotic manipulation.
\newblock \emph{arXiv preprint arXiv:1806.10293}, 2018.

\bibitem[Kapetanakis and Kudenko(2003)]{Kapetanakis2003coordinate}
S.~Kapetanakis and D.~Kudenko.
\newblock Improving on the reinforcement learning of coordination in cooperative multi-agent systems.
\newblock 04 2003.

\bibitem[Kapturowski et~al.(2018)Kapturowski, Ostrovski, Quan, Munos, and Dabney]{kapturowski2018recurrent}
S.~Kapturowski, G.~Ostrovski, J.~Quan, R.~Munos, and W.~Dabney.
\newblock Recurrent experience replay in distributed reinforcement learning.
\newblock 2018.

\bibitem[Kapturowski et~al.(2022)Kapturowski, Campos, Jiang, Rakićević, van Hasselt, Blundell, and Badia]{kapturowski2022humanlevel}
S.~Kapturowski, V.~Campos, R.~Jiang, N.~Rakićević, H.~van Hasselt, C.~Blundell, and A.~P. Badia.
\newblock Human-level atari 200x faster, 2022.

\bibitem[Kim et~al.(2019)Kim, Moon, Hostallero, Kang, Lee, Son, and Yi]{kim2019learning}
D.~Kim, S.~Moon, D.~Hostallero, W.~J. Kang, T.~Lee, K.~Son, and Y.~Yi.
\newblock Learning to schedule communication in multi-agent reinforcement learning, 2019.

\bibitem[Kim et~al.(2022)Kim, Riemer, Liu, Foerster, Everett, Sun, Tesauro, and How]{kim2022influencing}
D.-K. Kim, M.~Riemer, M.~Liu, J.~N. Foerster, M.~Everett, C.~Sun, G.~Tesauro, and J.~P. How.
\newblock Influencing long-term behavior in multiagent reinforcement learning, 2022.

\bibitem[Kim et~al.(2020)Kim, Park, and Sung]{kim2020communication}
W.~Kim, J.~Park, and Y.~Sung.
\newblock Communication in multi-agent reinforcement learning: Intention sharing.
\newblock In \emph{International Conference on Learning Representations}, 2020.

\bibitem[Kipf and Welling(2017)]{kipf2017semisupervised}
T.~N. Kipf and M.~Welling.
\newblock Semi-supervised classification with graph convolutional networks, 2017.

\bibitem[Kok et~al.(2005)Kok, Hoen, Bakker, and Vlassis]{kok2005utile}
J.~R. Kok, E.~J. Hoen, B.~Bakker, and N.~Vlassis.
\newblock Utile coordination: Learning interdependencies among cooperative agents.
\newblock In \emph{EEE Symp. on Computational Intelligence and Games, Colchester, Essex}, pages 29--36, 2005.

\bibitem[Koller and Milch(2003)]{koller2003multi}
D.~Koller and B.~Milch.
\newblock Multi-agent influence diagrams for representing and solving games.
\newblock \emph{Games and economic behavior}, 45\penalty0 (1):\penalty0 181--221, 2003.

\bibitem[Kong et~al.(2017)Kong, Xin, Liu, and Wang]{kong2017revisiting}
X.~Kong, B.~Xin, F.~Liu, and Y.~Wang.
\newblock Revisiting the master-slave architecture in multi-agent deep reinforcement learning, 2017.

\bibitem[Korpelevich(1976)]{Korpelevich1976TheEM}
G.~M. Korpelevich.
\newblock The extragradient method for finding saddle points and other problems.
\newblock 1976.
\newblock URL \url{https://api.semanticscholar.org/CorpusID:118602977}.

\bibitem[Kostrikov et~al.(2021)Kostrikov, Nair, and Levine]{kostrikov2021offline}
I.~Kostrikov, A.~Nair, and S.~Levine.
\newblock Offline reinforcement learning with implicit q-learning, 2021.

\bibitem[Koul(2019)]{magym}
A.~Koul.
\newblock ma-gym: Collection of multi-agent environments based on openai gym.
\newblock \url{https://github.com/koulanurag/ma-gym}, 2019.

\bibitem[Koutsoupias and Papadimitriou(1999)]{koutsoupias1999worst}
E.~Koutsoupias and C.~Papadimitriou.
\newblock Worst-case equilibria.
\newblock In \emph{Annual symposium on theoretical aspects of computer science}, pages 404--413. Springer, 1999.

\bibitem[Kovalev et~al.(2023)Kovalev, Beznosikov, Sadiev, Persiianov, Richtárik, and Gasnikov]{kovalev2023optimalalgorithmsdecentralizedstochastic}
D.~Kovalev, A.~Beznosikov, A.~Sadiev, M.~Persiianov, P.~Richtárik, and A.~Gasnikov.
\newblock Optimal algorithms for decentralized stochastic variational inequalities, 2023.
\newblock URL \url{https://arxiv.org/abs/2202.02771}.

\bibitem[Kraemer and Banerjee(2016)]{Kramer2016ctde}
L.~Kraemer and B.~Banerjee.
\newblock Multi-agent reinforcement learning as a rehearsal for decentralized planning.
\newblock \emph{Neurocomputing}, 190:\penalty0 82--94, 2016.
\newblock ISSN 0925-2312.
\newblock \doi{https://doi.org/10.1016/j.neucom.2016.01.031}.
\newblock URL \url{https://www.sciencedirect.com/science/article/pii/S0925231216000783}.

\bibitem[Kreps and Wilson(1982)]{kreps1982sequential}
D.~M. Kreps and R.~Wilson.
\newblock Sequential equilibria.
\newblock \emph{Econometrica: Journal of the Econometric Society}, pages 863--894, 1982.

\bibitem[Kulkarni et~al.(2016)Kulkarni, Saeedi, Gautam, and Gershman]{kulkarni2016deep}
T.~D. Kulkarni, A.~Saeedi, S.~Gautam, and S.~J. Gershman.
\newblock Deep successor reinforcement learning, 2016.

\bibitem[Kumar et~al.(2020)Kumar, Zhou, Tucker, and Levine]{kumar2020conservative}
A.~Kumar, A.~Zhou, G.~Tucker, and S.~Levine.
\newblock Conservative q-learning for offline reinforcement learning, 2020.

\bibitem[Kurach et~al.(2020)Kurach, Raichuk, Stańczyk, Zając, Bachem, Espeholt, Riquelme, Vincent, Michalski, Bousquet, and Gelly]{kurach2020google}
K.~Kurach, A.~Raichuk, P.~Stańczyk, M.~Zając, O.~Bachem, L.~Espeholt, C.~Riquelme, D.~Vincent, M.~Michalski, O.~Bousquet, and S.~Gelly.
\newblock Google research football: A novel reinforcement learning environment, 2020.

\bibitem[Kurutach et~al.(2018)Kurutach, Clavera, Duan, Tamar, and Abbeel]{kurutach2018modelensemble}
T.~Kurutach, I.~Clavera, Y.~Duan, A.~Tamar, and P.~Abbeel.
\newblock Model-ensemble trust-region policy optimization, 2018.

\bibitem[Laffont and Martimort(2009)]{laffont2009theory}
J.-J. Laffont and D.~Martimort.
\newblock The theory of incentives: the principal-agent model.
\newblock In \emph{The theory of incentives}. Princeton university press, 2009.

\bibitem[Lai et~al.(2020)Lai, Zha, Li, and Hu]{lai2020dual}
K.-H. Lai, D.~Zha, Y.~Li, and X.~Hu.
\newblock Dual policy distillation.
\newblock \emph{arXiv preprint arXiv:2006.04061}, 2020.

\bibitem[Lan et~al.(2021)Lan, Srinivasa, Wang, and Zheng]{lan2021warpdrive}
T.~Lan, S.~Srinivasa, H.~Wang, and S.~Zheng.
\newblock Warpdrive: Extremely fast end-to-end deep multi-agent reinforcement learning on a gpu, 2021.

\bibitem[Lanctot(2013)]{lanctot2013monte}
M.~Lanctot.
\newblock \emph{Monte Carlo sampling and regret minimization for equilibrium computation and decision-making in large extensive form games}.
\newblock University of Alberta (Canada), 2013.

\bibitem[Lanctot et~al.(2017)Lanctot, Zambaldi, Gruslys, Lazaridou, Tuyls, Perolat, Silver, and Graepel]{lanctot2017unified}
M.~Lanctot, V.~Zambaldi, A.~Gruslys, A.~Lazaridou, K.~Tuyls, J.~Perolat, D.~Silver, and T.~Graepel.
\newblock A unified game-theoretic approach to multiagent reinforcement learning, 2017.

\bibitem[Lasota et~al.(2017)Lasota, Fong, and Shah]{Lasota2017ASO}
P.~A. Lasota, T.~Fong, and J.~A. Shah.
\newblock A survey of methods for safe human-robot interaction.
\newblock \emph{Found. Trends Robotics}, 5:\penalty0 261--349, 2017.
\newblock URL \url{https://api.semanticscholar.org/CorpusID:207179253}.

\bibitem[Lauer and Riedmiller(2000)]{Lauer2000}
M.~Lauer and M.~A. Riedmiller.
\newblock An algorithm for distributed reinforcement learning in cooperative multi-agent systems.
\newblock page 535–542, 2000.

\bibitem[Lazaridou et~al.(2018)Lazaridou, Hermann, Tuyls, and Clark]{lazaridou2018emergence}
A.~Lazaridou, K.~M. Hermann, K.~Tuyls, and S.~Clark.
\newblock Emergence of linguistic communication from referential games with symbolic and pixel input.
\newblock \emph{arXiv preprint arXiv:1804.03984}, 2018.

\bibitem[Lechner et~al.(2023)Lechner, Yin, Seyde, Wang, Xiao, Hasani, Rountree, and Rus]{lechner2023gigastep}
M.~Lechner, L.~Yin, T.~Seyde, T.-H. Wang, W.~Xiao, R.~Hasani, J.~Rountree, and D.~Rus.
\newblock Gigastep - one billion steps per second multi-agent reinforcement learning.
\newblock In \emph{Advances in Neural Information Processing Systems}, 2023.
\newblock URL \url{https://openreview.net/forum?id=UgPAaEugH3}.

\bibitem[Levine et~al.(2020)Levine, Kumar, Tucker, and Fu]{Levine2020OfflineRL}
S.~Levine, A.~Kumar, G.~Tucker, and J.~Fu.
\newblock Offline reinforcement learning: Tutorial, review, and perspectives on open problems.
\newblock \emph{CoRR}, abs/2005.01643, 2020.
\newblock URL \url{https://arxiv.org/abs/2005.01643}.

\bibitem[Lewis(2008)]{lewis2008convention}
D.~Lewis.
\newblock \emph{Convention: A philosophical study}.
\newblock John Wiley \& Sons, 2008.

\bibitem[Li et~al.(2021)Li, Gupta, Morales, Allen, and Kochenderfer]{li2021deep}
S.~Li, J.~K. Gupta, P.~Morales, R.~Allen, and M.~J. Kochenderfer.
\newblock Deep implicit coordination graphs for multi-agent reinforcement learning, 2021.

\bibitem[Li et~al.(2022)Li, Xie, and Lu]{li2022dae}
Y.~Li, G.~Xie, and Z.~Lu.
\newblock Difference advantage estimation for multi-agent policy gradients.
\newblock 162:\penalty0 13066--13085, 17--23 Jul 2022.
\newblock URL \url{https://proceedings.mlr.press/v162/li22w.html}.

\bibitem[Lillicrap et~al.(2015)Lillicrap, Hunt, Pritzel, Heess, Erez, Tassa, Silver, and Wierstra]{lillicrap2015ddpg}
T.~P. Lillicrap, J.~J. Hunt, A.~Pritzel, N.~Heess, T.~Erez, Y.~Tassa, D.~Silver, and D.~Wierstra.
\newblock Continuous control with deep reinforcement learning.
\newblock \emph{arXiv preprint arXiv:1509.02971}, 2015.

\bibitem[Lin et~al.(2021)Lin, Huh, Stauffer, Lim, and Isola]{lin2021learning}
T.~Lin, J.~Huh, C.~Stauffer, S.~N. Lim, and P.~Isola.
\newblock Learning to ground multi-agent communication with autoencoders.
\newblock \emph{Advances in Neural Information Processing Systems}, 34:\penalty0 15230--15242, 2021.

\bibitem[Littman(1994)]{Littman1994Minimaxq}
M.~L. Littman.
\newblock Markov games as a framework for multi-agent reinforcement learning.
\newblock page 157–163, 1994.

\bibitem[Littman(2001{\natexlab{a}})]{Littman2001}
M.~L. Littman.
\newblock Value-function reinforcement learning in markov games.
\newblock \emph{Cognitive Systems Research}, 2\penalty0 (1):\penalty0 55--66, 2001{\natexlab{a}}.
\newblock ISSN 1389-0417.
\newblock \doi{https://doi.org/10.1016/S1389-0417(01)00015-8}.
\newblock URL \url{https://www.sciencedirect.com/science/article/pii/S1389041701000158}.

\bibitem[Littman(2001{\natexlab{b}})]{littman2001ffq}
M.~L. Littman.
\newblock Friend-or-foe q-learning in general-sum games.
\newblock page 322–328, 2001{\natexlab{b}}.

\bibitem[Liu et~al.(2023)Liu, Pu, Pan, Yi, Liang, and Zhang]{liu2023lazy}
B.~Liu, Z.~Pu, Y.~Pan, J.~Yi, Y.~Liang, and D.~Zhang.
\newblock Lazy agents: a new perspective on solving sparse reward problem in multi-agent reinforcement learning.
\newblock In \emph{International Conference on Machine Learning}, pages 21937--21950. PMLR, 2023.

\bibitem[Liu et~al.(2018)Liu, Feng, Mao, Zhou, Peng, and Liu]{liu2018actiondepedent}
H.~Liu, Y.~Feng, Y.~Mao, D.~Zhou, J.~Peng, and Q.~Liu.
\newblock Action-depedent control variates for policy optimization via stein's identity, 2018.

\bibitem[Liu et~al.(2021)Liu, Jain, Yeh, and Schwing]{liu2021cooperative}
I.-J. Liu, U.~Jain, R.~A. Yeh, and A.~G. Schwing.
\newblock Cooperative exploration for multi-agent deep reinforcement learning, 2021.

\bibitem[Loomes et~al.(2003)Loomes, Starmer, and Sugden]{loomes2003anomalies}
G.~Loomes, C.~Starmer, and R.~Sugden.
\newblock Do anomalies disappear in repeated markets?
\newblock \emph{The Economic Journal}, 113\penalty0 (486):\penalty0 C153--C166, 2003.

\bibitem[Lowe et~al.(2019)Lowe, Foerster, Boureau, Pineau, and Dauphin]{lowe2019pitfalls}
R.~Lowe, J.~Foerster, Y.-L. Boureau, J.~Pineau, and Y.~Dauphin.
\newblock On the pitfalls of measuring emergent communication, 2019.

\bibitem[Lowe et~al.(2020)Lowe, Wu, Tamar, Harb, Abbeel, and Mordatch]{lowe2020multiagent}
R.~Lowe, Y.~Wu, A.~Tamar, J.~Harb, P.~Abbeel, and I.~Mordatch.
\newblock Multi-agent actor-critic for mixed cooperative-competitive environments, 2020.

\bibitem[Luo et~al.(2021)Luo, Xu, Li, Tian, Darrell, and Ma]{luo2021algorithmic}
Y.~Luo, H.~Xu, Y.~Li, Y.~Tian, T.~Darrell, and T.~Ma.
\newblock Algorithmic framework for model-based deep reinforcement learning with theoretical guarantees, 2021.

\bibitem[Luong et~al.(2019)Luong, Hoang, Gong, Niyato, Wang, Liang, and Kim]{Luong2019Networking}
N.~C. Luong, D.~T. Hoang, S.~Gong, D.~Niyato, P.~Wang, Y.-C. Liang, and D.~I. Kim.
\newblock Applications of deep reinforcement learning in communications and networking: A survey.
\newblock \emph{IEEE Communications Surveys and Tutorials}, 21\penalty0 (4):\penalty0 3133--3174, 2019.
\newblock \doi{10.1109/COMST.2019.2916583}.

\bibitem[Lyu and Amato(2020)]{lyu2020likelihood}
X.~Lyu and C.~Amato.
\newblock Likelihood quantile networks for coordinating multi-agent reinforcement learning, 2020.

\bibitem[Ma et~al.(2021)Ma, Jayaraman, and Bastani]{Ma2021CODRL}
Y.~Ma, D.~Jayaraman, and O.~Bastani.
\newblock Conservative offline distributional reinforcement learning.
\newblock 34:\penalty0 19235--19247, 2021.
\newblock URL \url{https://proceedings.neurips.cc/paper_files/paper/2021/file/a05d886123a54de3ca4b0985b718fb9b-Paper.pdf}.

\bibitem[Mahajan et~al.(2020)Mahajan, Rashid, Samvelyan, and Whiteson]{mahajan2020maven}
A.~Mahajan, T.~Rashid, M.~Samvelyan, and S.~Whiteson.
\newblock Maven: Multi-agent variational exploration, 2020.

\bibitem[Makoviychuk et~al.(2021)Makoviychuk, Wawrzyniak, Guo, Lu, Storey, Macklin, Hoeller, Rudin, Allshire, Handa, and State]{makoviychuk2021isaac}
V.~Makoviychuk, L.~Wawrzyniak, Y.~Guo, M.~Lu, K.~Storey, M.~Macklin, D.~Hoeller, N.~Rudin, A.~Allshire, A.~Handa, and G.~State.
\newblock Isaac gym: High performance gpu-based physics simulation for robot learning, 2021.

\bibitem[Marsh and Onof(2008)]{marsh2008stigmergic}
L.~Marsh and C.~Onof.
\newblock Stigmergic epistemology, stigmergic cognition.
\newblock \emph{Cognitive Systems Research}, 9\penalty0 (1-2):\penalty0 136--149, 2008.

\bibitem[Matari{\'c}(1994)]{Matari1994LearningTB}
M.~J. Matari{\'c}.
\newblock Learning to behave socially.
\newblock 1994.

\bibitem[Matari{\'c}(1997)]{Matari1997ReinforcementLI}
M.~J. Matari{\'c}.
\newblock Reinforcement learning in the multi-robot domain.
\newblock \emph{Autonomous Robots}, 4:\penalty0 73--83, 1997.
\newblock URL \url{https://api.semanticscholar.org/CorpusID:14610547}.

\bibitem[Matignon et~al.(2007)Matignon, Laurent, and Le~Fort-Piat]{Matignon2007}
L.~Matignon, G.~J. Laurent, and N.~Le~Fort-Piat.
\newblock Hysteretic q-learning : an algorithm for decentralized reinforcement learning in cooperative multi-agent teams.
\newblock pages 64--69, 2007.
\newblock \doi{10.1109/IROS.2007.4399095}.

\bibitem[Matignon et~al.(2012)Matignon, Laurent, and Le~Fort-Piat]{matignon2012independent}
L.~Matignon, G.~J. Laurent, and N.~Le~Fort-Piat.
\newblock Independent reinforcement learners in cooperative markov games: a survey regarding coordination problems.
\newblock \emph{The Knowledge Engineering Review}, 27\penalty0 (1):\penalty0 1–31, 2012.
\newblock \doi{10.1017/S0269888912000057}.

\bibitem[Maynard~Smith(1976)]{maynard1976evolution}
J.~Maynard~Smith.
\newblock Evolution and the theory of games.
\newblock \emph{American scientist}, 64\penalty0 (1):\penalty0 41--45, 1976.

\bibitem[McAleer et~al.(2021)McAleer, Lanier, Wang, Baldi, and Fox]{mcaleer2021xdo}
S.~McAleer, J.~B. Lanier, K.~A. Wang, P.~Baldi, and R.~Fox.
\newblock Xdo: A double oracle algorithm for extensive-form games.
\newblock \emph{Advances in Neural Information Processing Systems}, 34:\penalty0 23128--23139, 2021.

\bibitem[McMahan et~al.(2003)McMahan, Gordon, and Blum]{mcmahan2003planning}
H.~B. McMahan, G.~J. Gordon, and A.~Blum.
\newblock Planning in the presence of cost functions controlled by an adversary.
\newblock In \emph{Proceedings of the 20th International Conference on Machine Learning (ICML-03)}, pages 536--543, 2003.

\bibitem[Melo and Veloso(2009)]{melo2009learning}
F.~S. Melo and M.~Veloso.
\newblock Learning of coordination: Exploiting sparse interactions in multiagent systems.
\newblock In \emph{Proceedings of The 8th International Conference on Autonomous Agents and Multiagent Systems-Volume 2}, pages 773--780. Citeseer, 2009.

\bibitem[Melo and Veloso(2011)]{melo2011decentralized}
F.~S. Melo and M.~Veloso.
\newblock Decentralized mdps with sparse interactions.
\newblock \emph{Artificial Intelligence}, 175\penalty0 (11):\penalty0 1757--1789, 2011.

\bibitem[Mirsky et~al.(2022)Mirsky, Carlucho, Rahman, Fosong, Macke, Sridharan, Stone, and Albrecht]{mirsky2022survey}
R.~Mirsky, I.~Carlucho, A.~Rahman, E.~Fosong, W.~Macke, M.~Sridharan, P.~Stone, and S.~V. Albrecht.
\newblock A survey of ad hoc teamwork research, 2022.

\bibitem[Mnih et~al.(2013)Mnih, Kavukcuoglu, Silver, Graves, Antonoglou, Wierstra, and Riedmiller]{mnih2013playing}
V.~Mnih, K.~Kavukcuoglu, D.~Silver, A.~Graves, I.~Antonoglou, D.~Wierstra, and M.~Riedmiller.
\newblock Playing atari with deep reinforcement learning.
\newblock \emph{arXiv preprint arXiv:1312.5602}, 2013.

\bibitem[Mnih et~al.(2016)Mnih, Badia, Mirza, Graves, Lillicrap, Harley, Silver, and Kavukcuoglu]{mnih2016asynchronous}
V.~Mnih, A.~P. Badia, M.~Mirza, A.~Graves, T.~P. Lillicrap, T.~Harley, D.~Silver, and K.~Kavukcuoglu.
\newblock Asynchronous methods for deep reinforcement learning, 2016.

\bibitem[Mokhtari et~al.(2019)Mokhtari, Ozdaglar, and Pattathil]{mokhtari2019unified}
A.~Mokhtari, A.~Ozdaglar, and S.~Pattathil.
\newblock A unified analysis of extra-gradient and optimistic gradient methods for saddle point problems: Proximal point approach, 2019.

\bibitem[Monaco and Sabarwal(2016)]{monaco2016games}
A.~J. Monaco and T.~Sabarwal.
\newblock Games with strategic complements and substitutes.
\newblock \emph{Economic Theory}, 62:\penalty0 65--91, 2016.

\bibitem[Myerson(1985)]{Myerson1985Intro}
R.~B. Myerson.
\newblock An introduction to game theory.
\newblock 1985.
\newblock URL \url{https://www.kellogg.northwestern.edu/research/math/papers/623.pdf}.

\bibitem[Nair et~al.(2005)Nair, Varakantham, Tambe, and Yokoo]{nair2005}
R.~Nair, P.~Varakantham, M.~Tambe, and M.~Yokoo.
\newblock Networked distributed pomdps: A synthesis of distributed constraint optimization and pomdps.
\newblock \emph{Proceedings of the National Conference on Artificial Intelligence}, 1:\penalty0 133--139, 01 2005.

\bibitem[Nash(1951)]{Nash1951NONCOOPERATIVEG}
J.~F. Nash.
\newblock Non-cooperative games.
\newblock \emph{Classics in Game Theory}, 1951.
\newblock URL \url{https://api.semanticscholar.org/CorpusID:264793164}.

\bibitem[Nashed and Zilberstein(2022)]{nashed2022survey}
S.~Nashed and S.~Zilberstein.
\newblock A survey of opponent modeling in adversarial domains.
\newblock \emph{Journal of Artificial Intelligence Research}, 73:\penalty0 277--327, 2022.

\bibitem[Ndousse et~al.(2021)Ndousse, Eck, Levine, and Jaques]{ndousse2021emergent}
K.~K. Ndousse, D.~Eck, S.~Levine, and N.~Jaques.
\newblock Emergent social learning via multi-agent reinforcement learning.
\newblock In \emph{International conference on machine learning}, pages 7991--8004. PMLR, 2021.

\bibitem[Nekoei et~al.(2023)Nekoei, Zhao, Rajendran, Liu, and Chandar]{nekoei2023towards}
H.~Nekoei, X.~Zhao, J.~Rajendran, M.~Liu, and S.~Chandar.
\newblock Towards few-shot coordination: Revisiting ad-hoc teamplay challenge in the game of hanabi.
\newblock In \emph{Conference on Lifelong Learning Agents}, pages 861--877. PMLR, 2023.

\bibitem[Ng et~al.(1999)Ng, Harada, and Russell]{ng1999policy}
A.~Y. Ng, D.~Harada, and S.~Russell.
\newblock Policy invariance under reward transformations: Theory and application to reward shaping.
\newblock In \emph{Icml}, volume~99, pages 278--287, 1999.

\bibitem[Nisan et~al.(2007)Nisan, Roughgarden, Tardos, and Vazirani]{Nisan2007AlgGT}
N.~Nisan, T.~Roughgarden, E.~Tardos, and V.~V. Vazirani.
\newblock Algorithmic game theory.
\newblock 2007.

\bibitem[Omidshafiei et~al.(2017)Omidshafiei, Pazis, Amato, How, and Vian]{omidshafiei2017deep}
S.~Omidshafiei, J.~Pazis, C.~Amato, J.~P. How, and J.~Vian.
\newblock Deep decentralized multi-task multi-agent reinforcement learning under partial observability, 2017.

\bibitem[Omidshafiei et~al.(2018)Omidshafiei, Kim, Liu, Tesauro, Riemer, Amato, Campbell, and How]{omidshafiei2018learning}
S.~Omidshafiei, D.-K. Kim, M.~Liu, G.~Tesauro, M.~Riemer, C.~Amato, M.~Campbell, and J.~P. How.
\newblock Learning to teach in cooperative multiagent reinforcement learning, 2018.

\bibitem[Orabona(2023)]{orabona2023modern}
F.~Orabona.
\newblock A modern introduction to online learning, 2023.

\bibitem[Osborne(2009)]{Osborne2003AnIT}
M.~J. Osborne.
\newblock An introduction to game theory.
\newblock 2009.

\bibitem[Palmer(2020)]{palmer2020thesis}
G.~Palmer.
\newblock \emph{Independent Learning Approaches: Overcoming Multi-Agent Learning Pathologies In Team-Games}.
\newblock PhD thesis, University of Liverpool, 2020.

\bibitem[Palmer et~al.(2018)Palmer, Tuyls, Bloembergen, and Savani]{palmer2018lenient}
G.~Palmer, K.~Tuyls, D.~Bloembergen, and R.~Savani.
\newblock Lenient multi-agent deep reinforcement learning, 2018.

\bibitem[Palmer et~al.(2019)Palmer, Savani, and Tuyls]{palmer2019negative}
G.~Palmer, R.~Savani, and K.~Tuyls.
\newblock Negative update intervals in deep multi-agent reinforcement learning, 2019.

\bibitem[Pan et~al.(2022)Pan, Huang, Ma, and Xu]{pan2022plan}
L.~Pan, L.~Huang, T.~Ma, and H.~Xu.
\newblock Plan better amid conservatism: Offline multi-agent reinforcement learning with actor rectification.
\newblock In \emph{International conference on machine learning}, pages 17221--17237. PMLR, 2022.

\bibitem[Panait and Luke(2005)]{Panait2005CooperativeML}
L.~Panait and S.~Luke.
\newblock Cooperative multi-agent learning: The state of the art.
\newblock \emph{Autonomous Agents and Multi-Agent Systems}, 11:\penalty0 387--434, 2005.
\newblock URL \url{https://api.semanticscholar.org/CorpusID:19706}.

\bibitem[Papoudakis et~al.(2019)Papoudakis, Christianos, Rahman, and Albrecht]{papoudakis2019dealing}
G.~Papoudakis, F.~Christianos, A.~Rahman, and S.~V. Albrecht.
\newblock Dealing with non-stationarity in multi-agent deep reinforcement learning, 2019.

\bibitem[Peng et~al.(2017)Peng, Wen, Yang, Yuan, Tang, Long, and Wang]{peng2017multiagent}
P.~Peng, Y.~Wen, Y.~Yang, Q.~Yuan, Z.~Tang, H.~Long, and J.~Wang.
\newblock Multiagent bidirectionally-coordinated nets: Emergence of human-level coordination in learning to play starcraft combat games.
\newblock \emph{arXiv preprint arXiv:1703.10069}, 2017.

\bibitem[Phan et~al.(2023)Phan, Ritz, Altmann, Zorn, Nüßlein, Kölle, Gabor, and Linnhoff-Popien]{phan2023attention}
T.~Phan, F.~Ritz, P.~Altmann, M.~Zorn, J.~Nüßlein, M.~Kölle, T.~Gabor, and C.~Linnhoff-Popien.
\newblock Attention-based recurrence for multi-agent reinforcement learning under stochastic partial observability, 2023.

\bibitem[Prashanth et~al.(2022)Prashanth, Fu, et~al.]{prashanth2022risk}
L.~Prashanth, M.~C. Fu, et~al.
\newblock Risk-sensitive reinforcement learning via policy gradient search.
\newblock \emph{Foundations and Trends{\textregistered} in Machine Learning}, 15\penalty0 (5):\penalty0 537--693, 2022.

\bibitem[Premack and Woodruff(1978)]{Premack1978tom}
D.~Premack and G.~Woodruff.
\newblock Does the chimpanzee have a theory of mind?
\newblock \emph{Behavioral and Brain Sciences}, 1\penalty0 (4):\penalty0 515–526, 1978.
\newblock \doi{10.1017/S0140525X00076512}.

\bibitem[Qu et~al.(2021)Qu, Wierman, and Li]{qu2021scalable}
G.~Qu, A.~Wierman, and N.~Li.
\newblock Scalable reinforcement learning for multi-agent networked systems, 2021.

\bibitem[Raileanu et~al.(2018)Raileanu, Denton, Szlam, and Fergus]{raileanu2018modeling}
R.~Raileanu, E.~Denton, A.~Szlam, and R.~Fergus.
\newblock Modeling others using oneself in multi-agent reinforcement learning.
\newblock In \emph{International conference on machine learning}, pages 4257--4266. PMLR, 2018.

\bibitem[Rakhlin and Sridharan(2013)]{rakhlin2013optimization}
A.~Rakhlin and K.~Sridharan.
\newblock Optimization, learning, and games with predictable sequences, 2013.

\bibitem[Rashid et~al.(2018)Rashid, Samvelyan, de~Witt, Farquhar, Foerster, and Whiteson]{rashid2018qmix}
T.~Rashid, M.~Samvelyan, C.~S. de~Witt, G.~Farquhar, J.~Foerster, and S.~Whiteson.
\newblock Qmix: Monotonic value function factorisation for deep multi-agent reinforcement learning, 2018.

\bibitem[Recht(2019)]{recht2019tour}
B.~Recht.
\newblock A tour of reinforcement learning: The view from continuous control.
\newblock \emph{Annual Review of Control, Robotics, and Autonomous Systems}, 2\penalty0 (1):\penalty0 253--279, 2019.

\bibitem[Resnick(2005)]{Resnick2005Thesis}
M.~Resnick.
\newblock Beyond the centralized mindset--explorations in massively-parallel microworlds.
\newblock 08 2005.

\bibitem[Robinson(1951)]{Robinson1951FicitiousPlay}
J.~J. Robinson.
\newblock An iterative method of solving a game.
\newblock \emph{Classics in Game Theory}, 1951.

\bibitem[Ross et~al.(2011)Ross, Gordon, and Bagnell]{ross2011reduction}
S.~Ross, G.~J. Gordon, and J.~A. Bagnell.
\newblock A reduction of imitation learning and structured prediction to no-regret online learning, 2011.

\bibitem[Rutherford et~al.(2023)Rutherford, Ellis, Gallici, Cook, Lupu, Ingvarsson, Willi, Khan, de~Witt, Souly, Bandyopadhyay, Samvelyan, Jiang, Lange, Whiteson, Lacerda, Hawes, Rocktaschel, Lu, and Foerster]{flair2023jaxmarl}
A.~Rutherford, B.~Ellis, M.~Gallici, J.~Cook, A.~Lupu, G.~Ingvarsson, T.~Willi, A.~Khan, C.~S. de~Witt, A.~Souly, S.~Bandyopadhyay, M.~Samvelyan, M.~Jiang, R.~T. Lange, S.~Whiteson, B.~Lacerda, N.~Hawes, T.~Rocktaschel, C.~Lu, and J.~N. Foerster.
\newblock Jaxmarl: Multi-agent rl environments in jax.
\newblock \emph{arXiv preprint arXiv:2311.10090}, 2023.

\bibitem[Salani{\'e}(2005)]{salanie2005economics}
B.~Salani{\'e}.
\newblock \emph{The economics of contracts: a primer}.
\newblock MIT press, 2005.

\bibitem[Samvelyan et~al.(2019)Samvelyan, Rashid, de~Witt, Farquhar, Nardelli, Rudner, Hung, Torr, Foerster, and Whiteson]{samvelyan19smac}
M.~Samvelyan, T.~Rashid, C.~S. de~Witt, G.~Farquhar, N.~Nardelli, T.~G.~J. Rudner, C.-M. Hung, P.~H.~S. Torr, J.~Foerster, and S.~Whiteson.
\newblock {The} {StarCraft} {Multi}-{Agent} {Challenge}.
\newblock \emph{CoRR}, abs/1902.04043, 2019.

\bibitem[Santos et~al.(2021)Santos, Ribeiro, Sardinha, and Melo]{Santos2021adhoc}
P.~Santos, J.~Ribeiro, A.~Sardinha, and F.~Melo.
\newblock Ad hoc teamwork in the presence of non-stationary teammates.
\newblock 09 2021.

\bibitem[Schaul et~al.(2015)Schaul, Quan, Antonoglou, and Silver]{schaul2015prioritized}
T.~Schaul, J.~Quan, I.~Antonoglou, and D.~Silver.
\newblock Prioritized experience replay.
\newblock \emph{arXiv preprint arXiv:1511.05952}, 2015.

\bibitem[Schaul et~al.(2016)Schaul, Quan, Antonoglou, and Silver]{schaul2016prioritized}
T.~Schaul, J.~Quan, I.~Antonoglou, and D.~Silver.
\newblock Prioritized experience replay, 2016.

\bibitem[Schulman et~al.(2017{\natexlab{a}})Schulman, Levine, Moritz, Jordan, and Abbeel]{schulman2017trust}
J.~Schulman, S.~Levine, P.~Moritz, M.~I. Jordan, and P.~Abbeel.
\newblock Trust region policy optimization, 2017{\natexlab{a}}.

\bibitem[Schulman et~al.(2017{\natexlab{b}})Schulman, Wolski, Dhariwal, Radford, and Klimov]{schulman2017proximal}
J.~Schulman, F.~Wolski, P.~Dhariwal, A.~Radford, and O.~Klimov.
\newblock Proximal policy optimization algorithms, 2017{\natexlab{b}}.

\bibitem[Schulman et~al.(2018)Schulman, Moritz, Levine, Jordan, and Abbeel]{schulman2018highdimensional}
J.~Schulman, P.~Moritz, S.~Levine, M.~Jordan, and P.~Abbeel.
\newblock High-dimensional continuous control using generalized advantage estimation, 2018.

\bibitem[Shah et~al.(2023{\natexlab{a}})Shah, Sridhar, Bhorkar, Hirose, and Levine]{shah2022gnm}
D.~Shah, A.~Sridhar, A.~Bhorkar, N.~Hirose, and S.~Levine.
\newblock {GNM: A General Navigation Model to Drive Any Robot}.
\newblock In \emph{International Conference on Robotics and Automation (ICRA)}, 2023{\natexlab{a}}.
\newblock URL \url{https://arxiv.org/abs/2210.03370}.

\bibitem[Shah et~al.(2023{\natexlab{b}})Shah, Sridhar, Dashora, Stachowicz, Black, Hirose, and Levine]{shah2023vint}
D.~Shah, A.~Sridhar, N.~Dashora, K.~Stachowicz, K.~Black, N.~Hirose, and S.~Levine.
\newblock Vi{NT}: A foundation model for visual navigation.
\newblock In \emph{7th Annual Conference on Robot Learning}, 2023{\natexlab{b}}.
\newblock URL \url{https://arxiv.org/abs/2306.14846}.

\bibitem[Shalev-Shwartz et~al.(2016)Shalev-Shwartz, Shammah, and Shashua]{shalevshwartz2016safe}
S.~Shalev-Shwartz, S.~Shammah, and A.~Shashua.
\newblock Safe, multi-agent, reinforcement learning for autonomous driving, 2016.

\bibitem[Shalev-Shwartz et~al.(2012)]{shalev2012online}
S.~Shalev-Shwartz et~al.
\newblock Online learning and online convex optimization.
\newblock \emph{Foundations and Trends{\textregistered} in Machine Learning}, 4\penalty0 (2):\penalty0 107--194, 2012.

\bibitem[Shao et~al.(2022)Shao, Lou, Zhang, Jiang, He, and Ji]{shao2022self}
J.~Shao, Z.~Lou, H.~Zhang, Y.~Jiang, S.~He, and X.~Ji.
\newblock Self-organized group for cooperative multi-agent reinforcement learning.
\newblock \emph{Advances in Neural Information Processing Systems}, 35:\penalty0 5711--5723, 2022.

\bibitem[Shapley(1952)]{Shapley1952}
L.~S. Shapley.
\newblock A value for n-person games.
\newblock 1952.
\newblock \doi{10.7249/P0295}.

\bibitem[Shapley(1953)]{Shapley1953StochasticGame}
L.~S. Shapley.
\newblock Stochastic games*.
\newblock \emph{Proceedings of the National Academy of Sciences}, 39\penalty0 (10):\penalty0 1095--1100, 1953.
\newblock \doi{10.1073/pnas.39.10.1095}.
\newblock URL \url{https://www.pnas.org/doi/abs/10.1073/pnas.39.10.1095}.

\bibitem[Shoham and Leyton-Brown(2008)]{Shoham2008MAS}
Y.~Shoham and K.~Leyton-Brown.
\newblock Multiagent systems: Algorithmic, game-theoretic, and logical foundations.
\newblock 2008.

\bibitem[Shoham et~al.(2003)Shoham, Powers, and Grenager]{Shoham2003MultiAgentRL}
Y.~Shoham, R.~Powers, and T.~Grenager.
\newblock Multi-agent reinforcement learning:a critical survey.
\newblock 2003.

\bibitem[Shoham et~al.(2007)Shoham, Powers, and Grenager]{Shoham2007MultiagentLearning}
Y.~Shoham, R.~Powers, and T.~Grenager.
\newblock If multi-agent learning is the answer, what is the question?
\newblock \emph{Artificial Intelligence}, 171:\penalty0 365--377, 05 2007.
\newblock \doi{10.1016/j.artint.2006.02.006}.

\bibitem[Silva and Costa(2019)]{silva2018transfer}
F.~Silva and A.~Costa.
\newblock A survey on transfer learning for multiagent reinforcement learning systems.
\newblock \emph{Journal of Artificial Intelligence Research}, 64, 03 2019.
\newblock \doi{10.1613/jair.1.11396}.

\bibitem[Silver et~al.(2014)Silver, Lever, Heess, Degris, Wierstra, and Riedmiller]{silver2014dpg}
D.~Silver, G.~Lever, N.~Heess, T.~Degris, D.~Wierstra, and M.~Riedmiller.
\newblock Deterministic policy gradient algorithms.
\newblock 32\penalty0 (1):\penalty0 387--395, 22--24 Jun 2014.
\newblock URL \url{https://proceedings.mlr.press/v32/silver14.html}.

\bibitem[Singh et~al.(2018)Singh, Jain, and Sukhbaatar]{singh2018learning}
A.~Singh, T.~Jain, and S.~Sukhbaatar.
\newblock Learning when to communicate at scale in multiagent cooperative and competitive tasks, 2018.

\bibitem[Singh et~al.(2000)Singh, Kearns, and Mansour]{Singh2000iga}
S.~Singh, M.~Kearns, and Y.~Mansour.
\newblock Nash convergence of gradient dynamics in general-sum games.
\newblock page 541–548, 2000.

\bibitem[Spanoudakis and Moraitis(2010)]{spanoudakis2010using}
N.~Spanoudakis and P.~Moraitis.
\newblock Using aseme methodology for model-driven agent systems development.
\newblock In \emph{International workshop on agent-oriented software engineering}, pages 106--127. Springer, 2010.

\bibitem[Stone and Veloso(2000)]{Stone2000MAS}
P.~Stone and M.~Veloso.
\newblock Multiagent systems: A survey from a machine learning perspective.
\newblock \emph{Autonomous Robots}, 8, 05 2000.
\newblock \doi{10.1023/A:1008942012299}.

\bibitem[Stone et~al.(2010)Stone, Kaminka, Kraus, and Rosenschein]{Stone2010adhoc}
P.~Stone, G.~A. Kaminka, S.~Kraus, and J.~S. Rosenschein.
\newblock Ad hoc autonomous agent teams: Collaboration without pre-coordination.
\newblock page 1504–1509, 2010.

\bibitem[Sukhbaatar et~al.(2016)Sukhbaatar, Szlam, and Fergus]{sukhbaatar2016learning}
S.~Sukhbaatar, A.~Szlam, and R.~Fergus.
\newblock Learning multiagent communication with backpropagation, 2016.

\bibitem[Sun et~al.(2021)Sun, Lee, and Lee]{sun2021dfac}
W.-F. Sun, C.-K. Lee, and C.-Y. Lee.
\newblock Dfac framework: Factorizing the value function via quantile mixture for multi-agent distributional q-learning, 2021.

\bibitem[Sunehag et~al.(2017)Sunehag, Lever, Gruslys, Czarnecki, Zambaldi, Jaderberg, Lanctot, Sonnerat, Leibo, Tuyls, and Graepel]{sunehag2017valuedecomposition}
P.~Sunehag, G.~Lever, A.~Gruslys, W.~M. Czarnecki, V.~Zambaldi, M.~Jaderberg, M.~Lanctot, N.~Sonnerat, J.~Z. Leibo, K.~Tuyls, and T.~Graepel.
\newblock Value-decomposition networks for cooperative multi-agent learning, 2017.

\bibitem[Sutton(1990)]{sutton1990dyna}
R.~S. Sutton.
\newblock Integrated architectures for learning, planning, and reacting based on approximating dynamic programming.
\newblock pages 216--224. Elsevier, 1990.

\bibitem[Sutton and Barto(2018)]{Sutton2018RL}
R.~S. Sutton and A.~G. Barto.
\newblock Reinforcement learning: An introduction.
\newblock 2018.

\bibitem[Sutton et~al.(1999{\natexlab{a}})Sutton, McAllester, Singh, and Mansour]{Sutton1999AC}
R.~S. Sutton, D.~McAllester, S.~Singh, and Y.~Mansour.
\newblock Policy gradient methods for reinforcement learning with function approximation.
\newblock 12, 1999{\natexlab{a}}.
\newblock URL \url{https://proceedings.neurips.cc/paper_files/paper/1999/file/464d828b85b0bed98e80ade0a5c43b0f-Paper.pdf}.

\bibitem[Sutton et~al.(1999{\natexlab{b}})Sutton, Precup, and Singh]{sutton1999between}
R.~S. Sutton, D.~Precup, and S.~Singh.
\newblock Between mdps and semi-mdps: A framework for temporal abstraction in reinforcement learning.
\newblock \emph{Artificial intelligence}, 112\penalty0 (1-2):\penalty0 181--211, 1999{\natexlab{b}}.

\bibitem[Tammelin(2014)]{tammelin2014solving}
O.~Tammelin.
\newblock Solving large imperfect information games using cfr+, 2014.

\bibitem[Tampuu et~al.(2015)Tampuu, Matiisen, Kodelja, Kuzovkin, Korjus, Aru, Aru, and Vicente]{tampuu2015multiagent}
A.~Tampuu, T.~Matiisen, D.~Kodelja, I.~Kuzovkin, K.~Korjus, J.~Aru, J.~Aru, and R.~Vicente.
\newblock Multiagent cooperation and competition with deep reinforcement learning, 2015.

\bibitem[Tan(1997)]{Tan1997independent}
M.~Tan.
\newblock Multi-agent reinforcement learning: Independent vs. cooperative agents, 1997.

\bibitem[Tchappi et~al.(2018)Tchappi, Galland, Kamla, and Kamgang]{tchappi2018brief}
I.~H. Tchappi, S.~Galland, V.~C. Kamla, and J.~C. Kamgang.
\newblock A brief review of holonic multi-agent models for traffic and transportation systems.
\newblock \emph{Procedia computer science}, 134:\penalty0 137--144, 2018.

\bibitem[Terry et~al.(2021)Terry, Black, Grammel, Jayakumar, Hari, Sullivan, Santos, Dieffendahl, Horsch, Perez-Vicente, et~al.]{terry2021pettingzoo}
J.~Terry, B.~Black, N.~Grammel, M.~Jayakumar, A.~Hari, R.~Sullivan, L.~S. Santos, C.~Dieffendahl, C.~Horsch, R.~Perez-Vicente, et~al.
\newblock Pettingzoo: Gym for multi-agent reinforcement learning.
\newblock \emph{Advances in Neural Information Processing Systems}, 34:\penalty0 15032--15043, 2021.

\bibitem[Theraulaz and Bonabeau(1999)]{theraulaz1999brief}
G.~Theraulaz and E.~Bonabeau.
\newblock A brief history of stigmergy.
\newblock \emph{Artificial life}, 5\penalty0 (2):\penalty0 97--116, 1999.

\bibitem[Tian et~al.(2023)Tian, Kuang, Liu, and Wang]{tian2023learning}
Q.~Tian, K.~Kuang, F.~Liu, and B.~Wang.
\newblock Learning from good trajectories in offline multi-agent reinforcement learning.
\newblock In \emph{Proceedings of the AAAI Conference on Artificial Intelligence}, volume~37, pages 11672--11680, 2023.

\bibitem[Touati and Ollivier(2021)]{touati2021learning}
A.~Touati and Y.~Ollivier.
\newblock Learning one representation to optimize all rewards, 2021.

\bibitem[Treutlein et~al.(2021)Treutlein, Dennis, Oesterheld, and Foerster]{treutlein2021new}
J.~Treutlein, M.~Dennis, C.~Oesterheld, and J.~Foerster.
\newblock A new formalism, method and open issues for zero-shot coordination.
\newblock In \emph{International Conference on Machine Learning}, pages 10413--10423. PMLR, 2021.

\bibitem[Tseng et~al.(2022)Tseng, Wang, Lin, and Isola]{tseng2022offline}
W.-C. Tseng, T.-H.~J. Wang, Y.-C. Lin, and P.~Isola.
\newblock Offline multi-agent reinforcement learning with knowledge distillation.
\newblock \emph{Advances in Neural Information Processing Systems}, 35:\penalty0 226--237, 2022.

\bibitem[Tucker et~al.(2018)Tucker, Bhupatiraju, Gu, Turner, Ghahramani, and Levine]{tucker2018mirage}
G.~Tucker, S.~Bhupatiraju, S.~Gu, R.~E. Turner, Z.~Ghahramani, and S.~Levine.
\newblock The mirage of action-dependent baselines in reinforcement learning, 2018.

\bibitem[Tumer and Agogino(2007)]{tumer2007distributed}
K.~Tumer and A.~Agogino.
\newblock Distributed agent-based air traffic flow management.
\newblock In \emph{Proceedings of the 6th international joint conference on Autonomous agents and multiagent systems}, pages 1--8, 2007.

\bibitem[Tuyls and Weiss(2012)]{Tuyls2012MAL}
K.~Tuyls and G.~Weiss.
\newblock Multiagent learning: Basics, challenges, and prospects.
\newblock \emph{Ai Magazine}, 33:\penalty0 41--52, 12 2012.
\newblock \doi{10.1609/aimag.v33i3.2426}.

\bibitem[Tuyls et~al.(2006)Tuyls, Hoen, and Vanschoenwinkel]{tuyls2006evolutionary}
K.~Tuyls, P.~J.~T. Hoen, and B.~Vanschoenwinkel.
\newblock An evolutionary dynamical analysis of multi-agent learning in iterated games.
\newblock \emph{Autonomous Agents and Multi-Agent Systems}, 12:\penalty0 115--153, 2006.

\bibitem[Tversky and Kahneman(1992)]{tversky1992advances}
A.~Tversky and D.~Kahneman.
\newblock Advances in prospect theory: Cumulative representation of uncertainty.
\newblock \emph{Journal of Risk and uncertainty}, 5:\penalty0 297--323, 1992.

\bibitem[Van~Hasselt et~al.(2016)Van~Hasselt, Guez, and Silver]{van2016deep}
H.~Van~Hasselt, A.~Guez, and D.~Silver.
\newblock Deep reinforcement learning with double q-learning.
\newblock 30\penalty0 (1), 2016.

\bibitem[van Hasselt et~al.(2018)van Hasselt, Doron, Strub, Hessel, Sonnerat, and Modayil]{vanhasselt2018deep}
H.~van Hasselt, Y.~Doron, F.~Strub, M.~Hessel, N.~Sonnerat, and J.~Modayil.
\newblock Deep reinforcement learning and the deadly triad, 2018.

\bibitem[Vlassis et~al.(2004)Vlassis, Elhorst, and Kok]{Vlassis2004Maxplus}
N.~Vlassis, R.~Elhorst, and J.~Kok.
\newblock Anytime algorithms for multiagent decision making using coordination graphs.
\newblock 1:\penalty0 953--957 vol.1, 2004.
\newblock \doi{10.1109/ICSMC.2004.1398426}.

\bibitem[Walke et~al.(2023)Walke, Black, Lee, Kim, Du, Zheng, Zhao, Hansen-Estruch, Vuong, He, Myers, Fang, Finn, and Levine]{walke2023bridgedata}
H.~Walke, K.~Black, A.~Lee, M.~J. Kim, M.~Du, C.~Zheng, T.~Zhao, P.~Hansen-Estruch, Q.~Vuong, A.~He, V.~Myers, K.~Fang, C.~Finn, and S.~Levine.
\newblock Bridgedata v2: A dataset for robot learning at scale.
\newblock In \emph{Conference on Robot Learning (CoRL)}, 2023.

\bibitem[Wang et~al.(2022)Wang, Zhang, Hu, Wang, Zhang, Gao, Hao, Lv, and Fan]{wang2022individual}
L.~Wang, Y.~Zhang, Y.~Hu, W.~Wang, C.~Zhang, Y.~Gao, J.~Hao, T.~Lv, and C.~Fan.
\newblock Individual reward assisted multi-agent reinforcement learning.
\newblock In \emph{International Conference on Machine Learning}, pages 23417--23432. PMLR, 2022.

\bibitem[Wang et~al.(2020{\natexlab{a}})Wang, Gupta, Mahajan, Peng, Whiteson, and Zhang]{wang2020rode}
T.~Wang, T.~Gupta, A.~Mahajan, B.~Peng, S.~Whiteson, and C.~Zhang.
\newblock Rode: Learning roles to decompose multi-agent tasks.
\newblock \emph{arXiv preprint arXiv:2010.01523}, 2020{\natexlab{a}}.

\bibitem[Wang et~al.(2020{\natexlab{b}})Wang, Han, Wang, Dong, and Zhang]{wang2020offpolicy}
Y.~Wang, B.~Han, T.~Wang, H.~Dong, and C.~Zhang.
\newblock Off-policy multi-agent decomposed policy gradients, 2020{\natexlab{b}}.

\bibitem[Wang et~al.(2016{\natexlab{a}})Wang, Bapst, Heess, Mnih, Munos, Kavukcuoglu, and de~Freitas]{wang2016sample}
Z.~Wang, V.~Bapst, N.~Heess, V.~Mnih, R.~Munos, K.~Kavukcuoglu, and N.~de~Freitas.
\newblock Sample efficient actor-critic with experience replay.
\newblock \emph{arXiv preprint arXiv:1611.01224}, 2016{\natexlab{a}}.

\bibitem[Wang et~al.(2016{\natexlab{b}})Wang, Schaul, Hessel, Hasselt, Lanctot, and Freitas]{wang2016dueling}
Z.~Wang, T.~Schaul, M.~Hessel, H.~Hasselt, M.~Lanctot, and N.~Freitas.
\newblock Dueling network architectures for deep reinforcement learning.
\newblock pages 1995--2003, 2016{\natexlab{b}}.

\bibitem[Watkins(1989)]{Watkins1989Qlearning}
C.~Watkins.
\newblock Learning from delayed rewards.
\newblock 01 1989.

\bibitem[Watkins and Dayan(1992)]{watkins1992q}
C.~J. Watkins and P.~Dayan.
\newblock Q-learning.
\newblock \emph{Machine learning}, 8:\penalty0 279--292, 1992.

\bibitem[Watter et~al.(2015)Watter, Springenberg, Boedecker, and Riedmiller]{Watter2015Embed}
M.~Watter, J.~T. Springenberg, J.~Boedecker, and M.~A. Riedmiller.
\newblock Embed to control: {A} locally linear latent dynamics model for control from raw images.
\newblock \emph{CoRR}, abs/1506.07365, 2015.
\newblock URL \url{http://arxiv.org/abs/1506.07365}.

\bibitem[Weber et~al.(2018)Weber, Racanière, Reichert, Buesing, Guez, Rezende, Badia, Vinyals, Heess, Li, Pascanu, Battaglia, Hassabis, Silver, and Wierstra]{weber2018imaginationaugmented}
T.~Weber, S.~Racanière, D.~P. Reichert, L.~Buesing, A.~Guez, D.~J. Rezende, A.~P. Badia, O.~Vinyals, N.~Heess, Y.~Li, R.~Pascanu, P.~Battaglia, D.~Hassabis, D.~Silver, and D.~Wierstra.
\newblock Imagination-augmented agents for deep reinforcement learning, 2018.

\bibitem[Wei et~al.(2021)Wei, Lee, Zhang, and Luo]{wei2021linear}
C.-Y. Wei, C.-W. Lee, M.~Zhang, and H.~Luo.
\newblock Linear last-iterate convergence in constrained saddle-point optimization, 2021.

\bibitem[Wei and Luke(2016)]{Wei2016lenient}
E.~Wei and S.~Luke.
\newblock Lenient learning in independent-learner stochastic cooperative games.
\newblock \emph{Journal of Machine Learning Research}, 17\penalty0 (84):\penalty0 1--42, 2016.
\newblock URL \url{http://jmlr.org/papers/v17/15-417.html}.

\bibitem[Weinberg and Rosenschein(2004)]{Weinberg2004}
M.~Weinberg and J.~Rosenschein.
\newblock Best-response multiagent learning in non-stationary environments.
\newblock pages 506--513, 2004.

\bibitem[Whiteson(2020)]{whiteson2020talk}
S.~Whiteson.
\newblock Factored value functions for cooperative multi-agent reinforcement learning.
\newblock \emph{MIT Embodied Intelligence Seminars}, 2020.

\bibitem[Williams(1988)]{Williams1988Reinforce}
R.~J. Williams.
\newblock Toward a theory of reinforcement-learning connectionist systems.
\newblock Technical Report NU-CCS-88-3, Northeastern University, College of Computer Science, 1988.

\bibitem[Williams(1992)]{Williams1992PG}
R.~J. Williams.
\newblock Simple statistical gradient-following algorithms for connectionist reinforcement learning.
\newblock \emph{Mach. Learn.}, 8\penalty0 (3–4):\penalty0 229–256, may 1992.
\newblock ISSN 0885-6125.
\newblock \doi{10.1007/BF00992696}.
\newblock URL \url{https://doi.org/10.1007/BF00992696}.

\bibitem[Wilson(2000)]{wilson2000sociobiology}
E.~O. Wilson.
\newblock \emph{Sociobiology}.
\newblock Harvard University Press, 2000.

\bibitem[Wolpert and Tumer(2001)]{wolpert2001optimal}
D.~H. Wolpert and K.~Tumer.
\newblock Optimal payoff functions for members of collectives.
\newblock \emph{Advances in Complex Systems}, 4\penalty0 (02n03):\penalty0 265--279, 2001.

\bibitem[Wu et~al.(2017)Wu, Mansimov, Liao, Grosse, and Ba]{wu2017scalable}
Y.~Wu, E.~Mansimov, S.~Liao, R.~Grosse, and J.~Ba.
\newblock Scalable trust-region method for deep reinforcement learning using kronecker-factored approximation, 2017.

\bibitem[Xiao et~al.(2022)Xiao, Tan, and Amato]{xiao2022asynchronous}
Y.~Xiao, W.~Tan, and C.~Amato.
\newblock Asynchronous actor-critic for multi-agent reinforcement learning.
\newblock \emph{Advances in Neural Information Processing Systems}, 35:\penalty0 4385--4400, 2022.

\bibitem[Xu et~al.(2018)Xu, Li, Tian, Sonobe, ichi Kawarabayashi, and Jegelka]{xu2018representation}
K.~Xu, C.~Li, Y.~Tian, T.~Sonobe, K.~ichi Kawarabayashi, and S.~Jegelka.
\newblock Representation learning on graphs with jumping knowledge networks, 2018.

\bibitem[Yang et~al.(2019)Yang, Zhao, Lin, Qin, Bian, and Liu]{yang2019fully}
D.~Yang, L.~Zhao, Z.~Lin, T.~Qin, J.~Bian, and T.-Y. Liu.
\newblock Fully parameterized quantile function for distributional reinforcement learning.
\newblock \emph{Advances in neural information processing systems}, 32:\penalty0 6193--6202, 2019.

\bibitem[Yang et~al.(2022)Yang, Zhao, Hu, Zhou, Zhu, and Li]{yang2022ldsa}
M.~Yang, J.~Zhao, X.~Hu, W.~Zhou, J.~Zhu, and H.~Li.
\newblock Ldsa: Learning dynamic subtask assignment in cooperative multi-agent reinforcement learning.
\newblock \emph{Advances in Neural Information Processing Systems}, 35:\penalty0 1698--1710, 2022.

\bibitem[Yang et~al.(2021{\natexlab{a}})Yang, Wang, Tang, Hao, Meng, Mao, Li, Liu, Chen, Hu, et~al.]{yang2021efficient}
T.~Yang, W.~Wang, H.~Tang, J.~Hao, Z.~Meng, H.~Mao, D.~Li, W.~Liu, Y.~Chen, Y.~Hu, et~al.
\newblock An efficient transfer learning framework for multiagent reinforcement learning.
\newblock \emph{Advances in neural information processing systems}, 34:\penalty0 17037--17048, 2021{\natexlab{a}}.

\bibitem[Yang et~al.(2021{\natexlab{b}})Yang, Ma, Li, Zheng, Zhang, Huang, Yang, and Zhao]{yang2021believe}
Y.~Yang, X.~Ma, C.~Li, Z.~Zheng, Q.~Zhang, G.~Huang, J.~Yang, and Q.~Zhao.
\newblock Believe what you see: Implicit constraint approach for offline multi-agent reinforcement learning.
\newblock \emph{Advances in Neural Information Processing Systems}, 34:\penalty0 10299--10312, 2021{\natexlab{b}}.

\bibitem[Yu et~al.(2022{\natexlab{a}})Yu, Velu, Vinitsky, Gao, Wang, Bayen, and Wu]{yu2022surprising}
C.~Yu, A.~Velu, E.~Vinitsky, J.~Gao, Y.~Wang, A.~Bayen, and Y.~Wu.
\newblock The surprising effectiveness of ppo in cooperative, multi-agent games, 2022{\natexlab{a}}.

\bibitem[Yu et~al.(2023)Yu, Gao, Liu, Xu, Tang, Yang, Wang, and Wu]{yu2023learning}
C.~Yu, J.~Gao, W.~Liu, B.~Xu, H.~Tang, J.~Yang, Y.~Wang, and Y.~Wu.
\newblock Learning zero-shot cooperation with humans, assuming humans are biased.
\newblock \emph{arXiv preprint arXiv:2302.01605}, 2023.

\bibitem[Yu et~al.(2022{\natexlab{b}})Yu, Kumar, Rafailov, Rajeswaran, Levine, and Finn]{yu2022combo}
T.~Yu, A.~Kumar, R.~Rafailov, A.~Rajeswaran, S.~Levine, and C.~Finn.
\newblock Combo: Conservative offline model-based policy optimization, 2022{\natexlab{b}}.

\bibitem[Yu et~al.(2022{\natexlab{c}})Yu, Jiang, Zhang, Jiang, and Lu]{yu2022model}
X.~Yu, J.~Jiang, W.~Zhang, H.~Jiang, and Z.~Lu.
\newblock Model-based opponent modeling.
\newblock \emph{Advances in Neural Information Processing Systems}, 35:\penalty0 28208--28221, 2022{\natexlab{c}}.

\bibitem[Zaïem and Bennequin(2019)]{zaïem2019learning}
M.~S. Zaïem and E.~Bennequin.
\newblock Learning to communicate in multi-agent reinforcement learning : A review, 2019.

\bibitem[Zhang and Shah(2014)]{Zhang2014}
C.~Zhang and J.~A. Shah.
\newblock Fairness in multi-agent sequential decision-making.
\newblock 27, 2014.
\newblock URL \url{https://proceedings.neurips.cc/paper_files/paper/2014/file/792c7b5aae4a79e78aaeda80516ae2ac-Paper.pdf}.

\bibitem[Zhang et~al.(2018)Zhang, Yang, Liu, Zhang, and Başar]{zhang2018fully}
K.~Zhang, Z.~Yang, H.~Liu, T.~Zhang, and T.~Başar.
\newblock Fully decentralized multi-agent reinforcement learning with networked agents, 2018.

\bibitem[Zhang et~al.(2019)Zhang, Vikram, Smith, Abbeel, Johnson, and Levine]{zhang2019solar}
M.~Zhang, S.~Vikram, L.~Smith, P.~Abbeel, M.~Johnson, and S.~Levine.
\newblock Solar: Deep structured representations for model-based reinforcement learning.
\newblock pages 7444--7453, 2019.

\bibitem[Zhang et~al.(2021)Zhang, Yang, An, and Zhang]{Zhang2021Coordinate}
Y.~Zhang, Q.~Yang, D.~An, and C.~Zhang.
\newblock Coordination between individual agents in multi-agent reinforcement learning.
\newblock \emph{Proceedings of the AAAI Conference on Artificial Intelligence}, 35\penalty0 (13):\penalty0 11387--11394, May 2021.
\newblock \doi{10.1609/aaai.v35i13.17357}.
\newblock URL \url{https://ojs.aaai.org/index.php/AAAI/article/view/17357}.

\bibitem[Zhao et~al.(2022)Zhao, Yang, Zhao, Hu, Zhou, Zhu, and Li]{zhao2022mcmarl}
J.~Zhao, M.~Yang, Y.~Zhao, X.~Hu, W.~Zhou, J.~Zhu, and H.~Li.
\newblock Mcmarl: Parameterizing value function via mixture of categorical distributions for multi-agent reinforcement learning, 2022.

\bibitem[Zheng et~al.(2017)Zheng, Yang, Cai, Zhang, Wang, and Yu]{zheng2017magent}
L.~Zheng, J.~Yang, H.~Cai, W.~Zhang, J.~Wang, and Y.~Yu.
\newblock Magent: A many-agent reinforcement learning platform for artificial collective intelligence, 2017.

\bibitem[Zheng et~al.(2021)Zheng, Chen, Wang, He, Hu, Chen, Fan, Gao, and Zhang]{zheng2021episodic}
L.~Zheng, J.~Chen, J.~Wang, J.~He, Y.~Hu, Y.~Chen, C.~Fan, Y.~Gao, and C.~Zhang.
\newblock Episodic multi-agent reinforcement learning with curiosity-driven exploration, 2021.

\bibitem[Zheng et~al.(2018)Zheng, Hao, and Zhang]{zheng2018weighted}
Y.~Zheng, J.~Hao, and Z.~Zhang.
\newblock Weighted double deep multiagent reinforcement learning in stochastic cooperative environments, 2018.

\bibitem[Zhou et~al.(2020)Zhou, Luo, Villella, Yang, Rusu, Miao, Zhang, Alban, Fadakar, Chen, Huang, Wen, Hassanzadeh, Graves, Chen, Zhu, Nguyen, Elsayed, Shao, Ahilan, Zhang, Wu, Fu, Rezaee, Yadmellat, Rohani, Nieves, Ni, Banijamali, Rivers, Tian, Palenicek, bou Ammar, Zhang, Liu, Hao, and Wang]{zhou2020smarts}
M.~Zhou, J.~Luo, J.~Villella, Y.~Yang, D.~Rusu, J.~Miao, W.~Zhang, M.~Alban, I.~Fadakar, Z.~Chen, A.~C. Huang, Y.~Wen, K.~Hassanzadeh, D.~Graves, D.~Chen, Z.~Zhu, N.~Nguyen, M.~Elsayed, K.~Shao, S.~Ahilan, B.~Zhang, J.~Wu, Z.~Fu, K.~Rezaee, P.~Yadmellat, M.~Rohani, N.~P. Nieves, Y.~Ni, S.~Banijamali, A.~C. Rivers, Z.~Tian, D.~Palenicek, H.~bou Ammar, H.~Zhang, W.~Liu, J.~Hao, and J.~Wang.
\newblock Smarts: Scalable multi-agent reinforcement learning training school for autonomous driving, 2020.

\bibitem[Zhu et~al.(2024)Zhu, Dastani, and Wang]{zhu2024survey}
C.~Zhu, M.~Dastani, and S.~Wang.
\newblock A survey of multi-agent deep reinforcement learning with communication.
\newblock \emph{Autonomous Agents and Multi-Agent Systems}, 38\penalty0 (1):\penalty0 4, 2024.

\bibitem[Zhu et~al.(2020)Zhu, Biyik, and Sadigh]{Zhu_2020}
Z.~Zhu, E.~Biyik, and D.~Sadigh.
\newblock Multi-agent safe planning with gaussian processes.
\newblock oct 2020.
\newblock \doi{10.1109/iros45743.2020.9341169}.
\newblock URL \url{https://doi.org/10.1109%2Firos45743.2020.9341169}.

\bibitem[Zinkevich(2003)]{Zinkevich2003giga}
M.~Zinkevich.
\newblock Online convex programming and generalized infinitesimal gradient ascent.
\newblock page 928–935, 2003.

\bibitem[Zinkevich et~al.(2007)Zinkevich, Johanson, Bowling, and Piccione]{Zinkevich2007regret}
M.~Zinkevich, M.~Johanson, M.~Bowling, and C.~Piccione.
\newblock Regret minimization in games with incomplete information.
\newblock 20, 2007.
\newblock URL \url{https://proceedings.neurips.cc/paper_files/paper/2007/file/08d98638c6fcd194a4b1e6992063e944-Paper.pdf}.

\bibitem[Zrnic et~al.(2022)Zrnic, Mazumdar, Sastry, and Jordan]{zrnic2022leads}
T.~Zrnic, E.~Mazumdar, S.~S. Sastry, and M.~I. Jordan.
\newblock Who leads and who follows in strategic classification?, 2022.

\end{thebibliography}

\end{document}